%% file: cta_mc_paper.tex
\makeatletter
\newif\iffinalversion\finalversionfalse
\newif\ifpreprintversion\preprintversionfalse
\newif\ifreviewversion\reviewversionfalse
\newif\ifwithtoc\withtocfalse
\newif\ifwithlinenumbers\withlinenumbersfalse
\makeatother

\finalversiontrue

\iffinalversion

  \documentclass[final,3p,times,twocolumn]{elsarticle} 
  \withtocfalse          
  \withlinenumbersfalse  
  \long\def\xblue#1{}   

\else

  \ifreviewversion
     \documentclass[preprint,review,12pt]{elsarticle}
     
     \long\def\xblue#1{}
     
     \withtocfalse
  \else
     \documentclass[preprint,12pt]{elsarticle}
     \withtoctrue
  \fi
  \ifpreprintversion
    \withlinenumbersfalse
    
    \long\def\xblue#1{}
    
  \else 
    \withlinenumberstrue
    
    \long\def\xblue#1{{\color{blue}{#1}}}
    
  \fi

\fi

\usepackage{natbib}
\usepackage{graphicx}
\usepackage{amssymb}
\usepackage{amsmath}
\usepackage{amsthm}
\usepackage{eurosym}
\usepackage{hyperref}
\usepackage{url}
\usepackage{aas_macros}
\usepackage{tablefootnote}

\newcommand{\mut}{\mu\mathrm{T}}
\newcommand{\rpsf}{R_{\mathrm{psf}}}
\newcommand{\rpix}{D_{\mathrm{pix}}}
\newcommand{\rnsb}{R_{\mathrm{nsb}}}
\newcommand{\snsb}{\Sigma_{\mathrm{nsb}}}
\newcommand{\aopt}{A_{\mathrm{opt}}}
\newcommand{\sspe}{\sigma_{\mathrm{spe}}}

\newcommand{\tth}{T_{\mathrm{th}}}

\newcommand{\bs}[1]{\boldsymbol{#1}}

\newcommand{\ntel}{\mathrm{N}_{\mathrm{tel}}}

\usepackage[usenames,dvipsnames]{color}

\newcommand{\newtext}[1]{{#1}}
\ifwithlinenumbers
\usepackage{lineno}
\fi

\graphicspath{{./}{./figures/}}

\makeatletter
\newif\ifwithtwocolumns\withtwocolumnsfalse
\if@twocolumn\withtwocolumnstrue\fi 
\makeatother
\iffinalversion\withtwocolumnstrue\fi

\def\hc#1{\leavevmode\hbox to \hsize{\hss #1\hss}\leavevmode}

\ifwithtwocolumns
  \def\includefigure#1{\hc{\resizebox{\columnwidth}{!}{\includegraphics{#1}}}}
\else
  \def\includefigure#1{\hc{\resizebox{11.6cm}{!}{\includegraphics{#1}}}}
\fi

\journal{Astroparticle Physics}

\begin{document}

\begin{frontmatter}

  \title{Monte Carlo Studies of medium-size telescope designs for the
    Cherenkov Telescope Array}

\author[inst-slac]{M.~Wood\corref{cor1}}\ead{mdwood@slac.stanford.edu}
\cortext[cor1]{Corresponding author}
\author[inst-slac]{T.~Jogler}\ead{tjogler@slac.stanford.edu}
\author[inst-stockholm]{J.~Dumm}\ead{jon.dumm@fysik.su.se}
\author[inst-slac,inst-erlangen]{S.~Funk}\ead{funk@slac.stanford.edu}

\address[inst-slac]{SLAC National Accelerator Laboratory, 2575 Sand
  Hill Road M/S 29, Menlo Park, CA 94025, USA}
\address[inst-stockholm]{Oskar Klein Centre and Dept. of Physics,
  Stockholm University, SE-10691 Stockholm, Sweden}
\address[inst-erlangen]{Erlangen Center for Astroparticle Physics (ECAP)
Friedrich-Alexander Universit\"at Erlangen-N\"urnberg
Erwin-Rommel Strasse 1, 91058 Erlangen, Germany}


\begin{abstract}
  We present studies for optimizing the next generation of
  ground-based imaging atmospheric Cherenkov telescopes (IACTs).
  Results focus on mid-sized telescopes (MSTs) for CTA, detecting very
  high energy gamma rays in the energy range from a few hundred GeV to
  a few tens of TeV.  We describe a novel, flexible detector Monte
  Carlo package, FAST (FAst Simulation for imaging air cherenkov
  Telescopes), that we use to simulate different array and telescope
  designs.  The simulation is somewhat simplified to allow for
  efficient exploration over a large telescope design parameter space.
  We investigate a wide range of telescope performance parameters
  including optical resolution, camera pixel size, and light
  collection area.  In order to ensure a comparison of the arrays at
  their maximum sensitivity, we analyze the simulations with the most
  sensitive techniques used in the field, such as maximum likelihood
  template reconstruction and boosted decision trees for background
  rejection.  Choosing telescope design parameters representative of
  the proposed Davies-Cotton (DC) and Schwarzchild-Couder (SC) MST
  designs, we compare the performance of the arrays by examining the
  gamma-ray angular resolution and differential point-source
  sensitivity.  We further investigate the array performance under a
  wide range of conditions, determining the impact of the number of
  telescopes, telescope separation, night sky background, and
  geomagnetic field.  We find a 30--40\% improvement in the gamma-ray
  angular resolution at all energies \newtext{when comparing arrays
    with an equal number of SC and DC telescopes}, significantly
  enhancing point-source sensitivity in the MST energy range.  We
  attribute the increase in point-source sensitivity to the improved
  optical point-spread function and smaller pixel size of the SC
  telescope design.
\end{abstract}


\end{frontmatter}

\ifwithtoc
\clearpage
\tableofcontents
\listoffigures 
\fi

\ifwithlinenumbers
\linenumbers
\fi


\iffinalversion

\input{section1.tex}
\input{section2.tex}

\input{section3.tex}

\input{section4.tex}

\input{section5.tex}
\input{acknowledgements.tex}

\else


\include{section1}
\include{section2}
\include{section3}
\include{section4}
\include{section5}
\include{acknowledgements} 



\fi

\bibliographystyle{mybst}
\bibliography{cta_mc_paper}

\end{document}

%% file: section1.tex
\section{Introduction}
\label{sec:introduction}

The ground-based imaging atmospheric Cherenkov telescope (IACT)
technique has led to significant progress in the field of very high
energy (VHE; $E>100$~GeV) gamma-ray astronomy over the last 25 years.
To date, 145 sources have been detected at VHE with $\sim$60 sources
discovered only in the last five
years\footnote{\tt{http://tevcat.uchicago.edu/}}.
IACTs allow us to study a wide range of scientific topics, many
uniquely accessible by VHE astronomy.  Current and future generations
of IACTs aim to probe the origins and acceleration processes of cosmic
rays~\citep{Bell201356, Acero2013276, Halzen2013155} and explore the
nature of black holes and their relativistic jets.  Other key
objectives include the search for dark matter, axion-like
particles~\citep{Doro2013189, Bergstrm201344}, and Lorentz invariance
violation~\citep{Ellis201350}. This will require
extensive observations on a number of source classes such as
pulsars and pulsar wind nebulae~\citep{deOaWilhelmi2013287}, galactic
binaries~\citep{Paredes2013301}, supernova
remnants~\citep{Aharonian201371}, active galactic
nuclei~\citep{Reimer2013103, Sol2013215}, and gamma-ray
bursts~\citep{Inoue2013252, Mszros2013134}.  The extragalactic sources
can be used as ``backlights'' to study the attenuation on the
extragalactic background light, useful for constraining star formation
history and other cosmological parameters such as the Hubble
constant~\citep{Dominguez:2013mfa}.

VHE gamma rays entering the Earth's atmosphere undergo $e^{+}e^{-}$
pair production, initiating electromagnetic cascades.  The
relativistic charged particles in the shower emit Cherenkov
ultraviolet and optical radiation, which is detectable at ground
level.  The majority of the emitted Cherenkov light is narrowly beamed
along the trajectory of the gamma-ray primary in a cone with an
opening angle of \newtext{$\sim1.0^\circ$}.  Due to the beaming effect, the
majority of the Cherenkov light falls within a Cherenkov light pool
with a diameter of 200--300~m and a nearly constant light density.  By
imaging the Cherenkov light emitted by the shower particles, IACTs
are able to reconstruct the direction and energy of the original gamma
ray and to distinguish gamma rays from the much more prevalent
cosmic-ray background.  High resolution imaging of the Cherenkov
shower offers significant benefits for IACTs by enabling a more
accurate measurement of the shower axis which has an intrinsic
transverse angular size of only a few arcminutes.  However the finite
shower width and stochastic fluctuations in the shower development
fundamentally limit the performance of IACTs.

The designs of IACTs are governed by a few key factors.  At low
energy, the number of Cherenkov photons compared to the night sky
background necessitates a large $\mathcal{O}$(10--20~m) mirror
diameter and high quantum efficiency camera.  The camera must also be
able to capture the signal very quickly since the duration of a
Cherenkov pulse can be as short as a few nanoseconds.  The optical
point-spread function (PSF) and camera pixel size should ideally be
suitably smaller than the angular dimension of the gamma-ray shower.
However the high cost-per-pixel of camera designs used in current
generation IACTs has generally dictated pixel sizes that are
significantly larger ($0.1^\circ$--$0.2^\circ$) than the angular size
of shower structure.  Multiple viewing angles of the same shower
offered by an array of telescopes drastically improves the
reconstruction performance and background rejection.  Finally, at high
energy, the sensitivity of IACTs is limited by \newtext{gamma-ray
  signal statistics}, requiring an array with a large effective
gamma-ray collection area.

The current generation of IACTs all have single-dish optical
systems.  These have small spherical mirror facets attached to either
a spherical dish (i.e. Davies-Cotton
(DC)~\citep{Davies:1957,Lewis:1990}) or a parabolic dish.  The
parabolic dish reduces the time spread of the Cherenkov signal but
introduces a larger off-axis optical PSF.  An intermediate design with
a spherical dish but a larger radius of curvature (intermediate-DC)
can be used to achieve an improved time spread while maintaining
off-axis performance~\citep{Bernlohr:2008kv, 2013APh....43..171B}.
These single-dish designs are appealing because they are relatively
inexpensive, mirror alignment is straightforward, and the optical PSF
at large field angles is better than that of monolithic spherical or
parabolic reflectors~\citep{Vassiliev:2006pw}.

The possibility of improving the PSF (especially off axis) and
reducing the plate scale of IACTs has driven the study of
Schwarzschild-Couder (SC) aplanatic telescopes with two aspheric
mirror surfaces\footnote{Though segmented, the mirror surfaces are
  often referred to as a singular mirror for brevity.}.  The improved
PSF across the field of view (FoV) allows for more accurate surveying
and mapping of extended sources.  The reduced plate scale is highly
compatible with new camera technologies such as Silicon
photomultipliers or multi-anode photomultiplier tubes.  These
technologies allow for a cost-effective, finely-pixelated image over a
large FoV.  Studies have been performed providing solutions for mirror
surfaces optimized to correct spherical and coma aberrations.  These
solutions are also isochronous, allowing for a short trigger
coincidence window \citep{Vassiliev:2007mj}.  The first SC prototype
is still being developed~\citep{Rousselle:2013nfa} and has several
challenges to overcome.  In particular, the tolerances of the
mechanical structure in the camera and mirror alignment system are
relatively stringent, which translates to a higher cost.  To provide
comparisons at a fixed cost, our SC simulations use a smaller mirror
area than that of the baseline DC design.

The Cherenkov Telescope Array (CTA) is an example of a next-generation
IACT observatory. CTA aims to surpass the current IACT systems such as
H.E.S.S.~\citep{Aharonian:2006pe}, MAGIC~\citep{MAGICref} and
VERITAS~\citep{VERITASref} by an order of magnitude in sensitivity and
enlarge the observable energy range from a few tens of GeV to beyond
one hundred TeV~\citep{2011ExA....32..193A}.  To achieve this broad
energy range and high sensitivity, CTA will incorporate telescopes of
three different sizes spread out over an area of $\sim$3~km$^2$.
Telescopes are denoted by their mirror diameter as large-size
telescopes (LSTs, $\sim$24~m), medium-size telescopes (MSTs,
$\sim$12~m), and small-size telescopes (SSTs, \newtext{$\sim$4~m in
  the current design}).  The baseline designs for the LST and MST both
feature a single reflector based on the DC optical design.  Telescope
designs based on dual-reflector SC optics are also being developed for
both medium- and small-sized telescopes.  The medium-size SC telescope
(SCT) would fill a similar role to the MST and predominantly
contribute to the sensitivity of CTA in the energy range between
100~GeV and 1~TeV.  In this paper we explore a range of telescope
models but focus primarly on the comparison of designs with
characteristics similar to the MST and SCT.  In the subsequent
discussion we use MST to refer to all telescope designs with a primary
mirror diameter of 9--12~m.  DC-MST and SC-MST are used to
specifically refer to telescopes with the imaging characteristics
similar to the MST and SCT designs, respectively.

The baseline design of CTA includes $\sim$four LSTs, $\sim$30 MSTs,
and $\sim$50 SSTs.  The sensitivity could be improved by a factor of
2--3 in the core energy range by expanding the MST array with an
additional 24--36 SCTs.  With these additional telescopes, the
combined MST and SCT array enters a new regime where the internal
effective area is comparable to the effective area of events landing
outside the array.  These so-called contained events have much
improved angular and energy resolution as well as background
rejection.  Extensive work is underway to optimize the design of CTA
for the wide range of science goals~\citep{2013APh....43..171B}.  The
scope of previous studies has been primarily on a straightforward
expansion of existing telescope designs to larger arrays.

In this paper, we describe a novel, flexible Monte Carlo simulation
and analysis chain.  We use them to evaluate the performance of
CTA-like arrays over a large range of telescope configurations and
design parameters.  Section 2 describes this simulation and the
simplified detector model.  In Section 3, we explain the analysis
chain, including a maximum likelihood shower reconstruction using
simulated templates.  This reconstruction was used for comparisons
between the maximum sensitivity for each array configuration.  In
Section 4, we show comparisons between possible CTA designs, focusing
primarily on the number of telescopes and the DC versus SC designs.
We conclude in Section 5.

%% file: section2.tex
\section{Simulation}
\label{sec:simulation}

We have studied the performance of a variety of array geometries and
telescope configurations for a hypothetical CTA site at an altitude of
2000~m.  Details of the site model and array geometry are described in
Sections \ref{subsec:shower_sim} and \ref{subsec:array_layout}.
Simulations of the telescope response were performed using a
simplified detector model described in Section \ref{subsec:det_model}.

\subsection{Air-Shower Simulations}\label{subsec:shower_sim}

Simulations of the gamma-ray and cosmic-ray air shower cascades were
performed with the CORSIKA \newtext{v6.99} Monte Carlo (MC) package
\citep{heck1998} and the \newtext{QGSJet II-03} hadronic interaction
model \citep{2006NuPhS.151..143O}.  We used a site model with an
elevation of 2000~m, a tropical atmospheric profile, and an equatorial
geomagnetic field configuration with $(B_{x},B_{z}) = (27.5~
\mu\mathrm{T},-15.0~\mu\mathrm{T})$.  This site model is identical to
the one used in \cite{2013APh....43..171B} and has similar
characteristics to the southern hemisphere sites proposed for CTA.

Gamma-ray showers were simulated as coming from a point on the sky at
20$^\circ$ zenith angle and 0$^\circ$ azimuth angle, as measured from
the local magnetic north over the energy range from 10~GeV to 30~TeV.
Protons and electrons were simulated with an isotropic distribution
that extends to 8$^\circ$ and 5$^\circ$ respectively from the
direction of the gamma-ray primary.  We use the spectral
parameterizations for proton and electron fluxes from
\citep{2008APh....30..149B}.  To account for the contribution of
heavier cosmic-ray nuclei we increase the proton flux by a factor 1.2.

\subsection{Array Geometry}\label{subsec:array_layout}

Proposed designs for CTA employ three telescope types (SST, MST, and
LST) with variable inter-telescope spacing from 120~m to more than
200~m \citep{2013APh....43..171B}.  The number of telescopes of each
type and their separations are chosen to optimize the differential
sensitivity over the full energy range of CTA.
\cite{2013APh....43..171B} found that two balanced arrays (arrays E
and I) that have 3--4 LSTs, 18--23 MSTs, and 30--50 SSTs \newtext{of
  7~m diameter} provide the best compromise in performance over the
full energy range of CTA while keeping the total cost of the array
within the projected CTA budget.

\begin{figure}[htb]
\includegraphics[width=.5\textwidth, keepaspectratio]{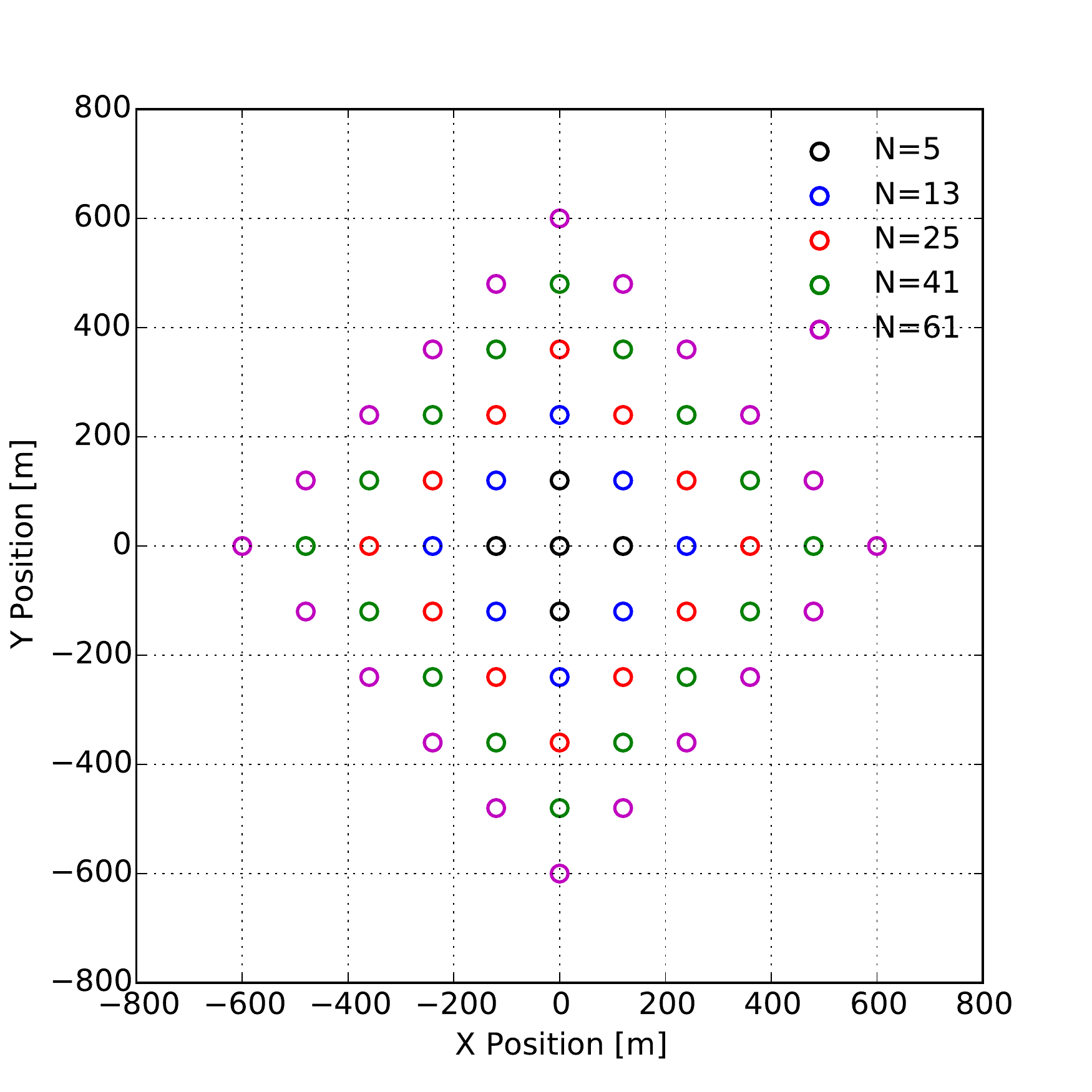}
\caption{\label{FIG:LAYOUT}Physical telescope positions for the five
  array geometries used for this study.  All geometries are composed
  of telescopes arranged on a uniform grid with 120~m spacing.  The
  smallest array is composed of five telescopes (black circles).  The
  larger arrays are constructed by the addition of successive rings of
  telescopes around the array boundary up to a maximum of 61
  telescopes in the baseline array geometry.}
\end{figure}

For this study we simulated an array geometry which is similar to the
one used for MSTs and LSTs in arrays E and I.  The array is composed
of 61 telescopes arranged on a grid with constant inter-telescope
spacing of 120~m (see Figure \ref{FIG:LAYOUT}).  A telescope spacing
of about 120~m is well motivated by the characteristic size of the
Cherenkov light pool for gamma-ray air showers and guarantees that
multiple telescopes will sample the shower within the shower light
pool.  Subsets of telescopes from the baseline array were used to
construct arrays with a reduced number of telescopes by removing
successive rings of telescopes along the array perimeter.  These
reduced arrays have between 5 and 41 telescopes and encompass arrays
that are similar in telescope number to both current IACT arrays ($\ntel=$
5) and the array designs currently considered for CTA ($\ntel=$ 25--41).
We also examined the performance of arrays with smaller and larger
inter-telescope separations (60~m--200~m) by rescaling the
inter-telescope separation of our baseline array.

All simulations were performed with homogeneous arrays composed of a
single telescope type.  We primarily consider telescope models with
mirror areas between the current MST and LST designs.  Because our
study is focused on the performance of arrays in the core CTA energy
range (100~GeV - 10~TeV) we did not consider SSTs.

\subsection{Detector Model}\label{subsec:det_model}

Simulations of IACT arrays have traditionally been performed with
highly detailed detector models that use optical ray-tracing to track
the trajectory and time of arrival of individual Cherenkov photons.
Because these models have a very large number of parameters, a brute
force optimization of the telescope design presents a significant
computational challenge.  In order to efficiently study the telescope
design parameter space, we have developed a simplified telescope
simulation tool, FAST (FAst Simulation for imaging air cherenkov
Telescopes), that is not tied to any particular mirror configuration
or camera technology.  In the FAST model, the telescope
characteristics are fully described by the following parameters:

\begin{itemize}
\item Effective light collection area: $\aopt$
\item 68\% containment radius of the optical PSF: $\rpsf$
\item Camera pixel size: $\rpix$
\item Effective camera trigger threshold: $\tth$
\item Single photo-electron (PE) charge resolution: $\sspe$
\item Pixel read-noise: $\sigma_{b}$
\item Effective integration window: $\Delta T$
\end{itemize}
While this simplified model lacks the level of detail provided by
other simulation tools, the performance of a realistic telescope
design can be approximated by an appropriate choice of these model
parameters.  In this section we describe in detail the implementation
of our model and how each of these parameters influence the telescope
response.

The geometrical model of the telescope consists of a primary mirror of
diameter $D$ with physical mirror area $A_{M} = \pi (D/2)^{2}$.  All
Cherenkov photons that intersect with the primary mirror surface are
propagated through the telescope simulation.  The photons collected by
the primary mirror are folded with a wavelength dependent photon
detection efficiency, $\epsilon(\lambda)$, that models losses from all
elements in the optical system and camera (mirrors, lightguides, and
photosensors).  Applying a detection probability to each collected
photon, we construct a list of detected photoelectrons (PEs) which are
used as input to the simulation of the telescope trigger, camera, and
optics.

We quantify the total light-collecting power of a telescope by its
effective light collection area, $\aopt(\lambda) =
A_{M}\epsilon(\lambda)$, the product of the physical mirror area with
the total photon detection efficiency at wavelength $\lambda$.  We
compute a wavelength-averaged effective area by folding
$\aopt(\lambda)$ with a model for the wavelength distribution of
Cherenkov light,

\begin{equation}\label{EQN:AOPT}
  \aopt = 
  \int_{\lambda_{0}}^{\lambda_{1}}P(\lambda,z)\aopt(\lambda)d\lambda,
\end{equation}
where 

\begin{equation}\label{EQN:PCHERENKOV}
  P(\lambda,z) \propto e^{-\tau(\lambda,z)}\lambda^{-2}
\end{equation}
is the normalized wavelength distribution of Cherenkov light at the
ground for an emission altitude $z$ and an optical depth for
atmospheric extinction $\tau(\lambda,z)$.  We use an atmospheric
extinction model generated with MODTRAN \citep{MODTRAN} for the
tropical atmosphere and an aerosol layer with a visibility of 50~km.
For all further evaluations of $\aopt$ we use $z = $10~km and an
integration over wavelength from 250~nm to 700~nm.

\begin{figure}[tb]
\includegraphics[width=.5\textwidth, keepaspectratio]{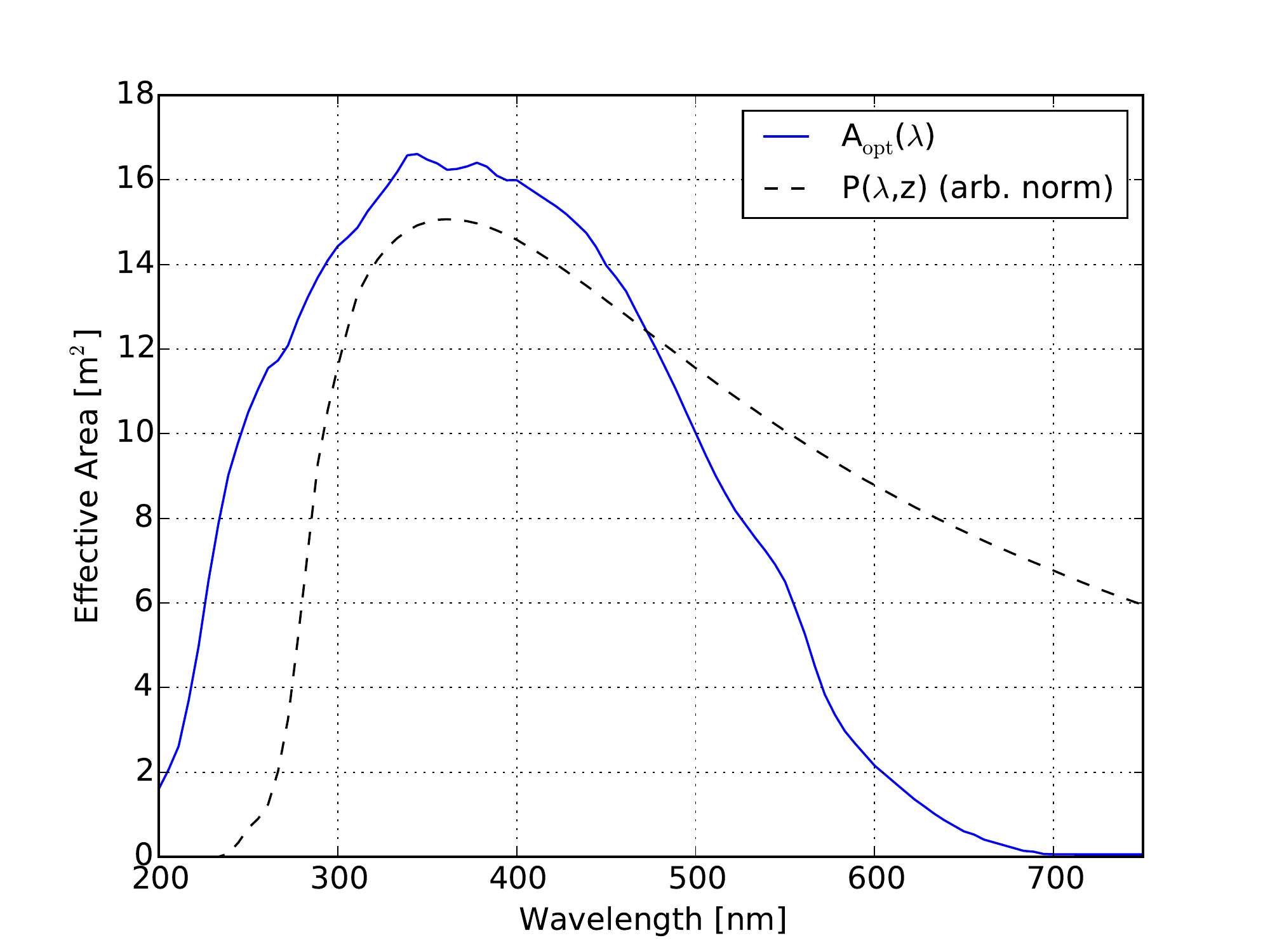}
\caption{\label{FIG:AOPT} Effective light collection area versus
  wavelength for the benchmark telescope model with $\aopt$ =
  11.18~m$^{2}$.  The dashed black line shows the spectral shape for
  Cherenkov light emitted at an elevation of 10~km after absorption by
  the atmosphere.}
\end{figure}

We define a benchmark telescope with a $D=12$~m primary diameter and a
total photon detection efficiency that includes losses from mirror
reflections and photosensor efficiency.  We use a photosensor model
with a spectral response that is characteristic of photomultiplier
tubes and has a peak efficiency of 24\% at 350~nm.  Losses from mirror
reflections are evaluated for a single optical surface using a
wavelength-dependent reflectivity with a peak efficiency of 89\% at
320~nm.  This reflectivity is similar to that of the aluminum and
aluminized glass mirrors used in current generation IACTs.  Figure
\ref{FIG:AOPT} shows the optical effective area of the telescope model
as a function of wavelength.  The effective light collection area of
our benchmark telescope is 11.18~m$^{2}$ which is representative of
medium-sized IACTs with $\sim$10~m aperture and 50--100~m$^{2}$ mirror
area.  The response of telescopes with larger or smaller light
collection areas is modeled using the same spectral response and
mirror area as the benchmark telescope model but scaling the photon
detection efficiency by the ratio $\aopt/11.18$~m$^2$.

The imaging response of the telescope optical system is simulated by
applying a model for the optical point-spread-function (PSF) to the
distribution of true photon arrival directions in the camera image
plane.  After applying a survival probability for detection, each
Cherenkov photon is assigned a random offset drawn from the optical
PSF.  We parameterize the optical PSF as a 2D Gaussian with a 68\%
containment radius, $\rpsf$, that is constant across the FoV.  We
consider values of $\rpsf$ between 0.02$^\circ$ and 0.08$^\circ$ which
is comparable to the range of PSF spot sizes for the CTA telescope
designs at both small and large field angles.  All telescopes are
simulated with an 8 deg FoV with a light collection area that is
constant with field angle.

\begin{figure*}[tb]
\includegraphics[width=.5\textwidth,keepaspectratio]{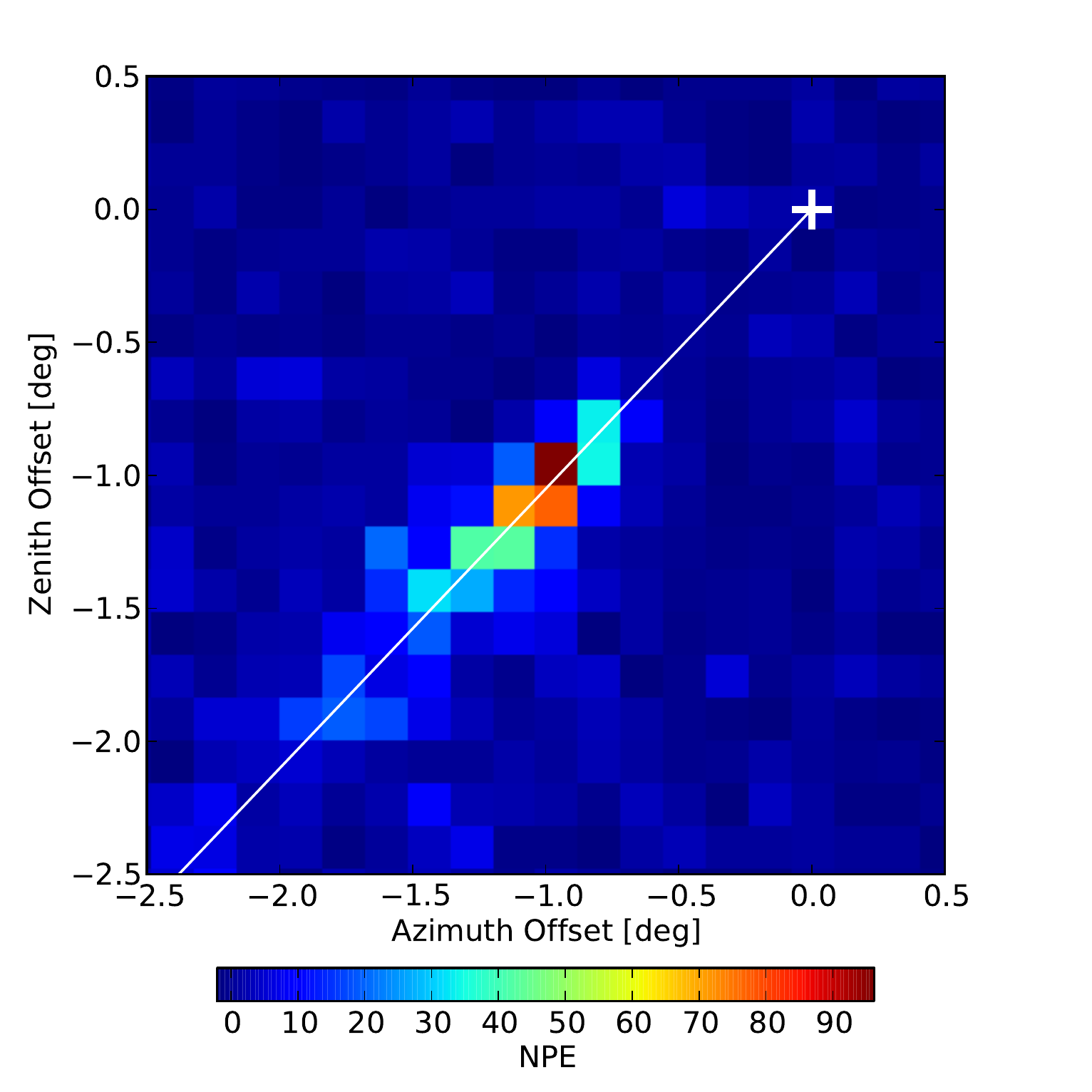}
\includegraphics[width=.5\textwidth,keepaspectratio]{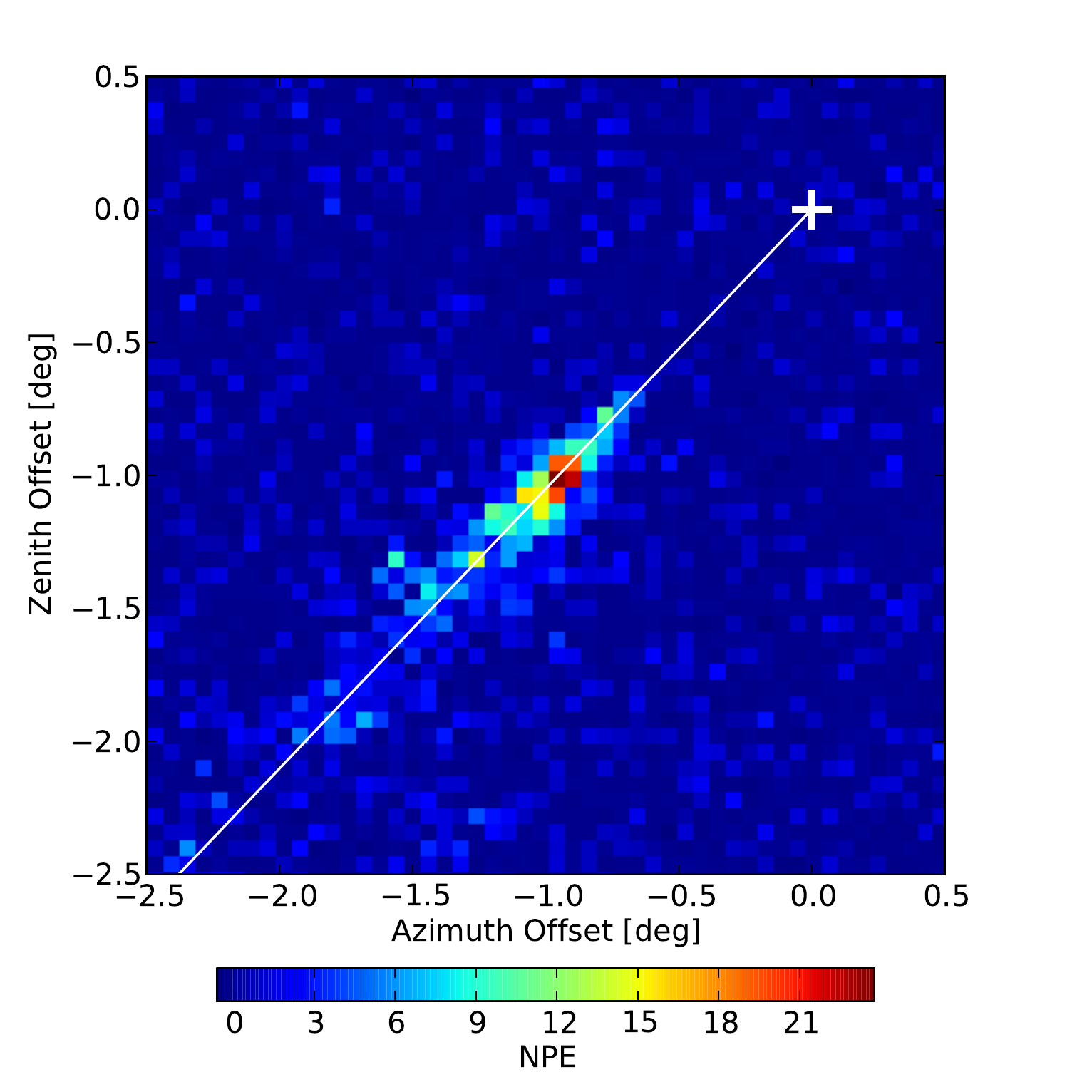}
\caption{\label{FIG:CAMERA_IMAGE} Camera images of the same 1~TeV
  gamma-ray shower with an impact distance of 120~m simulated with two
  different telescope pixel sizes: $\rpix = 0.16^\circ$
  (\textbf{left}) and $\rpix = 0.06^\circ$ (\textbf{right}).  Both
  telescope models have $\aopt$ = 11.18~m$^{2}$ and $\rpsf =
  0.02^\circ$.  The color scale denotes the measured signal amplitude
  in PEs for each pixel.  The white cross and solid line show the
  direction of the gamma-ray primary and the projection of its
  trajectory to the telescope image plane, respectively.}
\end{figure*}

Telescopes are simulated with a camera geometry composed of square
pixels of angular width $\rpix$ that uniformly tile the camera FoV.
Each pixel is assigned a time integrated signal that is the sum of the
detected Cherenkov photons, night-sky background (NSB) photons, and
detector noise.  The number of NSB photons is drawn from a Poisson
distribution where the average ($\mu_{b}$) is computed using an
implicit time integration window ($\Delta T$) of 16~ns.  The mean
number of NSB photons per pixel for a telescope with effective light
collection area $\aopt$ and pixel solid angle $\Delta\Omega$ is

\begin{equation}
\mu_{b} = \Delta T\Delta\Omega \epsilon
\int F_{\mathrm{nsb}}(\lambda) \aopt(\lambda) d\lambda,
\end{equation}
where $F_{\mathrm{nsb}}(\lambda)$ is the differential NSB flux versus
wavelength.  We use the NSB spectral model from
\citep{1998NewAR..42..503B} which is representative of the sky
brightness of an extragalactic observation field.  When folded with
the optical efficiency of our benchmark telescope model, the integral flux of
detected NSB photons is 365~MHz~deg$^{-2}$~m$^{-2}$.  Our benchmark
telescope model has
an NSB surface density in the image plane ($\snsb$) of 65.4~deg$^{-2}$
for an integration window of 16~ns.  We model the photosensor single
photoelectron response with a Gaussian with $\sspe = 0.4$~PE.  Each
channel is simulated with a Gaussian readout noise ($\sigma_{b}$) of
0.1~PE.  For the range of pixel sizes and optical throughputs
considered in this study, the readout noise is a subdominant component
of the pixel noise relative to NSB and is therefore not expected to
have a significant impact on the telescope performance.
Fig.~\ref{FIG:CAMERA_IMAGE} shows simulated camera images for
telescope models with two different pixel sizes observing the same
1~TeV gamma-ray shower.

The trigger system of an IACT array rejects noise-induced events while
maintaining high efficiency for cosmic-ray signals.  We simulate a
two-stage trigger system composed of a camera-level trigger for each
telescope and an array-level trigger that combines the camera triggers
of multiple telescopes to form the final trigger decision.  Camera
trigger designs used by current generation IACTs and envisioned for
CTA are generally based on a multi-level hierarchy whereby trigger
information from individual pixels or camera subfields is combined to
form the camera-level trigger decision
\citep{2004APh....22..285F,2006APh....25..391H,2011ExA....32..193A}.
The rate of accidental triggers is suppressed by requiring a time
coincidence of triggers from neighboring pixels or camera regions.

A useful quantity for characterizing the performance of different
camera trigger designs is the effective camera threshold, the true
gamma-ray image amplitude in PEs at which the camera trigger is 50\%
efficient.  \newtext{To first order the efficiency of the camera
  trigger depends only on the surface brightness of the Cherenkov
  shower image.  Because the angular size of the shower is only a weak
  function of distance and energy, we can approximate the response of
  a camera trigger by applying a fixed threshold on the true number of
  Cherenkov PEs in the camera FoV.}

\newtext{We simulate the camera trigger by applying a threshold $\tth$
  on the true number of Cherenkov PEs detected in the camera FoV.  For
  showers that trigger one or more telescopes, the array-level trigger
  is simulated requiring a multiplicity of at least two triggered
  telescopes.  The camera threshold provides a single parameter model
  that we use to explore influence of the trigger threshold on the
  array-level performance.} By calibrating $\tth$ to the effective
camera threshold of a given trigger design, we can also approximate
the trigger response that would be obtained with a more detailed
trigger simulation implementation.

Studies performed with the \texttt{sim\_telarray} detector simulation
package \citep{2008APh....30..149B} have shown that camera trigger
designs currently considered for the MSTs can achieve effective
trigger thresholds of 60--80~PE for a single telescope accidental
trigger rate of 1--10~kHz.  We adopted a trigger threshold of 60~PE for
our baseline telescope model with $\aopt = $ 11.18~m$^{2}$ which is
comparable to the effective threshold of the \textit{prod-2} MST model
\citep{Bernlohr:2013bla}.
To model the effective trigger threshold for telescopes with different
light collection areas, we used a simple scaling formula that
approximates the threshold needed to maintain a constant rate of
accidental triggers.  If the total pixel noise is dominated by NSB
photons, the rate of accidental triggers should be proportional to the
RMS fluctuations in the number of NSB photons collected in a trigger
pixel which scales as $\aopt^{1/2}$ if the \newtext{angular pixel
  size} is held fixed.  Telescopes with larger effective light
collection area achieve a lower trigger threshold through the
suppression of these NSB fluctuations relative to the signal amplitude
which increases linearly with $\aopt$.  We assign the effective
trigger threshold for a telescope with light collection area $\aopt$
as,

\begin{equation}\label{EQN:TRIGGER}
\tth = 60~\mathrm{PE}~\left(\frac{\aopt}{11.18~\mathrm{m}^2}\right)^{1/2}.
\end{equation}

For the studies presented in Section \ref{sec:results}, we consider a
benchmark array (M61) with 61 identical telescopes 
with $\aopt$ = 11.18~m$^{2}$, $\rpsf$ = 0.02$^\circ$, $\rpix$ =
0.06$^\circ$, and $\tth = $ 60~PE.  Our baseline telescope model is
representative of a generic medium-sized telescope design with SC-like
imaging characteristics.  In Section
\ref{subsec:benchmark_arrays_define} we additionally consider other
telescope models that were specifically chosen to match the
characteristics of the proposed CTA telescope designs.

\subsection{Simplifications of the Detector Simulation}

\newtext{The FAST package uses a highly simplified model of the
  telescope optics and camera.  This allows us to perform a more
  general exploration of the IACT design parameter space without
  focusing on the details of any specific optical design or camera
  technology.  These simplifications also make the FAST simulation
  much less computationally intensive than traditional simulation
  tools such as \texttt{sim\_telarray} which further facilitates the
  exploration of a large phase space of telescope and array design
  concepts.  When simulating comparable telescope designs with FAST
  and \texttt{sim\_telarray} we observe an order-of-magnitude
  reduction in computation time.  Here, we discuss the limitations of
  the approach taken in the FAST simulation package and describe the areas in
  which a more detailed simulation of the telescope camera and optics
  could potentially affect our results.}

\newtext{ FAST does not use raytracing to account for shadowing of the
  camera by telescope structure and assumes a simple 2D Gaussian PSF
  that is constant over the FoV.  Full raytracing simulations can be
  used to model effects such as shadowing by the telescope structure
  and light losses from gaps in the mirror surfaces.  Raytracing also
  allows a more realistic modeling of the telescope PSF.  Optical
  aberrations intrinsic to the design of IACTs introduce a strong
  field-angle dependence to both the size and shape of the PSF.  In
  the FAST telescope model, all effects that influence the optical
  performance of the telescope are folded into the effective optical
  area ($\aopt$) and the 68\% containment radius of the PSF ($\rpsf$).
  Shadowing and other light losses in the optical system are thus
  taken into account by a reduction in the effective optical area.
  The field-angle dependence of the PSF is studied by comparing the
  performance of telescope designs with the best and worst PSF at any
  field angle.  The performance of a real telescope should always fall
  between that of telescopes with smallest and largest PSF in the FoV.
  We did not study the effect of an asymmetric PSF but it is plausible
  to assume that the array performance with telescopes that have an
  asymmetric PSF can be estimated by enlarging the symmetric PSF in
  FAST.}

\newtext{FAST does not simulate the timing of signals and assumes an
  ideal data acquisition system that is only limited by the
  irreducible noise from NSB photons.
  Effects that distort the measurement of the pixel charge such as
  cross-talk, after-pulsing, timing jitter, non-linearity, and
  saturation might lead to a reduction in performance.  However these
  effects should affect all telescope types in a similar way in the
  sense that photon charges might be only partially reconstructed.
  Per the telescope performance requirements set forward by CTA, we
  expect the influence of these effects to be sub-dominant to the
  irreducible limitations on the IACT technique set by shower physics
  (NSB and shower fluctuations).  Many of these effects can be
  partially mitigated with pixel-level calibration or pre-processing
  analysis procedures.  For instance, suppression algorithms such as
  clipping of pixel signal amplitudes or the removal of isolated high
  PE pixels have been demonstrated as efficient techniques to reduce
  the impact of after-pulsing.  In principle these effects could also
  be approximated in the FAST detector model by increasing the
  pixel-level noise or worsening the charge resolution.
  We note that our analysis does not use shower timing parameters or
  shower development timing in any explicit way and thus is robust
  with respect to changes in timing requirements.}

\newtext{Another simplification in FAST is the trigger threshold
  decision logic that assumes any shower image above a certain
  threshold will trigger the telescope.  Realistic trigger electronics
  might have several characteristics that have to be simulated in
  detail but in the early planning state of an array our method is
  very valuable to give a solid estimate of the array performance as
  long as the trigger threshold is not chosen aggressively and the
  requirements of the telescope array are met. Indeed we show in
  Section \ref{subsec:simtel_comparison} that our telescope threshold
  might be too conservative given that more realistic simulations find
  enhanced performance near the energy threshold of simulated
  arrays. For relative comparisons of arrays our simplifications are
  unimportant. To estimate the influence on the absolute performance
  impacts of our simplified simulations we compare our results to a
  much more sophisticated detector simulation in Section
  \ref{subsec:simtel_comparison}.}

%% file: section3.tex
\section{Analysis}
\label{sec:analysis}

The analysis of the telescope image data is performed using well
established techniques for the analysis of IACT data.  The
analysis is performed in three stages: preparation of the telescope
images, reconstruction of the event properties, and training and
optimization of cuts.

\subsection{Image Cleaning and Parameterization}

The image analysis is applied to the telescope pixel amplitudes to
derive a set of telescope-level parameters which characterize the
distribution of light in each telescope.  Analysis of the telescope
image data begins with the application of an image cleaning analysis
that selects pixels that have a signal amplitude that is larger than
noise.  Traditionally image cleaning has been performed using
variations of a nearest-neighbor algorithm
\citep{1997APh.....8....1D}.  A search is performed for groups of
neighboring pixels which exceed a threshold defined in terms of the
absolute amplitude or the amplitude relative to the RMS noise in the
pixel.  These algorithms work well as long as the dimension of the
pixel is of the same order as the Cherenkov image size.  However in
the limit of small pixel sizes these algorithms will lose efficiency
for low energy showers where the signal is spread out over too many
pixels to be discernible above noise when only considering nearest
neighbors.

In order to circumvent the limitations of the nearest-neighbor pixel
algorithms, we use an \textit{Aperture} cleaning algorithm that
performs a smoothing over the camera with an angular scale
($R=0.12^{\circ}$) that is of the same order as the width of a
gamma-ray induced Cherenkov shower (0.1--0.2$^{\circ}$).

\begin{figure}[tb]
\includegraphics[width=.5\textwidth, keepaspectratio]
{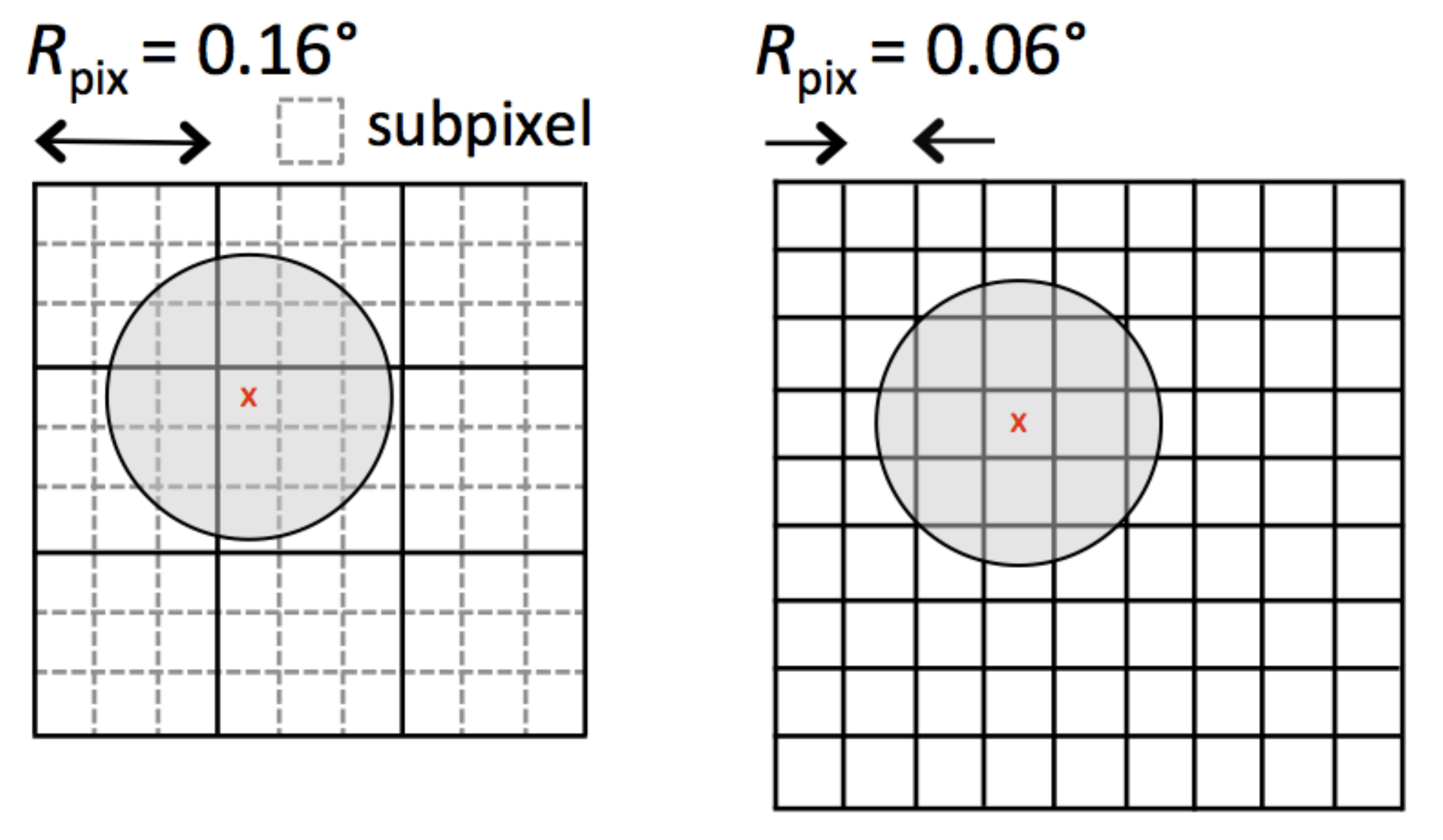}
\caption{\label{FIG:APERTURE_CLEANING}Illustration of the aperture
  cleaning algorithm on small camera subsections with
  $R=0.12^{\circ}$. In the DC-like case (\textit{left}), pixels are
  subdivided since they are large compared to the aperture.  Each
  subpixel is used as the center of an aperture for image intensity
  calculation.  This calculation is based on the number of PEs and the
  fraction of the pixel area within the aperture, normalized to the
  area of the aperture.  For the SC-like case (\textit{right}),
  smaller pixels do not require subdivision.  }
\end{figure}

In order to detect efficiently images that lie on pixel boundaries we
divide each pixel into $N \times N$ \textit{subpixels} where $N =
\lceil \rpix/0.06^\circ \rceil$.  We compute the image intensity in
the neighborhood of subpixel $i$ as

\begin{equation}
\bar{s}(R) = \sum_{j}s_{j}w_{i,j}(R),
\end{equation}
where $w_{i,j}(R)$ is the fraction of the solid angle of pixel $j$
contained within the circular aperture of radius $R$ centered on
subpixel $i$ (see Figure~\ref{FIG:APERTURE_CLEANING}).  The pixel
image threshold is defined relative to the expected noise within the
pixel aperture

\begin{equation}
\sigma(R) = \left(\sum_{j}(\sigma_b^2 + \mu_{b})w_{i,j}(R)\right)^{1/2}.
\end{equation}
For the present analysis we adopt an image threshold of
$\bar{s}/\sigma = 7$ which for our baseline array corresponds to an image
intensity of 319 PE deg$^{-2}$ and an integrated charge of 14.4 PE
within the cleaning aperture.  Any pixel for which one or more
subpixels exceeds the cleaning threshold is flagged as an image pixel.
The simulations do not include photodetector after-pulsing, which can
cause noise isolated in single pixels.  These may need to be
suppressed if the aperture cleaning method is applied in other
scenarios.  Telescope images are discarded at this point if fewer than
three image pixels are present.

The image cleaning is only used by the geometric reconstruction,
itself a seed for the likelihood reconstruction.  As such, a
relatively low threshold was chosen to maximize the reconstruction
efficiency for low-energy events.

Following the image cleaning analysis, an image analysis is applied to
the amplitudes of image pixels ($s_{j}$) to calculate a set of image
parameters that characterize the light distribution in the focal
plane.  \newtext{The image parameters include the total image
  size, $S$, the image centroid, the second central moments along the
  major and minor axes of the image denoted as length $l$ and width
  $w$, and the orientation of the major axis in the image plane.}

\subsection{Shower Reconstruction}
\label{subsec:recon}

The shower reconstruction determines a trajectory and energy for each
event by fitting a \textit{shower model} to the telescope image data.
The shower model parameters ($\bs{\theta}$) are the primary energy
($E$), the primary direction ($\mathbf{e}$), the primary impact
position ($\mathbf{R}$), and the atmospheric column depth of the first
interaction point ($\lambda$).  In an array of IACTs, each telescope
views the shower from a different perspective and provides an
independent constraint on the shower parameters.  By using image data
from multiple telescopes, one can perform a stereoscopic
reconstruction of the shower trajectory.  For the analysis algorithms
presented in this section, we assume on-axis observations of a
gamma-ray source in parallel pointing mode whereby the optical axes of
the telescopes in the array are aligned with the shower direction.
However the procedures described here can be also applied to the case
of non-aligned telescope pointing.

\begin{figure}[tb]
\centering
\includegraphics[width=.5\textwidth, keepaspectratio]
{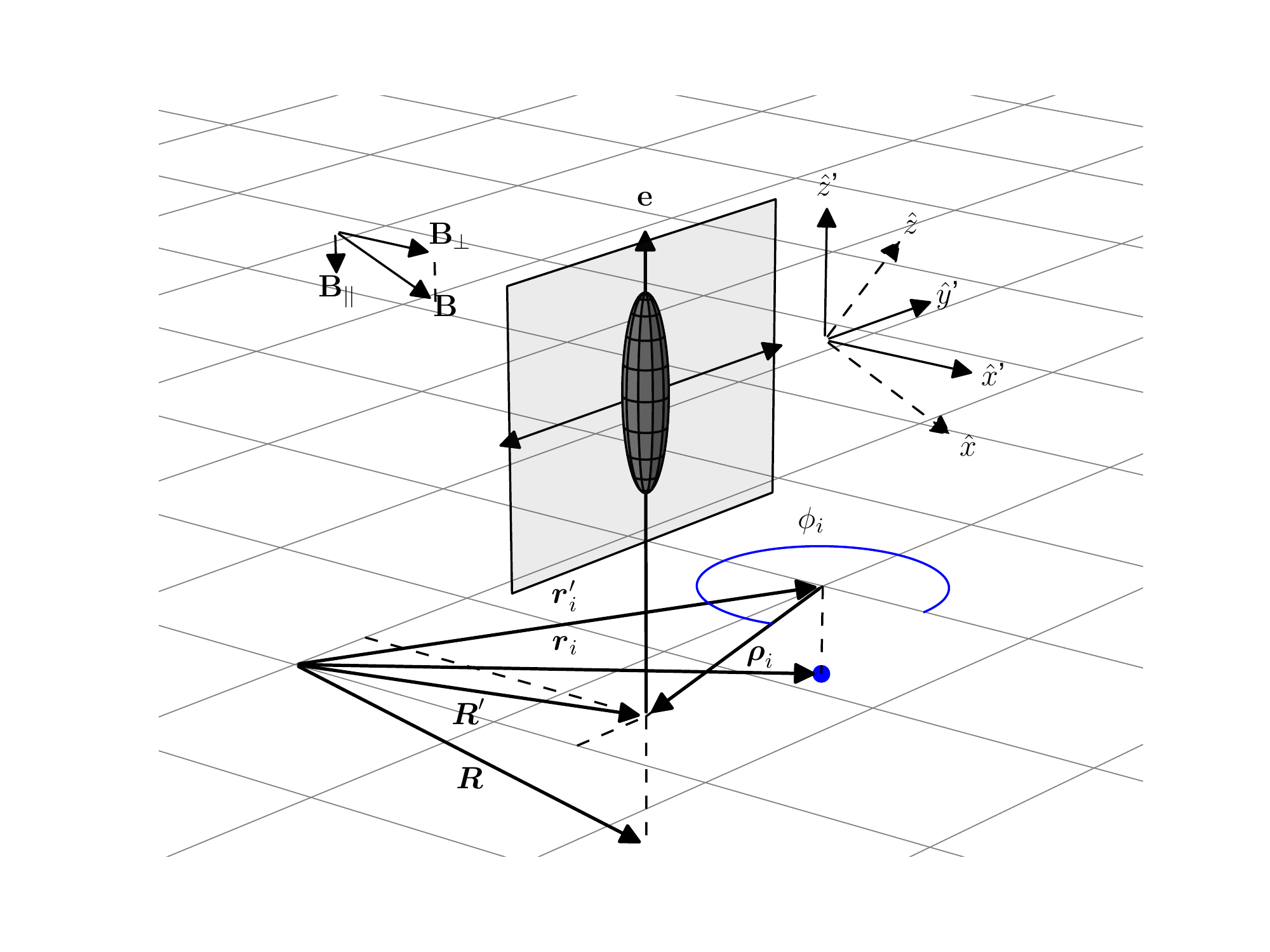}
\caption{\label{FIG:EVENT_GEOMETRY} Illustration of the geometry of a
  gamma-ray shower as shown in the shower coordinate system.  The
  gamma-ray trajectory is defined by its impact position
  $\mathbf{R}'$ in the $x'-y'$ plane and arrival direction
  $\mathbf{e}$.  The GF induces an elongation in the shower in the
  plane orthogonal to $\mathbf{B}_\perp$ (indicated by the grey shaded
  square).  The shower impact vector, $\bs{\rho}_{i}$, describes the
  position of the shower impact position relative to the telescope at
  $\mathbf{r}_{i}$ (closed blue circle).  The
  shower position angle, $\phi_{i}$, is defined by the angle between the shower
  impact vector and the $x'-$axis.}
\end{figure}

In presenting the implementation of the shower reconstruction
algorithms, we use a global coordinate system defined with the
$x$-axis parallel to the direction of magnetic north and the $z$-axis
perpendicular to the Earth's surface.  The positions of the array
telescopes are denoted by $\mathbf{r}_{i}$.  For the array layouts
considered for this study, the telescopes are arranged in a regular
grid in the $x$-$y$ plane with all telescopes located at the same
height above sea level ($z = 2000$~m).
Shower reconstruction is performed in a \textit{shower} coordinate
system with the $z$-axis aligned with the shower trajectory and defined
by the basis vectors:

\begin{equation}
\begin{split}
  \hat{z}' &= \mathbf{e}\\
  \hat{y}' &=
  \frac{(\mathbf{e}-\left(\hat{z}\cdot\mathbf{e}\right)\hat{z})}
  {\sqrt{1-(\hat{z}\cdot\mathbf{e})^2}}
  \times\hat{z}\\
  \hat{x}' &= \hat{z}' \times \hat{y}'
\end{split}
\end{equation}
We use $\mathbf{r}_{i}'$ and $\mathbf{R}'$ to represent the
projections of the telescope positions and the shower impact position
to the $x'-y'$ plane.  An illustration of the geometry of a shower is
shown in Figure \ref{FIG:EVENT_GEOMETRY}.  The shower impact vector,

\begin{equation}
\bs{\rho}_{i} = \mathbf{R} - \mathbf{r}_{i} - 
(\hat{z}'\cdot(\mathbf{R} - \mathbf{r}_{i}))\hat{z}',
\end{equation}
describes the location of the shower impact position relative to
telescope $i$ in the $x'-y'$ plane.  The shower impact distance
($\rho_{i} = |\bs{\rho}_{i}|$) is the distance of closest approach
between the shower and the telescope.

The geomagnetic field (GF) alters the development of the gamma-ray
shower by deflecting the charged particles in the electromagnetic
cascade.  The GF deflects particles in a plane perpendicular to their
trajectories with a strength proportional to the perpendicular
component of the GF vector.  For the shower particles that
predominantly contribute to the emitted Cherenkov light, the
perpendicular component is comparable to the GF vector component
perpendicular to the shower direction ($\mathbf{B}_{\perp} =
\mathbf{B} - (\mathbf{B}\cdot \hat{z}')\hat{z}'$).  Deflection of the
shower particles by the GF breaks the azimuthal symmetry of the shower
causing an elongation in the shower particle distribution in the plane
orthogonal to $\mathbf{B}_{\perp}$.

Due to the asymmetry in the shower development induced by the GF, the
Cherenkov light distribution observed by a telescope depends on both
the distance to the shower impact position ($\rho$) and the
orientation of the shower impact vector relative to the GF.  We
parameterize the shower orientation with respect to telescope $i$ by
the shower position angle ($\phi_{i}$) defined by

\begin{equation}
\cos\phi_{i} = \hat{x}'\cdot \frac{\bs{\rho}_{i}}{|\bs{\rho}_{i}|}.
\end{equation}
Telescopes with a shower position angle of 0$^\circ$ and 90$^\circ$
view the shower in the planes parallel and perpendicular to its
elongated axis respectively (see Figure 5).

The shower reconstruction is performed in two consecutive stages.  A
\textit{geometric} reconstruction algorithm is first used to obtain a
robust estimation of the shower parameters.  In this stage the shower
energy and interaction depth are initially assigned using
\textit{look-up tables}.  In the second stage the shower parameters
derived from the geometric reconstruction are refined using a
\textit{likelihood} reconstruction algorithm that performs a joint fit
to the image intensity in all telescopes.

\subsubsection{Geometric Reconstruction}
  
The geometric reconstruction algorithm is a 3-D stereoscopic
reconstruction technique based on the traditional Hillas image
parameterization of the shower images \citep{hillas1985}.  The emitted
Cherenkov light from a gamma-ray shower produces an approximately
elliptical distribution in the telescope focal plane with the major
axis of the ellipse aligned with the shower trajectory.
The projected shower trajectory as observed by a telescope with impact
vector $\bs{\rho}_{i}$ can be described by the equation

\begin{equation}
  \mathbf{e}_{s,i}(t) = \frac{\bs{\rho}_{i} +
    \mathbf{e}t}{|\bs{\rho}_{i} + \mathbf{e}t|},
\end{equation}
\newtext{where $t$ is the physical distance along the shower axis from
  its intersection with the shower plane.}  Each telescope that
observes the shower constrains the trajectory to lie in the plane
formed by the vectors $\hat{z}'$ and $\bs{\rho}_{i}$.  When multiple
telescope images are present, the intersection point of the projected
shower axes provides a unique solution for both the shower direction
($\mathbf{e}$) and its impact position in the shower plane
($\mathbf{R}'$).

The solution for the shower trajectory that best matches the projected
shower axes of all telescopes is found by minimizing a pair of
$\chi^{2}$-like parameters that independently optimize the shower
direction and core position.  In the case of the shower direction we
solve for the vector $\mathbf{e}$ that minimizes

\begin{equation}
\chi^2_{e}(\mathbf{e}) = 
\sum_{i} \kappa(S_{i},w_{i},l_{i})\Delta_{e,i}(\mathbf{e})^2,
\end{equation}
where $\Delta_{e,i}(\mathbf{e})$ is the distance of closest approach
between the major axis of the image ellipse and the shower direction
projected to the image plane of telescope $i$, and $\kappa$ is a
weighting function that controls the contribution of each telescope to
the total sum.  Images that are brighter and more elongated provide a
better constraint on the shower trajectory, and therefore we use as
our weighting function the product of the image size with square of
the image ellipse eccentricity,

\begin{equation}
\newtext{\kappa(S_{i},w_{i},l_{i}) = S_{i} \frac{l_{i}^2-w_{i}^{2}}{l_{i}^2}.}
\end{equation}

The shower core position is reconstructed by minimizing

\begin{equation}
\chi^2_{R}(\mathbf{R}) = \sum_{i} \kappa(S_{i},w_{i},l_{i})\Delta_{R,i}(\mathbf{R})^2,
\end{equation}
where
\begin{equation}
\Delta_{R,i}(\mathbf{R}) = \left|\bs{\rho}_{i}(\mathbf{R})-
\left(\bs{\rho}_{i}(\mathbf{R})\cdot
\mathbf{e}_{\rho,i}\right)\mathbf{e}_{\rho,i}\right|
\end{equation}
is the distance of closest of approach between the image axis of
telescope $i$ projected to the shower plane ($\mathbf{e}_{\rho,i}$)
and the core location.  After reconstruction of the shower trajectory,
the shower energy is reconstructed using look-up tables for the shower
energy as a function of the image size and impact distance from the
telescope.  The shower energy estimate is calculated from a weighted
average of telescope energy estimates given by

\begin{equation}
E = \left(\sum_{i} \sigma_{E}(S_{i},\rho_{i})\right)^{-1}
\sum_{i}\frac{E(S_{i},\rho_{i})}{\sigma_{E}(S_{i},\rho_{i})},
\end{equation}
where $E(S_{i},\rho_{i})$ and $\sigma_{E}(S_{i},\rho_{i})$ are
functions for the expectation value and standard deviation of the
shower energy derived from simulations.

\subsubsection{Likelihood Reconstruction}

The likelihood reconstruction performs a global fit to the telescope
image data using a model for the expected pixel amplitude
$\mu(\bs{\theta})$ as a function of the shower parameters
$\bs{\theta}$.  Pixel expectation values are evaluated from an image
template model, $I(\mathbf{e};\bs{\rho},\bs{\theta})$, a probability
distribution function for the image intensity in the direction
$\mathbf{e}$ as measured by a telescope that observes a shower with
parameters $\bs{\theta}$ and impact vector $\bs{\rho}$.  More
details on the generation of the image intensity model are presented
in Section \ref{SEC:IMAGE_TEMPLATES}.  The agreement between the
telescope image model and the data is evaluated by means of an array
likelihood function.  Shower parameters are determined by a
maximization of an array likelihood function.  Maximization of the
array likelihood as a function of shower fit parameters is performed
using a numerical non-linear optimization technique.  In order to
ensure stable fit convergence, the shower parameters are initially
seeded with a set of values derived by the geometric reconstruction
($\bs{\theta}_{ \rm{geo}}$).

We use a formulation of the array likelihood function which is similar
to the one presented in \cite{2009APh....32..231D}.  The array
likelihood is computed from a pixel-by-pixel comparison between the
observed and predicted image intensities.  The likelihood provides a
statistical model for the measured pixel signal ($s$) as a function of
input models for signal and background.  The measured pixel signal is
modeled as the sum of three components: Cherenkov signal photons, 
NSB photons, and Gaussian noise arising from
detector fluctuations.  The pixel likelihood function is

\begin{equation}\label{eqn:PIXEL_LIKELIHOOD}
L{_\text{pix}}(s|\mu(\bs{\theta}),\mu_{b},\sigma_{b},\sspe)
= \sum_{n} \frac{(\mu+\mu_{b})^{n}e^{-(\mu+\mu_{b})}}{n!}g(s,n),
\end{equation}
where $\mu$ is the model amplitude, $\mu_{b}$ is the NSB amplitude,
$\sigma_{b}$ is the standard deviation of the detector noise,
$\sspe$ is the width of the single PE response function, and

\begin{equation}
g(s,n) =
\frac{1}{\sqrt{2\pi\left(\sigma_{b}^2+n\sspe^2\right)}}
\exp\left[-\frac{(s-n)^2}{2\left(\sigma_{b}^2+n\sspe^2\right)}\right].
\end{equation}
The model amplitude for pixel $j$ in telescope $i$ is calculated by an
integration of the image template model over the pixel,

\begin{equation}
\mu_{ij}(\bs{\theta}) = 
\int_{\Omega_{ij}}I(\mathbf{e};\bs{\rho}_{i},\bs{\theta})d\Omega,
\end{equation}
where $\Omega_{ij}$ is the 2-D angular integration region.

The array likelihood is calculated from the product of the pixel
likelihoods in all telescopes,

\begin{equation}\label{eqn:ARRAY_LIKELIHOOD}
  L(\mathbf{s}|\bs{\mu}(\bs{\theta}),\mu_{b},\sigma_{b},\sigma_{\gamma}) 
  = \prod_{i,j} 
  L_{\text{pix}}(s_{ij}|\mu_{ij}(\bs{\theta}),\mu_{b},\sigma_{b},\sigma_{\gamma}),
\end{equation}
where $s_{i,j}$ and $\mu_{i,j}$ are the signal and model amplitude of
pixel $i$ in telescope $j$.  The set of pixels included in the
computation of Equation \ref{eqn:ARRAY_LIKELIHOOD} can encompass the
entire camera.  Unlike for the geometric reconstruction techniques,
each pixel is weighted by its expected contribution to the total image
intensity.  Therefore the inclusion of pixels on the shower periphery
does not significantly improve or degrade the reconstruction
performance.  Although the array likelihood can be calculated using
all pixels in the camera, using a smaller number of pixels
significantly reduces the computation time needed for the shower
likelihood optimization.  In order to select pixels that will provide
a useful constraint on the shower parameters, we choose a set of
pixels $\mathcal{P}$ in each telescope that satisfies the relation

\begin{equation}
\sum_{j \in \mathcal{P}} \mu_{j}(\bs{\theta}_{\rm{geo}}) \geq f\sum_{j} 
\mu_{j}(\bs{\theta}_{\rm{geo}}),
\end{equation}
where $\mu_{j}$ is the expected image intensity in pixel $j$ and $f$
is the fraction of the total image intensity.  We build the set
$\mathcal{P}$ by adding pixels in order of their expected intensity
until the total amplitude fraction exceeds $f$.  Having found that the
reconstruction performance is relatively insensitive to $f$ for values
$\gtrsim 0.75$, we use $f = 0.75$.

An underlying assumption of the likelihood formulation presented here
is that the shower can be treated as a continuous distribution of
particles.  However, because the electromagnetic cascade is a
stochastic process, non-statistical deviations from the image model
are expected due to fluctuations in the shower development.  These
deviations become especially important at low energies where the total
number of shower particles is small and the influence of the GF
becomes large.  These shower fluctuations will tend to worsen the
performance of the method relative to what would be expected in the
case of purely statistical fluctuations.

\subsubsection{Image Templates}\label{SEC:IMAGE_TEMPLATES}

The image model, $I(\mathbf{e};\bs{\rho},\bs{\theta})$, is the
probability distribution function for the measured telescope image
intensity in photoelectrons (PEs) versus direction, $\mathbf{e}$.  The
model is parameterized as a function of the shower properties (energy
and first interaction depth) and the impact position of the shower
relative to the telescope.
The model is generated by averaging the intensity of a large sample of
simulated showers generated at a sequence of fixed offsets, energies
and interaction depths.  The image templates for this study were
generated with the CORSIKA shower simulation package and the detector
simulation described in Section \ref{subsec:det_model}.  While the
image templates used for this study are MC-based we note that the
likelihood reconstruction can also be applied using templates
generated with semi-analytic shower models
\citep{2009APh....32..231D}.
\begin{figure}[tb]
\includefigure{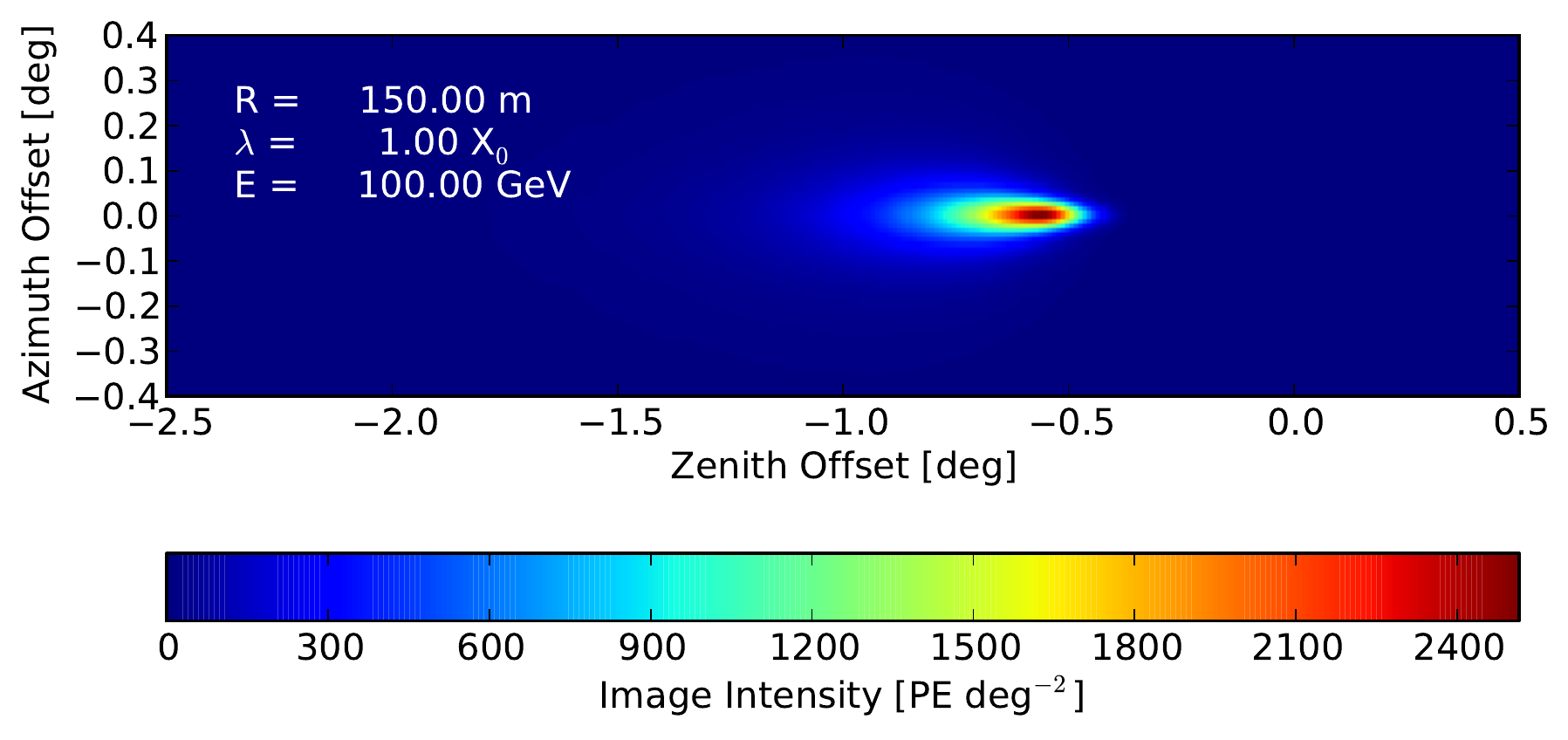}
\includefigure{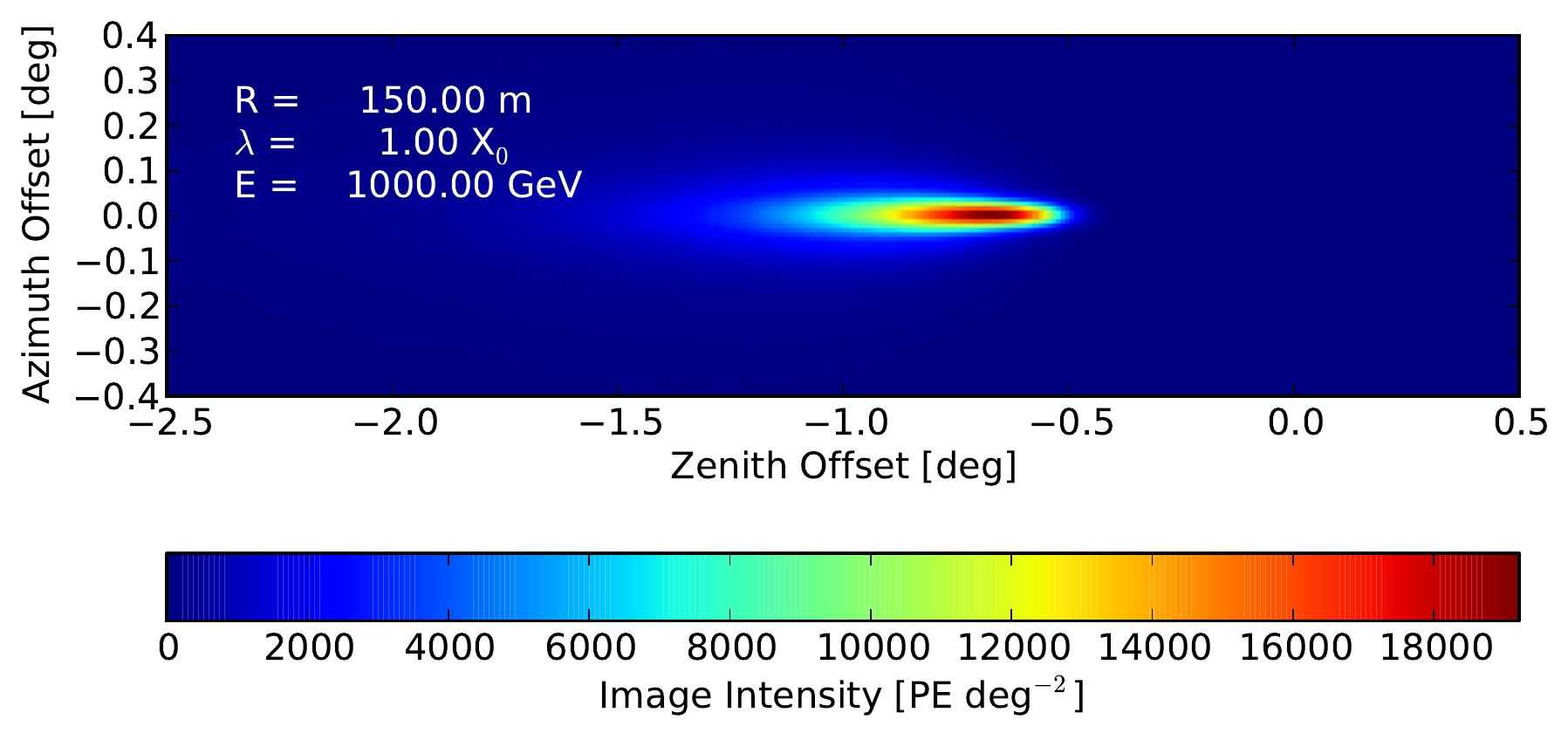}
\includefigure{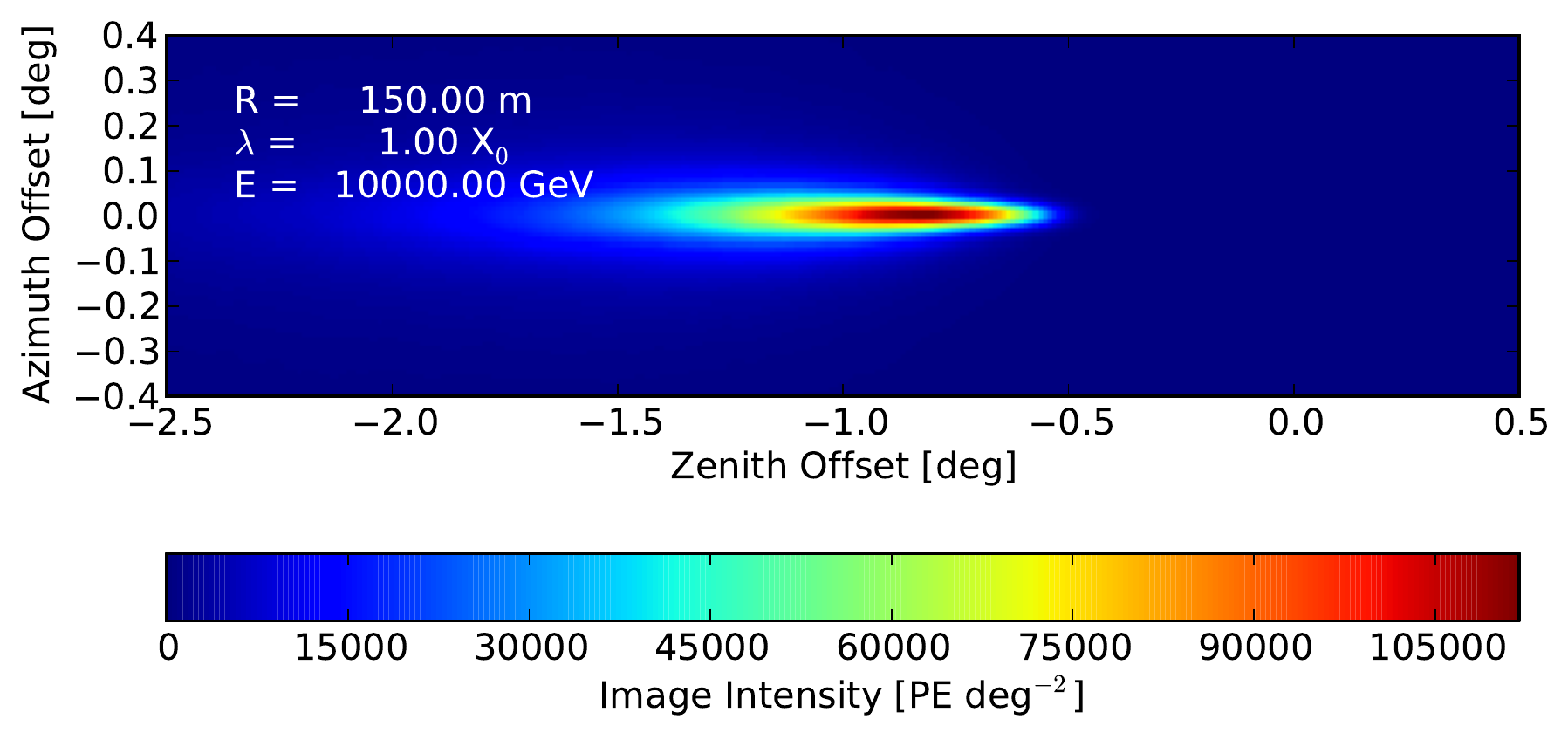}
\caption{\label{FIG:TEMPLATE_ENERGY}Image intensity templates for
  three different three gamma-ray energies (100 GeV, 1 TeV, and 10
  TeV) generated for a telescope model with $\rpsf = 0.02^\circ$ and
  $\aopt = 11.18$~m$^{2}$.  The images show the expectation for the
  measured intensity of Cherenkov light as a function of angular
  offset from the primary gamma-ray direction.  The image templates
  shown here are evaluated at an impact distance ($\rho$) of 150~m and
  a first interaction depth ($\lambda$) of 1~X$_{0}$.}
\end{figure}

Because the templates are produced from a simulation of the shower,
the image model incorporates all effects that influence the measured
image intensity including atmospheric attenuation, geomagnetic field,
telescope optics, and telescope detector response.  The image model is
a continuous distribution for the shower photons in the focal plane
and the same template can therefore be used to compute the image
intensity for cameras with arbitrary pixel geometry and field-of-view.
For this study, we use image templates computed for the baseline
telescope model with $D = 12$~m and $\aopt = 11.18$~m$^{2}$.  The
image intensity for other telescopes is calculated by rescaling the
image intensity by the ratio of the telescope light collection area to
the baseline telescope model.

The image intensity templates are stored on a six-dimensional grid:
\begin{itemize}
\item $\mathrm{log}_{10}$(Energy) and Interaction Depth ($\mathrm{log}_{10}E$ and $\lambda$)
\item Core Impact Distance and Position Angle ($\rho$ and $\phi$) 
\item Projected Zenith and Azimuth Offset in Image Template Coordinates 
\end{itemize}
\newtext{ The shower templates are defined in a coordinate system
  rotated by $\phi_{i}$ with respect to the shower axis such that the
  $x$-axis is aligned with the shower axis.  In order to keep the
  memory footprint of the full six-dimensional template at a
  manageable level, the image templates are only defined over an
  angular region within one degree of the shower axis.}

The expected image intensity is computed from the image template
sequence by a linear interpolation in the six-dimensional template
space.  The templates are also used to derive first derivatives of the
image intensity as a function of the shower parameters which are used
for calculation of the likelihood gradient.

Figure \ref{FIG:TEMPLATE_ENERGY} shows the image templates evaluated
for gamma-ray showers of three different energies.  The primary energy
affects both the total intensity of the shower image as well as its
shape.  Higher energy showers propagate further into the atmosphere
and result in shower images that are more extended along the shower
axis.  The core impact distance sets the geometrical perspective of
the telescope and selects the Cherenkov light emission from particles
with a specific range of angles with respect to the shower axis.
Showers observed inside the Cherenkov light pool ($\rho \lesssim
120$~m) appear both brighter and narrower as the telescope accepts
Cherenkov light from higher energy particles that are closely aligned
with the shower primary.  More distant showers are dimmer and
increasingly offset from the primary origin.  The interaction depth
sets the starting point of the shower and primarily influences the
displacement of the shower image along the shower axis.

\begin{figure*}[tb]
\includegraphics[width=.33\textwidth, keepaspectratio]{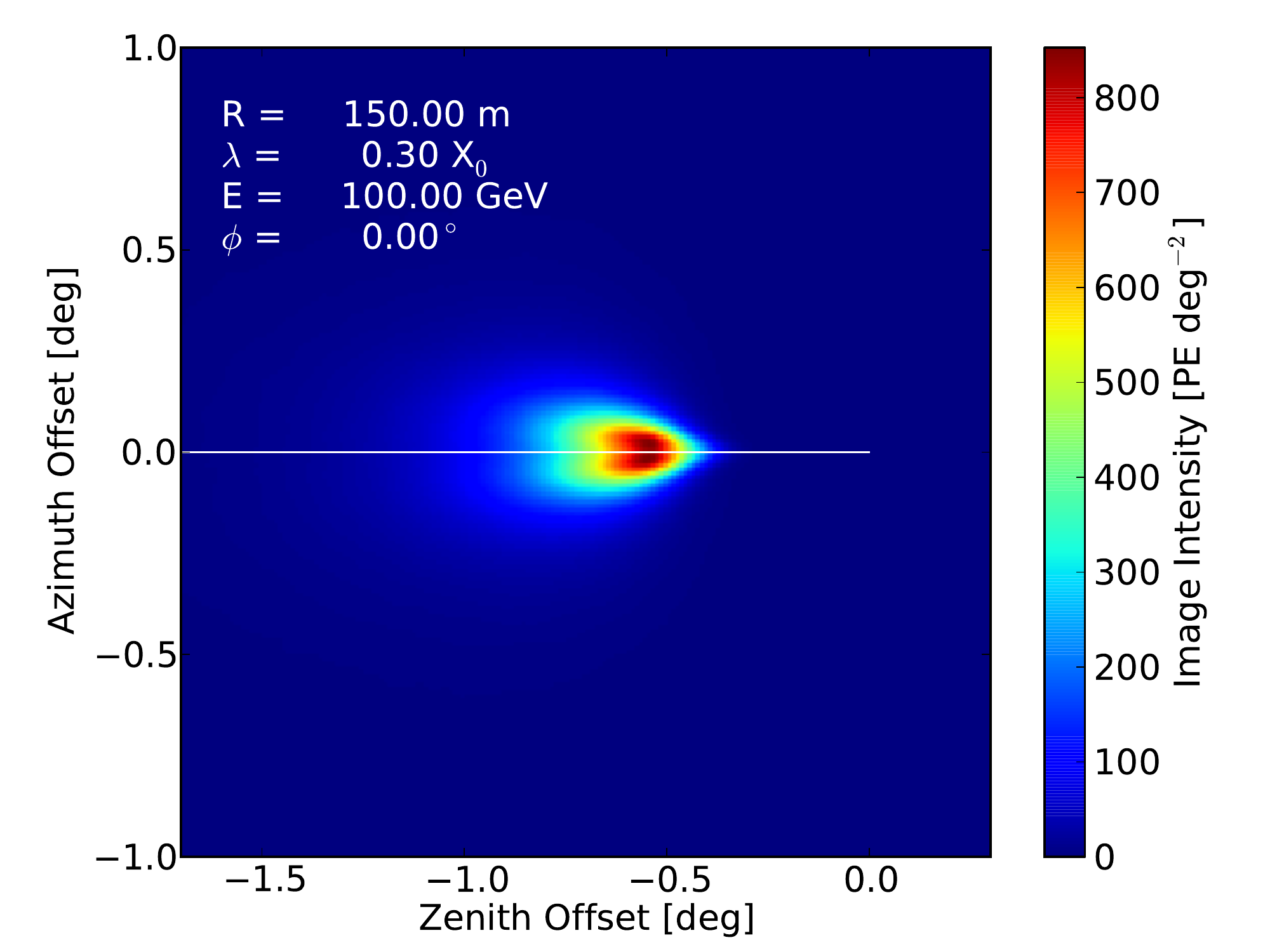}
 \includegraphics[width=.33\textwidth, keepaspectratio]{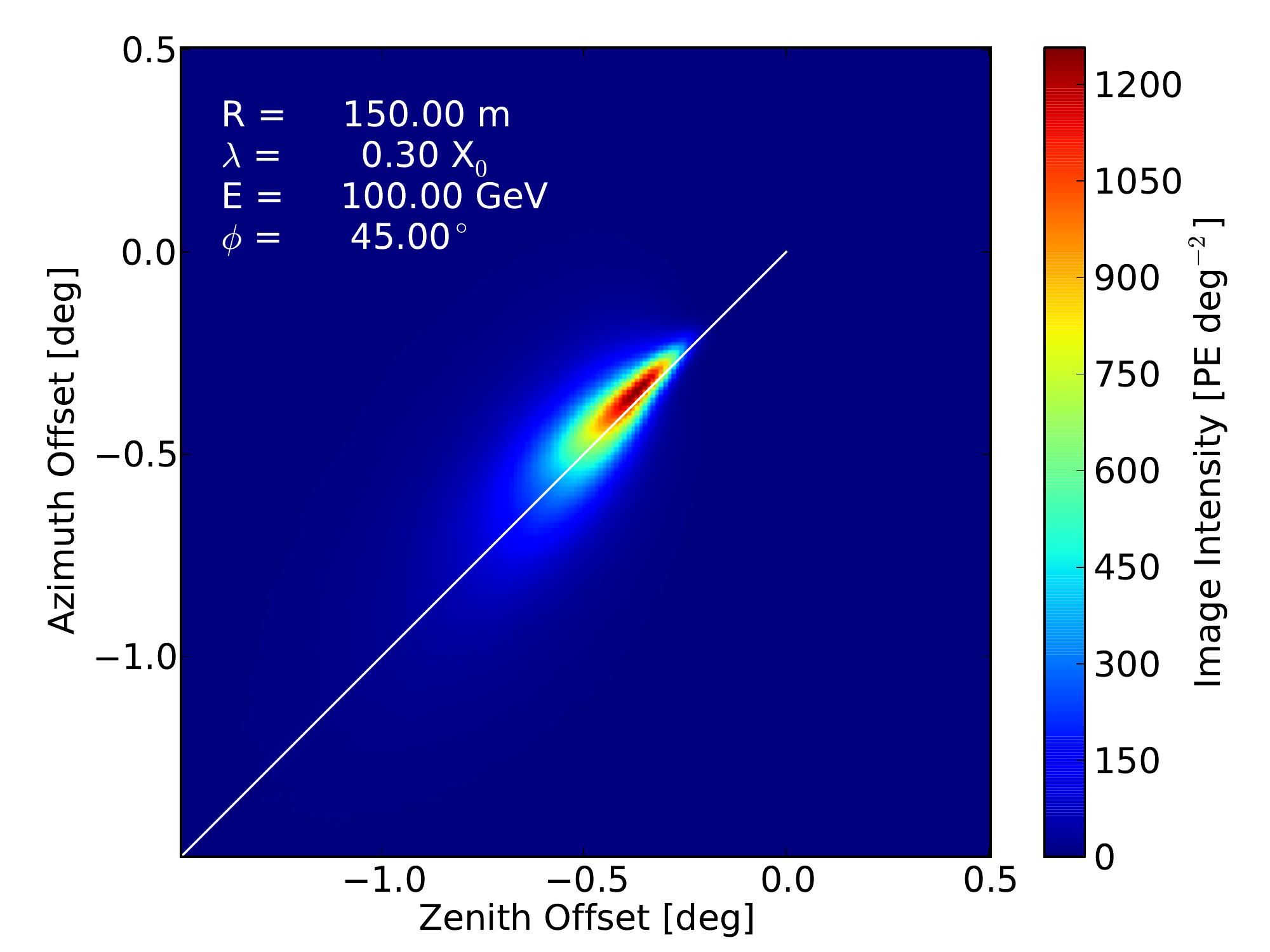}
\includegraphics[width=.33\textwidth, keepaspectratio]{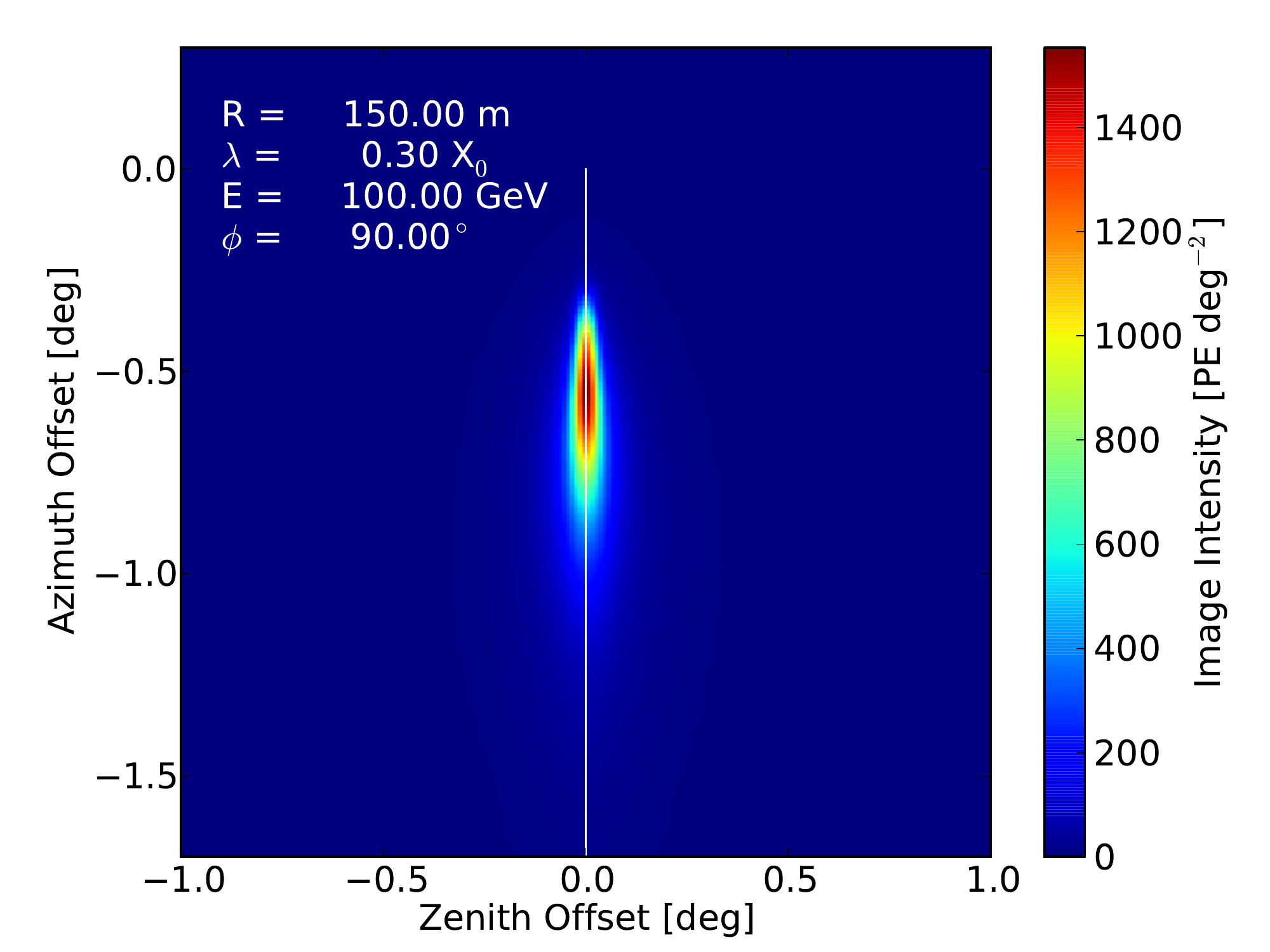}
\caption{\label{FIG:TEMPLATE_PHI}Image intensity templates as a
  function of angular offset from the primary gamma-ray direction for
  three values of the shower position angle: $\phi= 0^\circ$, $\phi=
  45^\circ$, $\phi= 90^\circ$. The templates shown are evaluated for a
  gamma-ray shower with an energy of 100~GeV, an impact distance of
  150~m, and an interaction depth of 0.3~X$_{0}$.  The solid white
  line in each image shows the projection of the primary trajectory to
  the image plane.}
\end{figure*}

In the absence of the GF, the shower template is symmetric with
respect to the shower position angle.  The GF breaks the axial symmetry
as the Lorentz force preferentially perturbs the trajectory of the
shower particles into the plane orthogonal to $\mathbf{B}_{\perp}$.  The GF effect is
especially pronounced for showers with small interaction depth for
which the average propagation distance between the first and second
interactions is large.  Figure \ref{FIG:TEMPLATE_PHI} illustrates the
impact of the GF on the image template for three values of the core
position angle ($\phi$): $0^\circ$ (parallel), $45^\circ$, and
$90^\circ$ (perpendicular).  The shower width monotonically \newtext{decreases}
as the shower position angle is increased from $0^\circ$ to $90^\circ$
reflecting the asymmetry in the shower development.  For intermediate
viewing angles ($\phi = 45^\circ$), the GF also causes a 
rotation of the image major axis relative to the shower axis.

\subsection{Gamma/Hadron Separation and Cut Optimization}\label{subsec:gamma_hadron}


The final stage of the event analysis determines parameters that can
be used for discrimination between cosmic- and gamma-ray initiated air
showers.  The Cherenkov images produced by cosmic-ray showers can
generally be distinguished from gamma-ray showers by their wider and
more irregular appearance.  Hadronic subshowers may also produce
isolated clusters of Cherenkov light in the telescope image plane.
\newtext{A widely used set of parameters for background discrimination
  are the so-called \textit{mean scaled parameters} defined already in
  the HEGRA collaboration \citep{1997APh.....8....1D} and extensively
  used by current generation IACT experiments (see e.g.,
  \citep{2006APh....25..380K}).  The mean scaled parameters provide a
  measure of the deviation between the observed and expected telescope
  image moments for a gamma-ray shower.}
Using a set of simulated gamma-ray showers, lookup tables for the mean
and standard deviation of the image moment parameters are produced as
a function of the telescope image size and telescope impact distance
(denoted here as $p(S,\rho)$ and $\sigma_{p}(S,\rho)$).  For a
telescope image parameter $p_{i}$ we define the array-level parameter
as

\begin{equation}
p = \left(\sum_{i} w_{i}\right)^{-1}
\sum_{i}w_{i}\frac{p_{i} - p(S_{i},\rho_{i})}{\sigma_{p}(S_{i},\rho_{i})},
\end{equation}
where the sum is over all telescopes with reconstructed image
parameters and $w_{i}$ is a weighting factor.  We use $w_{i} =
S_{i}/\sigma_{p}(S_{i},\rho_{i})$ assigning a larger weight to
telescopes with brighter images and a smaller expected dispersion in
the image parameter.

A second class of discriminant variables can be obtained by computing
a goodness-of-fit between the data and the image template model
evaluated at the best-fit shower parameters \citep{2009APh....32..231D}.
When considering Gaussian-distributed data the natural goodness-of-fit
parameter is the $\chi^{2}$ statistic.  
For the purposes of background discrimination, it is not critical to
have an exact model for the asymptotic distribution of the test
statistic as long as it provides good separation power between signal
and background.  To quantify the agreement between the measured and
expected pixel signals we define a $\chi^{2}$-like parameter which we
call the goodness-of-fit,

\begin{equation}\label{EQN:GOODNESS}
\mathcal{G} = \frac{1}{N}\sum_{i}\sum_{j \in \mathcal{P}_{i}}\frac{\left(s_{i,j}-\mu_{i,j}(\bs{\theta})\right)^2}{\mu_{i,j}(\bs{\theta})+\mu_{b}},
\end{equation}
where $\mathcal{P}_{i}$ is a set of pixels in telescope $i$ and $N$ is
the total number of pixels in the summation.  We found that the best
separating power was achieved by evaluating Equation
\ref{EQN:GOODNESS} using the set of telescope pixels that survive
image cleaning, which we refer to as the \textit{image
  goodness-of-fit}.

To maximally exploit the rejection power drawn from the ensemble of
event parameters we further make use of boosted decision trees (BDTs)
generated with the TMVA package \citep{2007physics...3039H}.  The use
of machine learning techniques have been shown to provide significant
improvement in overall background rejection power when applied to IACT
data \citep{2009APh....31..383O}.  We specifically use BDTs trained
with the \textit{GradBoost} algorithm with 200 trees, a maximum depth
of 8, and a shrinkage parameter of 0.1.  We train the decision tree
analysis using the following six parameters: mean scaled width, mean
scaled length, mean scaled displacement, array core distance, first
interaction depth, and image goodness-of-fit.  In order to avoid
overtraining we use a training data set that constitutes 20\% of the
total gamma-ray and proton Monte Carlo samples.

%% file: section4.tex
\section{Results}
\label{sec:results}

Using the simulation and analysis frameworks described in
Sections~\ref{sec:simulation} and \ref{sec:analysis}, we have explored
the influence of the telescope design on the sensitivity and gamma-ray
reconstruction performance of various array design concepts.
Section~\ref{subsec:metrics} outlines the performance metrics used for
comparison of the arrays.
Section~\ref{subsec:benchmark_arrays_define} defines a reference array
alongside several benchmark configurations which are representative of
realistic telescope and array configurations that will be chosen for
CTA.  In the subsequent sections we examine the influence of each
telescope parameter on the array performance.  In
Section~\ref{subsec:benchmark_arrays_compare} we study the performance
of the benchmark arrays.

\subsection{Performance Metrics and Cut Optimization}\label{subsec:metrics}

Our primary metric for the comparison of different array and telescope
designs is the differential gamma-ray point-source sensitivity
evaluated following the standard procedure for CTA-related studies
\citep{2013APh....43..171B}.  The differential sensitivity is
evaluated in a sequence of logarithmic bins of reconstructed energy
with a width of 0.2 dex (five bins per decade of energy).  In each
energy bin we calculate the expected number of signal and background
events within an energy-dependent aperture of radius $\theta$.  The
number of signal events is estimated assuming a gamma-ray point-source
in the center of the FoV.  The residual background rate is estimated
by scaling the \newtext{number of background events} reconstructed in
the inner $3^{\circ}$ of the camera to the gamma-ray extraction area.
In each bin we require a 5$\sigma$ excess above background and at
least 10 signal events.  The source significance is calculated using
the method of \cite{1983ApJ...272..317L} and a signal-free background
region with a solid angle five times larger than the signal aperture.
We further require a signal with a fractional amplitude above
background of 5\% in order to account for systematic errors in
background estimation.

Sensitivity to spatially extended gamma-ray sources is calculated
following the same procedure but with the gamma-ray signal spread out
uniformly over a disk with angular diameter $D$.  For a source with a
given flux, the diffuse source sensitivity is always worse than the
point-source sensitivity.  In the case of a point-source, the
sensitivity depends on both background rejection efficiency and the
PSF.  The diffuse-source sensitivity, however, depends primarily on
the background rejection efficiency and is nearly independent of the
PSF when $D$ is larger than the PSF.  

The quality of the gamma-ray reconstruction is estimated from the 
simulated gamma-ray PSF, shower core resolution, and energy resolution.
The most important of these quantities is the gamma-ray PSF as it
directly impacts the sensitivity to point-sources and the morphology
of extended gamma-ray sources.  We characterize the gamma-ray PSF by
the radius that contains 68\% of the distribution of reconstruction
errors (68\% containment radius).

When evaluating the performance of an array we apply several selection
criteria to reject both background events and gamma-ray events with
poor reconstruction quality.
\textit{Point-source} cuts are composed of two energy-dependent
selections on the gamma/hadron rejection parameter and aperture
radius, $\xi(E)$ and $\theta(E)$, parameterized as a cubic spline.
The shape of these parameterizations is independently optimized for
each array and exposure time to maximize the differential point-source
sensitivity versus energy.  At high energies \newtext{where the
  sensitivity} of IACT arrays transitions from being background- to
signal-limited, the optimal point-source sensitivity is obtained by
increasing the gamma-ray efficiency and including events with poorer
reconstruction quality and a higher background contamination
level. \textit{Diffuse-source} cuts are used when evaluating diffuse
source sensitivity and comprise the same selection on the gamma/hadron
parameter but with the aperture size fixed to the radius of the source
($\theta(E) = D/2$).

\textit{Reconstruction} cuts are used to define a homogeneous sample
of well-reconstructed showers with core locations within a predefined
fiducial area of the array.  Showers passing reconstruction cuts must
have an impact distance from the array center that is less than 1.2 times 
the distance from the center of the array to the nearest point along 
the array edge (as defined by the outer ring of telescopes).  Showers with core locations
near or within the array boundary (\textit{contained} events) are
sampled by a large number of telescopes that view the shower from
multiple perspectives and allow for a more precise stereoscopic
reconstruction of the shower trajectory.  In arrays with mean
telescope separations on par with the Cherenkov light pool size,
contained events will also have one or more telescopes that sample the
shower within its Cherenkov light pool, where the Cherenkov light from
the highest energy shower particles is visible.  The light emitted
from these particles provides a much better constraint on the shower
trajectory than the light emitted by lower energy shower particles.
Events outside the array boundary (\textit{uncontained} events) are
sampled by a smaller number of telescopes for which the viewing angles
are more closely aligned.  This results in a less precise
determination of the shower trajectory.


Reconstruction cuts provide a measure of the gamma-ray reconstruction
performance that can be evaluated independently of the source strength
and exposure time.  Relative to point-source cuts, reconstruction cuts
offer worse point-source sensitivity but a significantly better
gamma-ray PSF at high energies (above 1~TeV).  The improvement in the
gamma-ray PSF can be attributed to the removal of uncontained events
which are bright enough to trigger the array at high energies.  This
selection is very useful when studying strong sources to check
morphology and spectral features while not relying on the best
signal-to-noise ratio.

\subsection{Benchmark Arrays}\label{subsec:benchmark_arrays_define}

\begin{table}[tb]
  \caption{Geometrical characteristics and optical performance 
    of the camera and optical systems of the DC-MST, SC-MST,
    and LST telescope designs chosen for the \textit{prod-2} MC design
    study \citep{Bernlohr:2013bla}.  The off-axis PSF performance is
    evaluated at a field angle equal to 3/4 of the distance to the
    edge of the FoV.  \newtext{For the LST and DC-MST designs which have
      cameras composed of
      hexagonal pixels, the given value of $\rpix$ is the width of a
      square pixel with the same solid angle as the hexagonal pixel
      used in that design.}}
\label{TABLE:PROD1_TELESCOPE_PARAMETERS}
\centering
\begin{tabular}{l|ccc}
&LST&SC-MST&DC-MST\\
\hline
$\rpix$ [$^\circ$]&0.084&\newtext{0.067}&\newtext{0.171}\\
$A_{M}$ [m$^2$]&412&50&100\\
$\aopt$ [m$^2$]&52.5&7.29&13.65\\
$\rpsf$ [$^\circ$] (on-axis)&0.03&0.04&0.04\\
$\rpsf$ [$^\circ$] (off-axis)&0.12&0.04&0.08\\
FoV [$^\circ$]&4.5&8&8\\
\end{tabular}
\end{table}

The baseline CTA concept is an array of 50--100 telescopes distributed
over an area of \newtext{$\sim$4--5~km$^{2}$} and composed of small-,
medium-, and large-sized telescopes.  Previous simulation studies have
found that intermediate layouts with a few (3-4) large-sized, on the
order of 20 medium-sized, and 50--60 small-sized telescopes offers the
best tradeoff in performance over the full energy range of CTA
\citep{2013APh....43..171B}. Table
\ref{TABLE:PROD1_TELESCOPE_PARAMETERS} shows the primary
characteristics of the currently considered designs for the large- and
medium-sized telescopes.  The LST design is optimized for sensitivity
at gamma-ray energies below 100~GeV and features a large effective
mirror area which enables efficient triggering and reconstruction of
low energy showers.  The DC-MST and SC-MST are two alternative designs
for the medium-sized telescope that would provide sensitivity in the
core energy range of CTA (100~GeV--10~TeV).  The DC-MST is a single
dish telescope that is similar in overall design to current generation
IACTs \newtext{with a camera composed of hexagonal pixels with
  flat-to-flat spacing of 0.184$^\circ$.} \footnote{Equivalent in
  solid angle to square pixels with $\rpix=0.171^\circ$.}  The SC-MST
employs the dual-mirror Schwarzschild-Couder optical design and uses a
smaller pixel size of 0.067$^\circ$ to achieve higher resolution
imaging of the gamma-ray shower.  Ray-tracing simulations of the
SC-MST OS with realistic alignment tolerances for the mirrors and
camera focal plane have demonstrated an optical PSF with a 68\%
containment radius between 0.02$^\circ$ and 0.04$^\circ$
\citep{Rousselle:2013nfa}.  Although the array designs considered for
previous MC studies only incorporated DC-MSTs, the higher angular
resolution SC-MST provides a potentially compelling option for the
medium-sized CTA telescope.

\begin{table*}[tb]
  \caption{Number of telescopes and telescope model
    parameters of the benchmark array configurations used for this
    study.  All arrays are composed of telescopes arranged on a
    uniform grid with constant inter-telescope spacing of 120~m as 
    shown in Figure~\ref{FIG:LAYOUT}.  }
\label{TABLE:BENCHMARK_ARRAYS}
\centering
\begin{tabular}{l|cccccc}
Name&$\ntel$&$\aopt$ [m$^2$]&$\rpix$
[$^\circ$]&$\rpsf$ [$^\circ$]&$\tth$ [PE]&$\rnsb$ [MHz]\\
\hline
M61&61&11.18&0.06&0.02&60&14.7\\
M61SC&61&8.38&0.06&0.02&45&11.1\\
M61DC&61&14.91&0.16&0.08&80&139.5\\
M25DC&25&14.91&0.16&0.08&80&139.5\\
L5&5&47.15&0.06&0.02&123&61.9\\
L61&61&47.15&0.06&0.02&123&61.9\\
\end{tabular}
\end{table*}


We consider several different benchmark array configurations shown in
Table \ref{TABLE:BENCHMARK_ARRAYS} to explore the performance of
different array and telescope designs for CTA.  M61 is a reference
array configuration with an effective light collection area that is
intermediate between the DC- and SC-MST designs and an imaging
performance similar to the SC-MST.  M61SC is an array configuration
with the same imaging performance as M61 but with a reduced light
collection area that is more comparable to the SC-MST design.  M61DC
and M25DC are chosen to represent a 61 and 25 telescope array
respectively composed of telescopes of the DC-MST design.  The latter
configuration corresponds to the number of MSTs in the baseline CTA
design \citep{2013APh....43..171B}.  Arrays L5 and L61 are composed of
telescopes with an optical effective area comparable to the LST design
and an imaging performance similar to the SC-MST.

\begin{figure}[tb]
\includegraphics[width=.5\textwidth, keepaspectratio]
{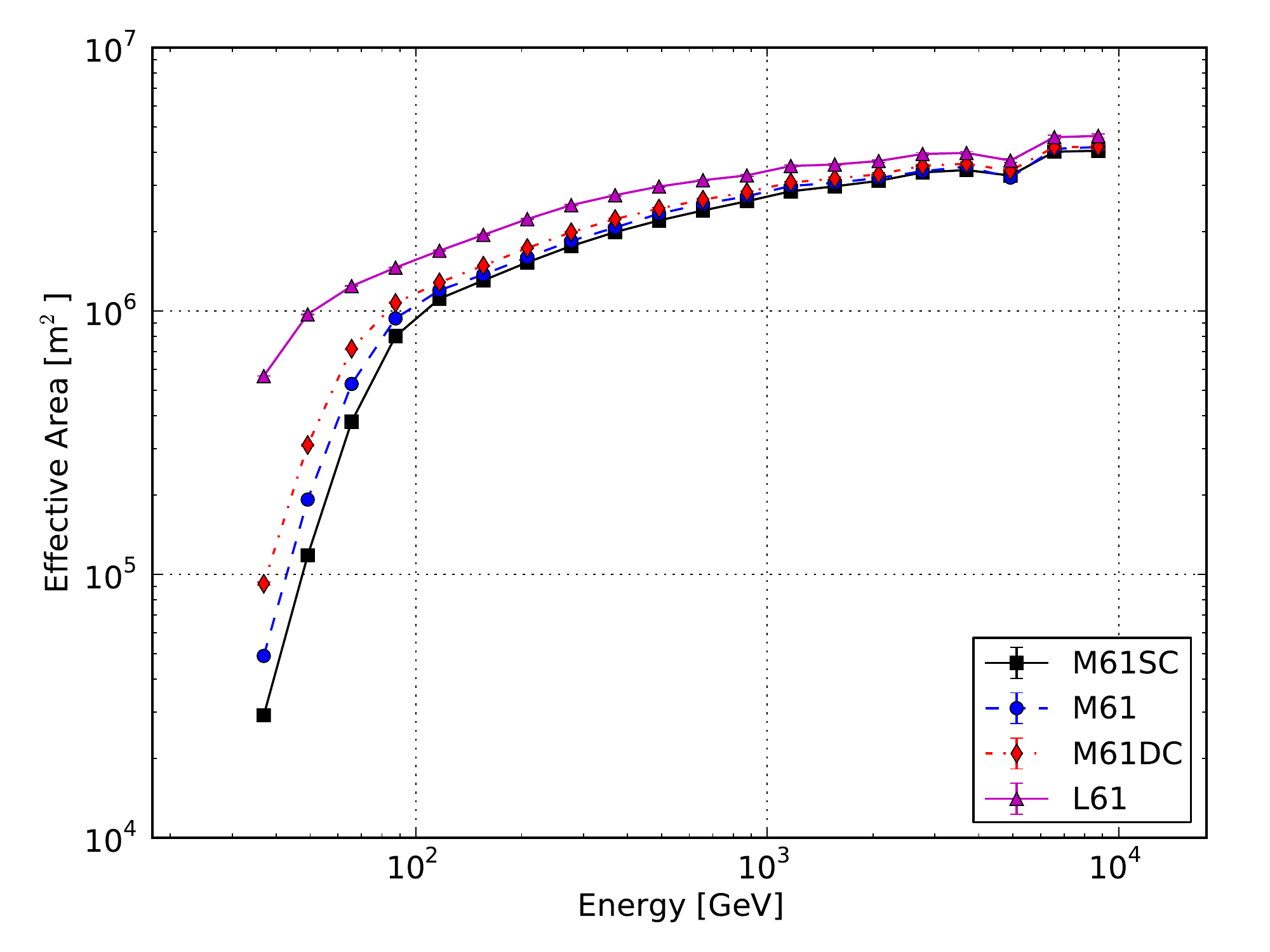}
\caption{\label{FIG:TRIGGER_EFFAREA}Trigger effective area versus
  gamma-ray energy for arrays M61SC, M61, M61DC, and L61.  The camera
  trigger thresholds ($\tth$) for these arrays are 45, 60, 80, and 123
  PE respectively. }
\end{figure}

We show in Figure \ref{FIG:TRIGGER_EFFAREA} the trigger effective area
for arrays M61, M61SC, M61DC, and L61.  The camera trigger threshold
of each array is set using the telescope effective light collection
area and Equation \ref{EQN:TRIGGER}.  The sharp downturn in the
effective area of the MST arrays around 100~GeV can be attributed to
the onset of the trigger energy threshold.  Below the trigger
threshold energy, the image amplitude of an average gamma-ray shower
is insufficient to trigger telescopes within the Cherenkov light pool.
At these energies only showers with large interaction depth can be
effectively recorded, and the total effective area is primarily
determined by the trigger efficiency for contained showers that impact
within the array perimeter.  At higher energies all of the arrays
become fully efficient for triggering contained showers and
differences in the effective area arise predominantly from the
efficiency for detecting showers around the array perimeter.  As the
shower energy increases, the area over which the arrays are fully
efficient continues to increase and eventually extends well beyond the
physical footprint of the array.  Relative differences in the
effective area for telescopes with different $\aopt$ are significantly
smaller at the highest energies as gains in the effective area only
come from showers around the array perimeter.

\subsection{Comparison with Other Telescope Simulations}\label{subsec:simtel_comparison}

The simplifications in the detector response of the FAST simulation
yield a much faster simulation code and enables the study of a broader
parameter space compared to more detailed telescope simulations.  Our
simplified telescope model also enables us to employ a shower
likelihood model which is nearly perfectly matched to the response
characteristics of the telescopes.  These simplifications allow us to
explore the theoretical limit of the performance achievable by an IACT
array when all characteristics of the telescope optics and camera are
accounted for in the event reconstruction.

\begin{figure*}[tb]
\includegraphics[width=.49\textwidth, keepaspectratio]{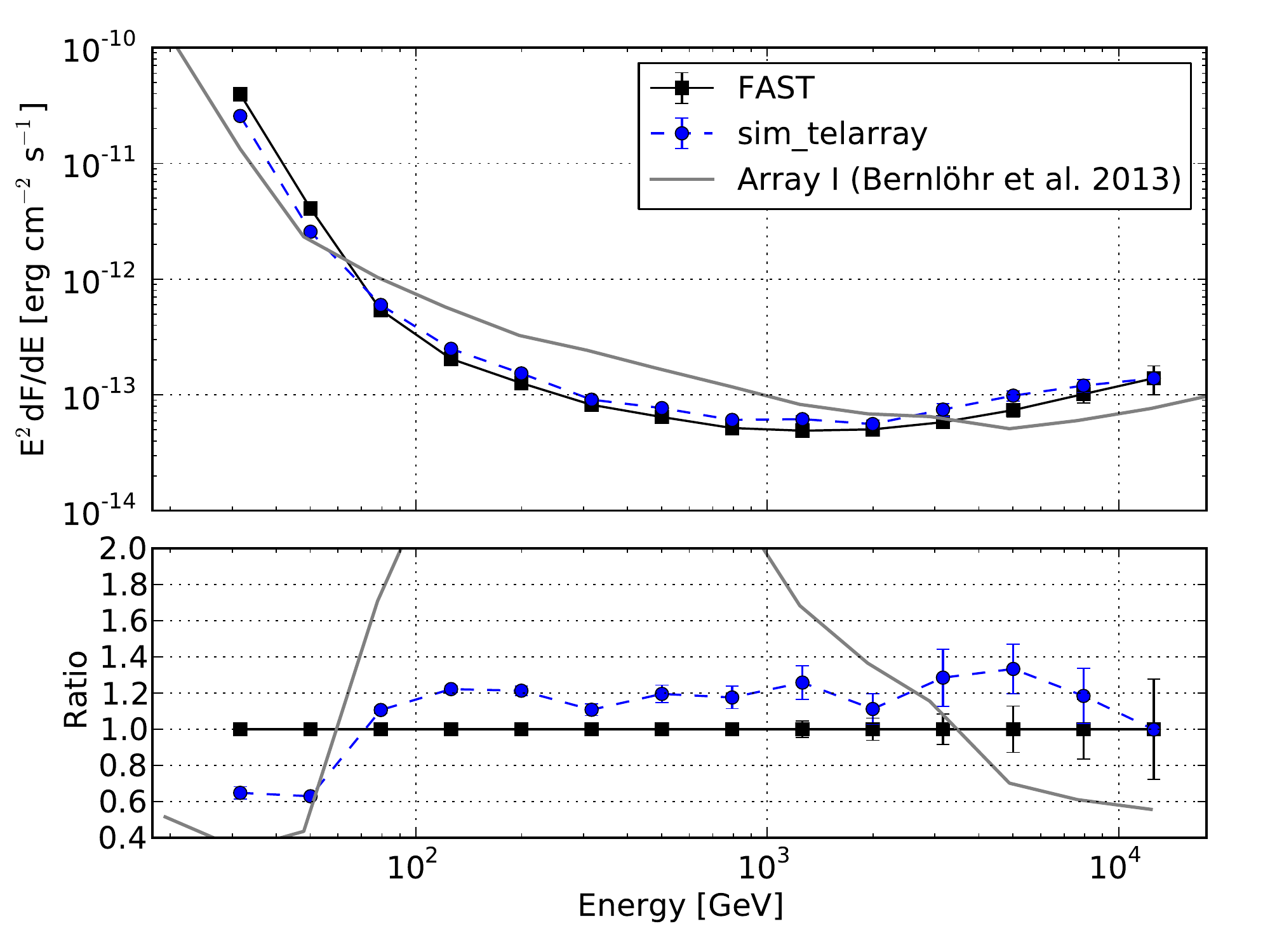}
\includegraphics[width=.49\textwidth, keepaspectratio]{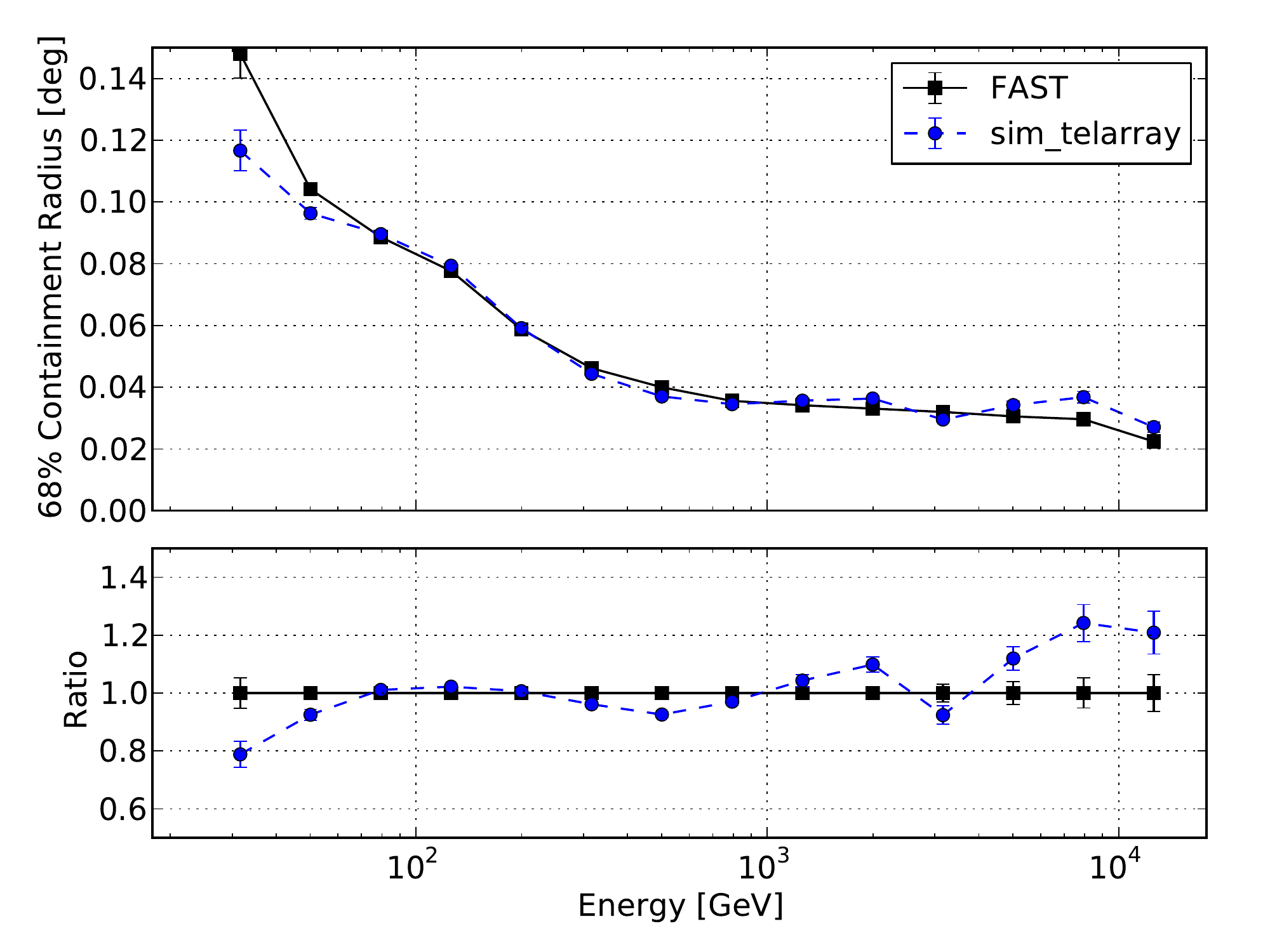}
\caption{\label{FIG:SIMTEL_COMP} Performance of a 61 telescope array
  simulated with FAST (black squares) and \texttt{sim\_telarray} (blue
  circles).  \textbf{Left:} Differential point-source sensitivity for
  a 50~h observation time.  Shown as the solid gray line is the
  differential sensitivity of Array~I from \cite{2013APh....43..171B}
  evaluated with the most sensitive analysis at each energy from the
  four alternative analyses presented in that work (MPIK, IFAE, SAM,
  and Paris-MVA).  \textbf{Right:} 68\% containment radius of the
  gamma-ray PSF after \textit{point-source cuts}. }
\end{figure*}

We have assessed the impact of the simplifications made in the FAST
tool on the derived point-source sensitivity and gamma-ray PSF by
comparing FAST against the well-tested \texttt{sim\_telarray} package.
We use both packages to simulate a 61 telescope array with the same
geometry as our benchmark array layout with 120~m inter-telescope
separation.  For the \texttt{sim\_telarray} simulation we use the
\textit{prod-2} SCT model \citep{Bernlohr:2013bla} with a trigger
pixel threshold of 3.1~PE.  For the FAST simulations, we use a
telescope model with the same performance characteristics as the
\textit{prod-2} SCT model shown in Table
\ref{TABLE:PROD1_TELESCOPE_PARAMETERS} and a camera trigger threshold
of 42~PE.  Relative to the telescope model used for the M61SC
benchmark array, the \textit{prod-2} SCT model has a larger pixel size
and optical point-spread function and a slightly smaller light
collection area.  For gamma-ray showers near the trigger threshold (E
$\sim$ 100~GeV), the \texttt{sim\_telarray} telescope model has a
slightly lower effective camera threshold as compared to the telescope
in our simplified simulations. The choice of a higher threshold for
our simulations was made to be conservative and limit possible
overestimations in sensitivity close to the threshold.

Fig.~\ref{FIG:SIMTEL_COMP} shows the comparison of the array
performance obtained when simulating the same array with
\texttt{sim\_telarray} and FAST.  We include in the same figure the
point-source sensitivity of Array~I from \cite{2013APh....43..171B}
which was simulated using \texttt{sim\_telarray} but with a
different array and telescope setup.  At energies above 75~GeV the
point-source sensitivity obtained with the FAST simulations is 20\%
better than the \texttt{sim\_telarray} simulations.
The gamma-ray PSF and gamma-ray reconstruction efficiency is similar
over the same energy range indicating that the improvement in
point-source sensitivity can be attributed to an enhanced gamma-hadron
separation in the FAST simulations.
Below 50~GeV the \texttt{sim\_telarray} simulations yield a 40\%
better sensitivity due to the slightly lower telescope trigger threshold.
Although we observe measurable differences in the array performance
when comparing our simulations with \texttt{sim\_telarray}, the
differences in point-source sensitivity are much smaller than the
differences between individual analysis packages that use the same
\texttt{sim\_telarray} simulations as input (see e.g. the comparison
of alternative analyses in \cite{2013APh....43..171B}).  Furthermore
the conclusions drawn in this work about the relative performance of
different array and telescope designs is most likely not affected by
these differences.  It is thus easy to scale our results and readily
compare them to other \texttt{sim\_telarray} results.

\subsection{Performance of the Likelihood Reconstruction}

\begin{figure*}[tb]
\includegraphics[width=.49\textwidth, keepaspectratio]{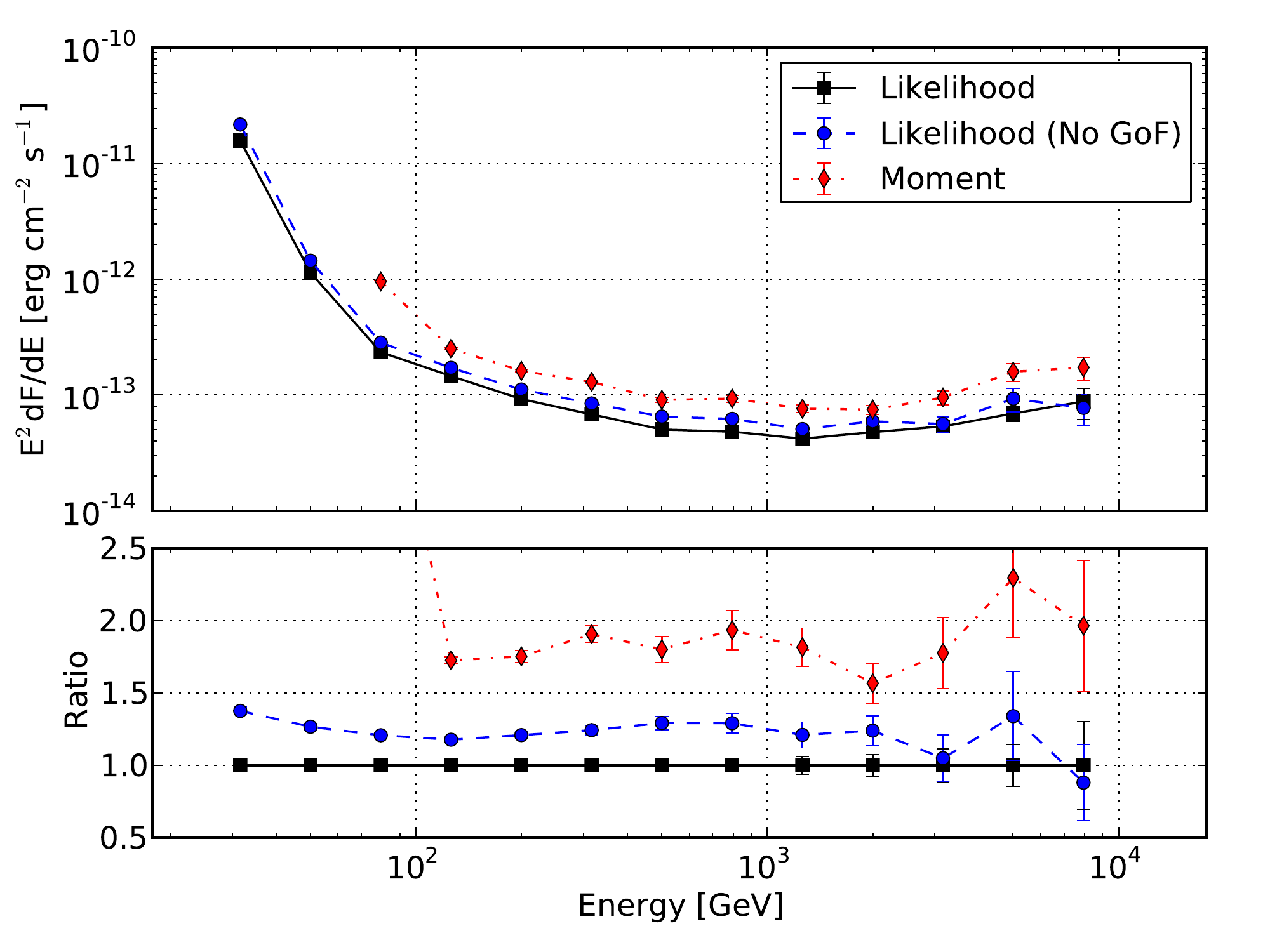}
\includegraphics[width=.49\textwidth, keepaspectratio]{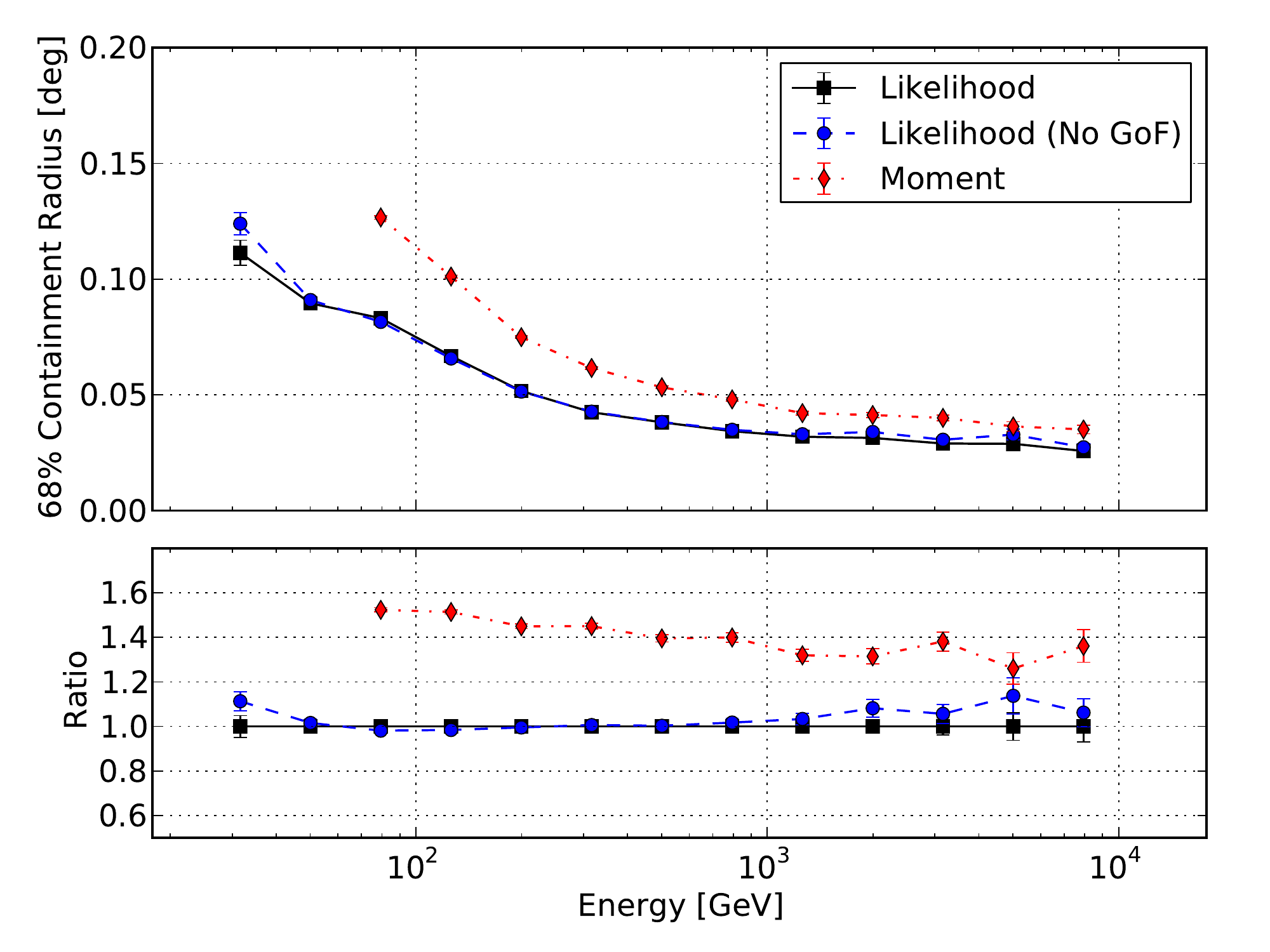}
\caption{\label{FIG:RECON_METHOD} Reconstruction performance and
  gamma-ray point-source sensitivity of array M61 obtained with
  different event reconstruction and analysis algorithms: likelihood
  (black, solid), likelihood without goodness-of-fit (blue, dashed),
  and moment (red, dot-dashed).  \textbf{Left:} Differential
  point-source sensitivity for a 50~h observation time.
  \textbf{Right:} 68\% containment radius of the gamma-ray PSF after
  \textit{point-source cuts}. }
\end{figure*}

Relative to moment-based reconstruction techniques, likelihood-based
reconstruction algorithms have been shown to provide better gamma-ray
angular resolution as well as improved separation between gamma-ray
and cosmic-ray induced showers
\citep{2006APh....25..195L,2009APh....32..231D,2013APh....43..171B}.
We assess the relative improvement from the likelihood approach by
comparing its performance with an analysis that uses only the
geometric trajectory reconstruction and moment-based image
parameterization described in Section \ref{subsec:recon} which we
refer to here as the \textit{moment} reconstruction.  Because the
moment reconstruction is more sensitive to the presence of noise
fluctuations in the image, we use a slightly higher cleaning threshold
($\bar{s}/\sigma = 9$) than the threshold used for the likelihood
analysis.  We use a BDT background discriminant trained with the same
settings described in Section \ref{subsec:gamma_hadron} but excluding
parameters derived from the likelihood analysis.

Figure~\ref{FIG:RECON_METHOD} shows the comparison of the point-source
sensitivity obtained with the moment analysis, the likelihood
analysis, and a likelihood analysis in which the goodness-of-fit (GOF)
parameter is excluded from the training of the decision tree.  With
the likelihood-based analysis, we find a factor of two improvement in
point-source sensitivity and a 30-40\% improvement in the gamma-ray
PSF over the full energy range.  As seen from the comparison between
the likelihood analyses performed with and without the GOF parameter,
the improvement in the point-source sensitivity is attributable to
gains in both the gamma-ray angular resolution and the background
rejection power.  The addition of the GOF parameter provides an
additional 30\% improvement in sensitivity.

We also observe that the energy threshold for the likelihood analysis is
considerably lower ($E_{\mathrm{th}} \simeq 50\textrm{ GeV}$) relative
to the moment reconstruction ($E_{\mathrm{th}} \simeq 100\textrm{
  GeV}$).  The improved performance of the likelihood analysis at low
energies can be attributed to both the higher image reconstruction
efficiency and the smaller bias of the likelihood energy estimator.
Because the likelihood reconstruction is insensitive to the inclusion
of pixels with small signals, the cleaning threshold can be optimized
to maximize the reconstruction efficiency for low-energy showers
without impacting the performance at higher energies.

\subsection{Influence of the Geomagnetic Field}\label{subsec:geomag}

\begin{figure*}[tb]
\includegraphics[width=.49\textwidth, keepaspectratio]{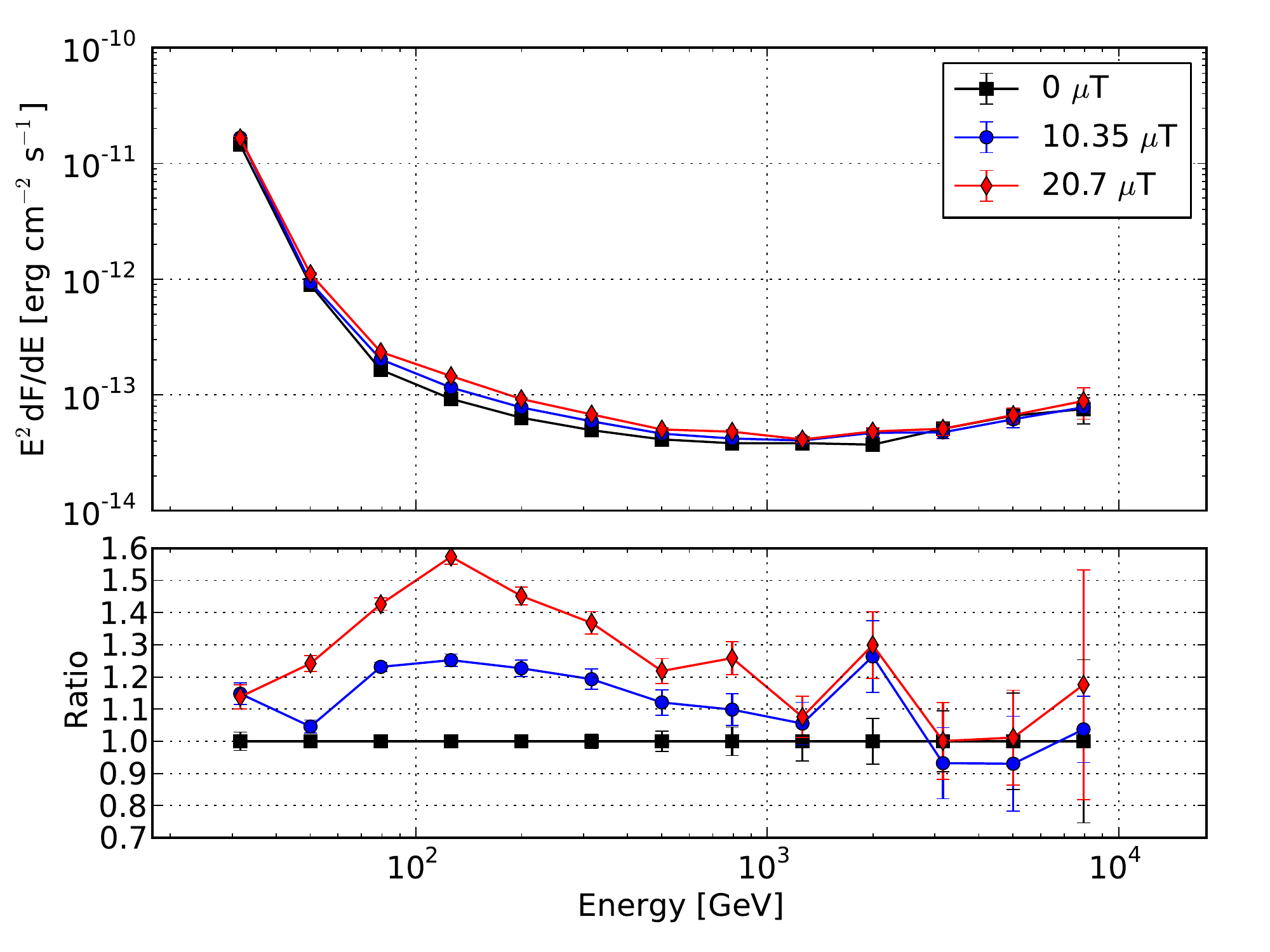}
\includegraphics[width=.49\textwidth, keepaspectratio]{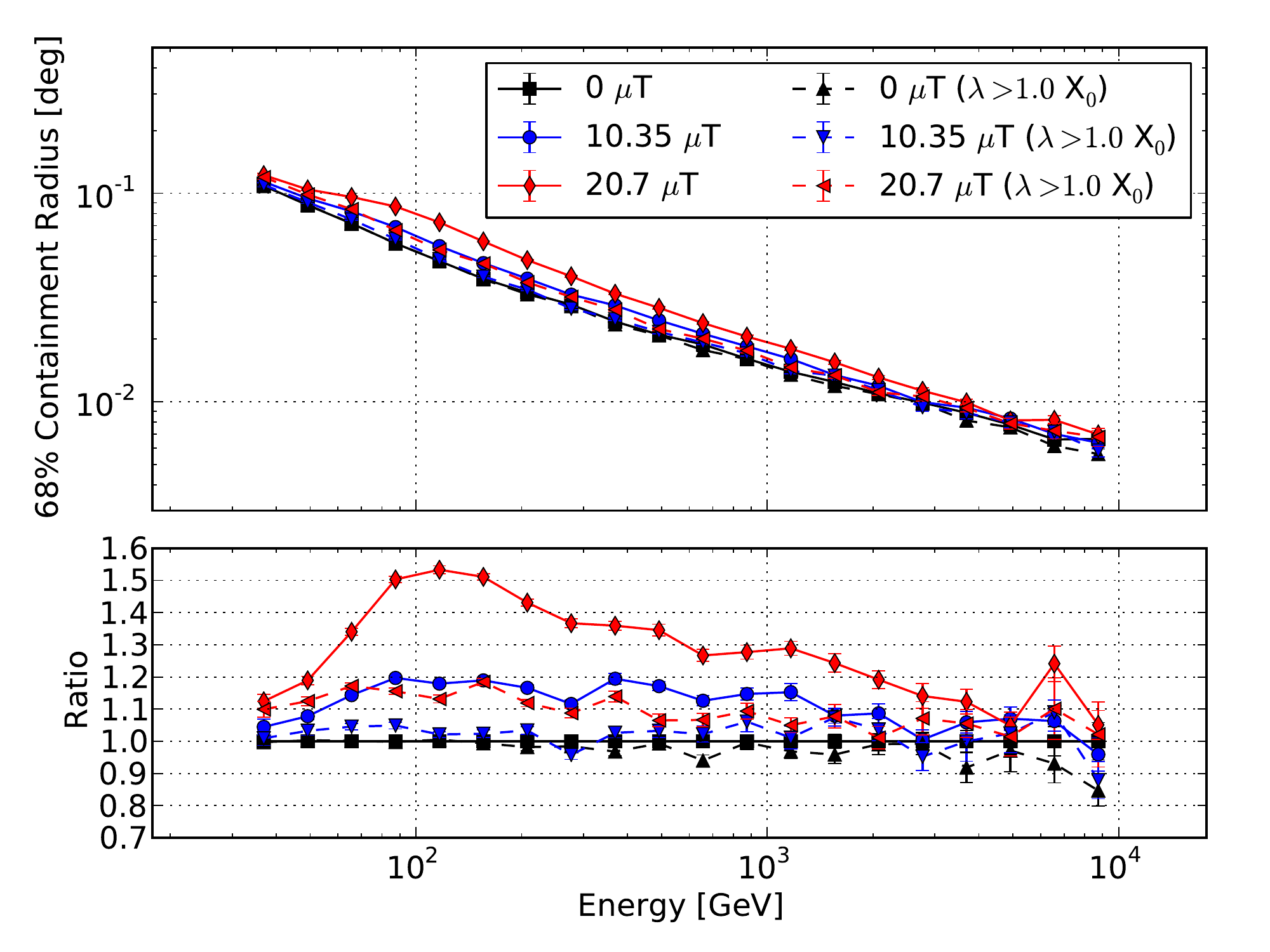}
\caption{\label{FIG:BFIELD} Performance of array M61 simulated with
  the equatorial GF ($B_\perp = 20.7~\mu \mathrm{T}$; red diamonds and
  solid line), a GF configuration with a reduced perpendicular
  component ($B_\perp = 10.35~\mu \mathrm{T}$; blue circles and solid
  line), and no GF (black squares and solid line).  \textbf{Left:}
  Differential point-source sensitivity for a 50~h observation time.
  \textbf{Right:} Gamma-ray angular resolution (68\% containment
  radius) after \textit{reconstruction} cuts.  Dashed curves show the
  same comparison for gamma-ray showers with an interaction depth
  ($\lambda$) greater than 1.0~X$_{0}$.}
\end{figure*}

\newtext{The deflection of charged particles in the EM shower by the
geomagnetic field (GF) can significantly distort the shapes of gamma-ray
images recorded by IACTs
\citep{1992JPhG...18L..55B,1999JPhG...25.1223C,2008NIMPA.595..572C}.}
The strength and orientation of the GF is thus an important
consideration for the selection of candidate sites for an IACT
observatory.  Its influence can be as large or larger than the site
elevation \citep{2013APh....45....1S}.  The magnitude of the induced
deflection is proportional to the perpendicular component of the GF
($B_\perp$) and therefore the strength of the GF effect depends on
both the magnitude of the GF vector as well as its orientation
relative to the shower trajectory.  Due to the asymmetry in the shower
shape induced by the GF, the distortion visible to a telescope also
depends on the orientation of the shower impact point relative to the
telescope position.  Telescopes with shower position angles close to
$0^\circ$ or $180^\circ$ see a larger GF effect as the GF-induced
elongation in the shower occurs primarily in the plane perpendicular
to the telescope pointing.

To obtain a realistic assessment of the GF effect for any given
observatory site would require simulations with many telescope
orientations as they occur for realistic observation profiles of
gamma-ray sources.  We did not carry out such simulations and instead
focused on the effect of the GF for a few representative values of
$B_\perp$.  Our baseline site configuration has $(B_{x},B_{z}) =
(27.5~\mut,-15.0~\mut)$ with $B_\perp = 20.7~\mut$ when observing a
shower with $Zn = 20^\circ$ and $Az = 0^\circ$.  To test the influence
of the GF strength we performed simulations of array M61 for two
additional site models: a site with $(B_{x},B_{z}) = (19.84~\mut,
-24.24~\mut)$ that has a perpendicular GF component that is half as
large as for our baseline site ($B_\perp = 10.35~\mut$) and a site
with no geomagnetic field.  \newtext{By limiting ourselves to these
  few cases we can only give a general guidance for effects of the
  magnetic field on any observable gamma-ray source. Depending on the
  specific source observation profile, the effect of the GF for an
  individual source might be different.}

The configurations we tested have a range of field strengths that are
comparable to the southern Hemisphere sites considered for CTA.  The
Namibian H.E.S.S. site and the Argentinian Leoncito sites have
$(B_{x},B_{z}) = (12.1~\mut,-25.5~\mut)$ and $(B_{x},B_{z}) =
(20.1~\mut,-12.2~\mut)$, respectively \citep{2013APh....45....1S}.
Because the strength and orientation of the GF is generally a slowly
varying function of the site latitude and longitude these two sites
provide a good representation of the expected GF effect for sites in
Africa and South America.  When observing a shower at $Zn = 20^\circ$
and $Az = 0^\circ$ the Namibian and Argentinian sites have
perpendicular components of $2.7~\mut$ and $14.7~\mut$.  However a
more realistic measure of the expected GF effect is the average
perpendicular component over the range of azimuth angles that a
gamma-ray source is observed.  The Namibian and Argentinian sites have
an average GF strength at $Zn = 20^\circ$ of $13.4~\mut$ and
$19.7~\mut$, respectively.

The comparison of the array performance for the three GF
configurations is presented in Figure~\ref{FIG:BFIELD}.
We find that the effect of the GF strength is strongest at 100~GeV
where the point-source sensitivity is reduced by 50\% when increasing
$B_\perp$ from $0~\mu\mathrm{T}$ to $20.7~\mu\mathrm{T}$.  We also
observe that the effect of the GF scales linearly with $B_\perp$ such
that the site configuration with $B_\perp = 10.35~\mut$ suffers
approximately half of the reduction in sensitivity relative to our
baseline site configuration.  Below energies of 100 GeV, the effect of
the GF is lessened because only gamma rays that convert deep in the
atmosphere can be efficiently reconstructed.  The lower the particle
interacts in the atmosphere the less it is affected by the GF.  At
higher gamma-ray energies the impact of the GF is lessened due to both
the higher energy of the secondary particles and the larger path
length in the atmosphere.  As shown in the right panel of
Figure~\ref{FIG:BFIELD} the GF worsens the point-source sensitivity
primarily by degrading the gamma-ray PSF.  For showers with
interaction depths larger than 1~X$_{0}$, differences in the gamma-ray
PSF between the different GF configurations are found to be less than
20\% illustrating that the influence of the GF increases with
decreasing interaction depth.

\subsection{Imaging Performance} 

\begin{figure*}[tb]
\includegraphics[width=.49\textwidth, keepaspectratio]{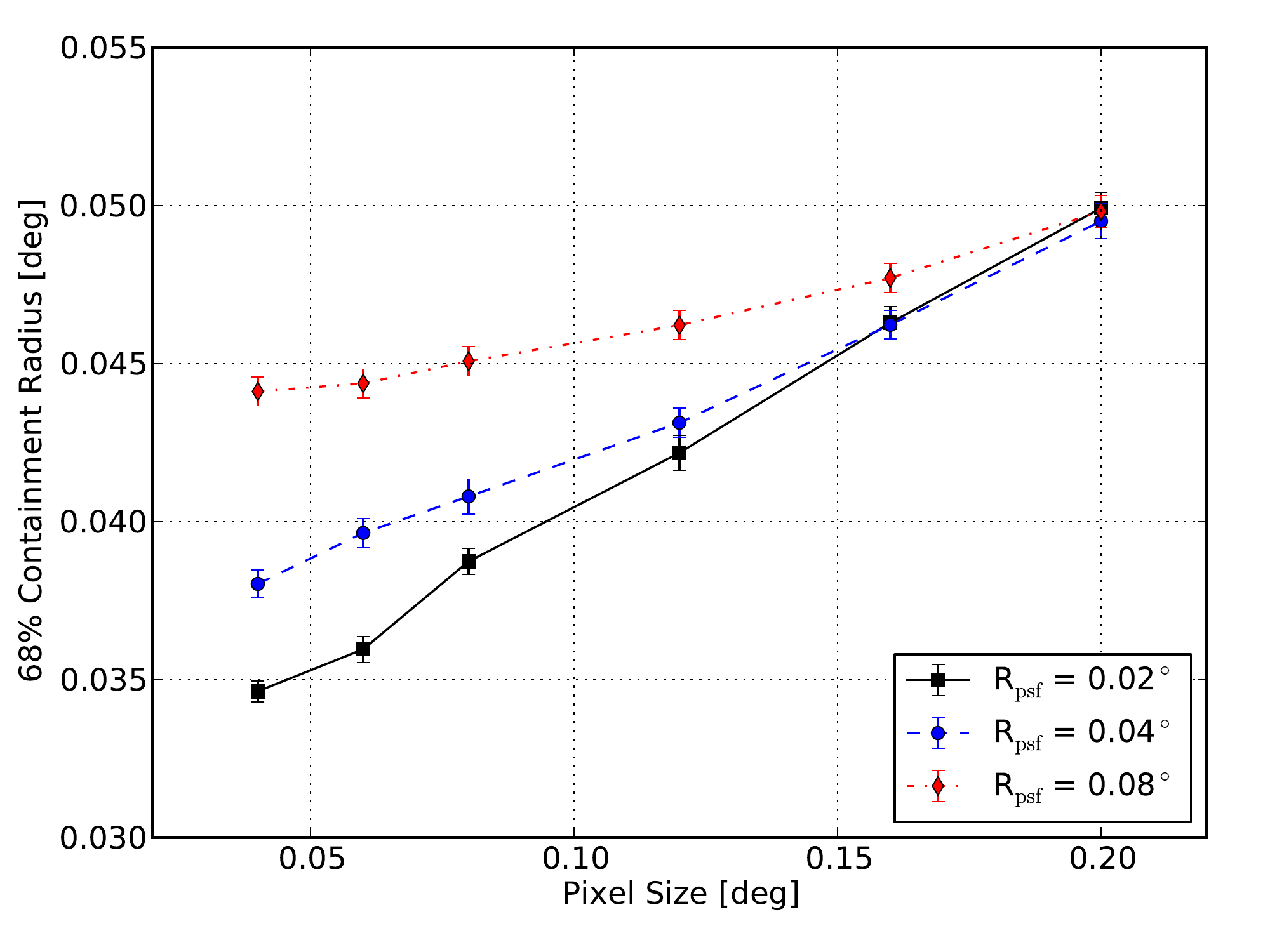}
\includegraphics[width=.49\textwidth, keepaspectratio]{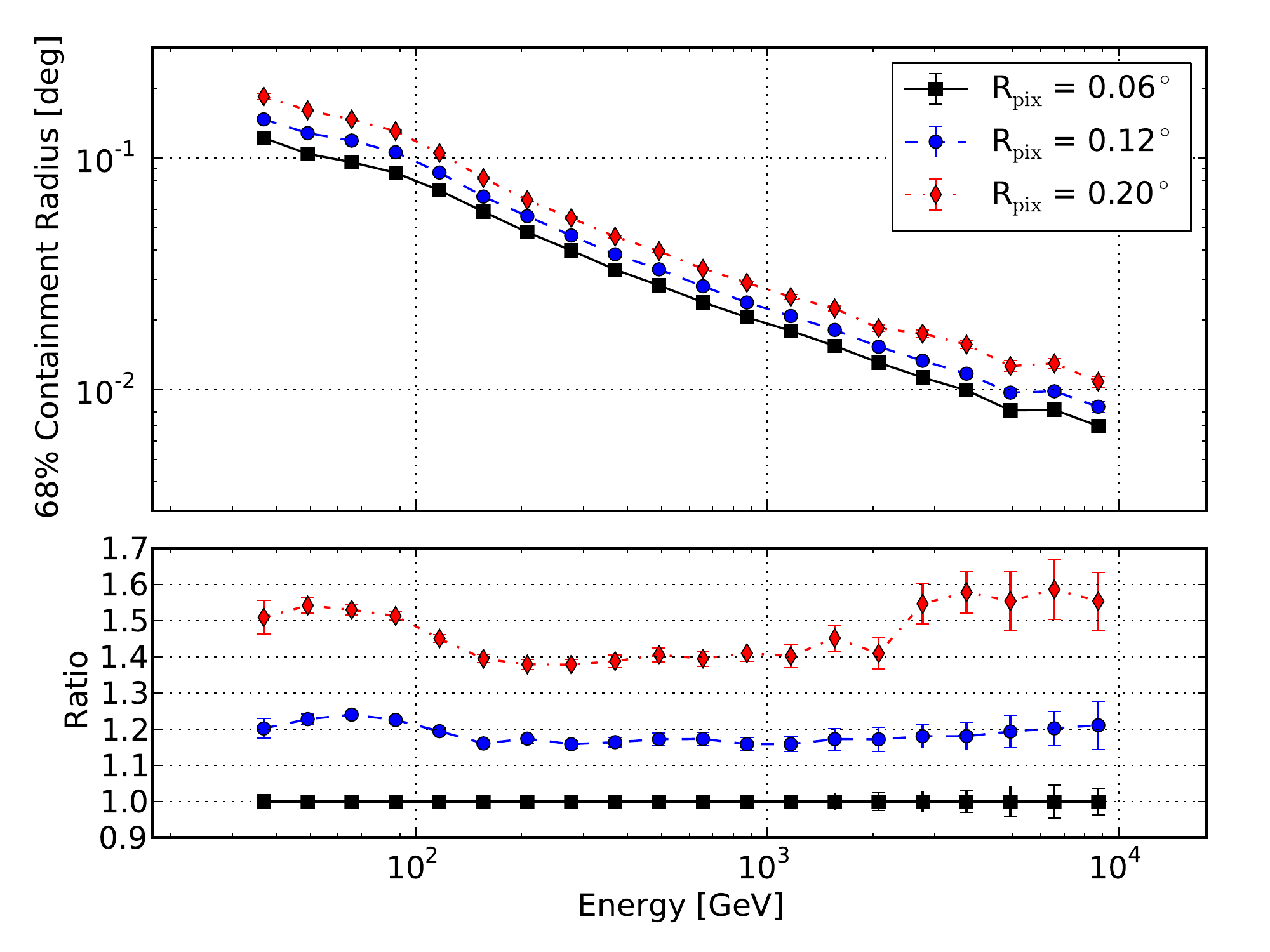}
\caption{\label{FIG:PIXEL_SIZE_PSF}\textbf{Left:} 68\% containment
  angle of the gamma-ray PSF at 317~GeV versus camera pixel size for
  telescope models with different optical PSFs ($\rpsf$): $0.02^\circ$
  (black squares), $0.04^\circ$ (blue circles), $0.08^\circ$ (red
  diamonds).  The gamma-ray PSF is evaluated after applying
  \textit{reconstruction cuts}.  The baseline configuration for all
  simulations is array M61.  \textbf{Right:} 68\% containment angle of
  the gamma-ray PSF versus gamma-ray energy for array M61 with $\rpsf
  = 0.02^\circ$ simulated with different telescope pixel sizes:
  $0.06^\circ$ (black squares), $0.12^\circ$ (blue circles),
  $0.20^\circ$ (red diamonds).}
\end{figure*}

The telescope design has a large impact on the resulting gamma-ray PSF
obtained with the complete array. The optical design of the individual
telescopes defines their achievable optical PSF and the camera design
determines how efficiently the optical PSF can be translated into an
improved gamma-ray PSF.  For a given optical PSF, the gamma-ray PSF can
be improved by reducing the camera pixel size.  
In the limit that the pixel size is much smaller than the PSF, the
improvement of the gamma-ray PSF saturates and a further reduction in
pixel size does not provide any measurable advantage but increases the
cost of the camera.  Thus the optimal tradeoff between performance and
cost is one in which the pixel size is appropriately matched to the
quality of the optical PSF.  Current generation IACTs have cameras
using pixel sizes from 0.1$^\circ$ to 0.16$^\circ$ and an optical PSF
at the center of the FoV which is considerably smaller than the pixel
size.  Here we explore a new parameter space for the IACT imaging
resolution by examining the performance of camera designs with pixel
sizes between 0.04$^\circ$ and 0.1$^\circ$.  Such designs begin to
resolve the core of the Cherenkov shower which has an intrinsic angular
size of $\sim$0.01$^\circ$.

The left panel of Fig.~\ref{FIG:PIXEL_SIZE_PSF} shows the gamma-ray
PSF versus pixel size for arrays with different optical PSFs.  For an
optical PSF of $0.08^{\circ}$ the gamma-ray PSF shows only a modest
improvement of $\sim10$\% when reducing the pixel size from
$0.2^{\circ}$ to $0.04^{\circ}$.  An optical PSF between
$0.02^{\circ}$ and $0.04^{\circ}$ is found to be critical to realize
the full improvement in gamma-ray PSF that can be achieved by reducing
the camera pixel size below $0.12^{\circ}$.  The improvement of the
gamma-ray PSF at different energies when reducing the pixel size is
shown in Fig.~\ref{FIG:PIXEL_SIZE_PSF}.  The gamma-ray PSF is
significantly better at all energies when reducing the pixel
size. There is a slight modulation seen in the improvement versus
energy more pronounced for larger pixel sizes. The smaller pixel size
performs best at low and high energies ($E<100\textrm{ GeV}$ and
$E>2.5\textrm{ TeV}$) while the improvement is less pronounced in the
intermediate energy range. An improvement of the gamma-ray PSF of
about 20\% in the full energy range by reducing the pixel diameter
from $0.12^{\circ}$ to $0.06^{\circ}$ demonstrates a realistic
difference between currently considered optical telescope designs for
CTA.

\begin{figure*}[tb]
\includegraphics[width=.49\textwidth]{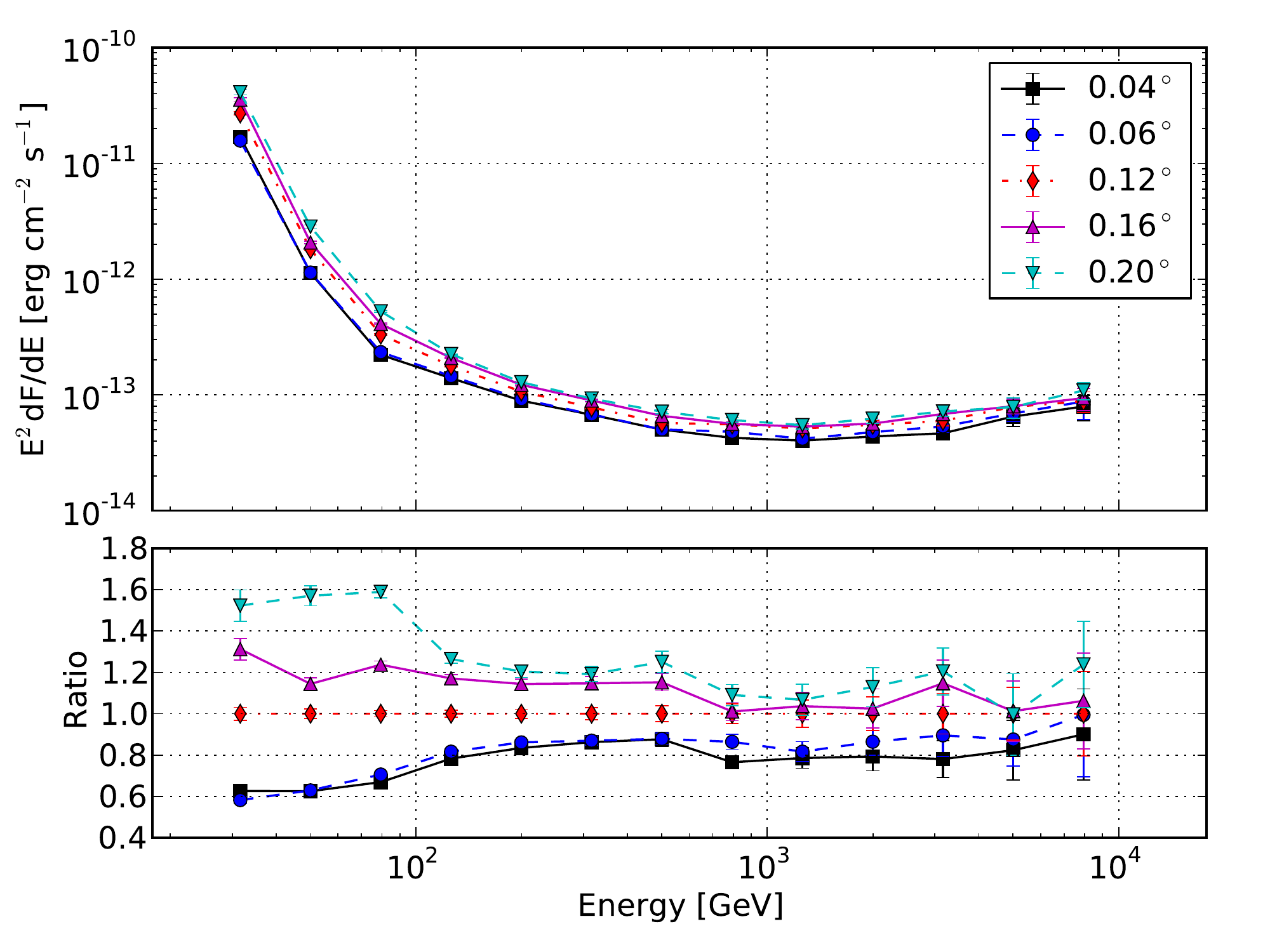}
\includegraphics[width=.49\textwidth]{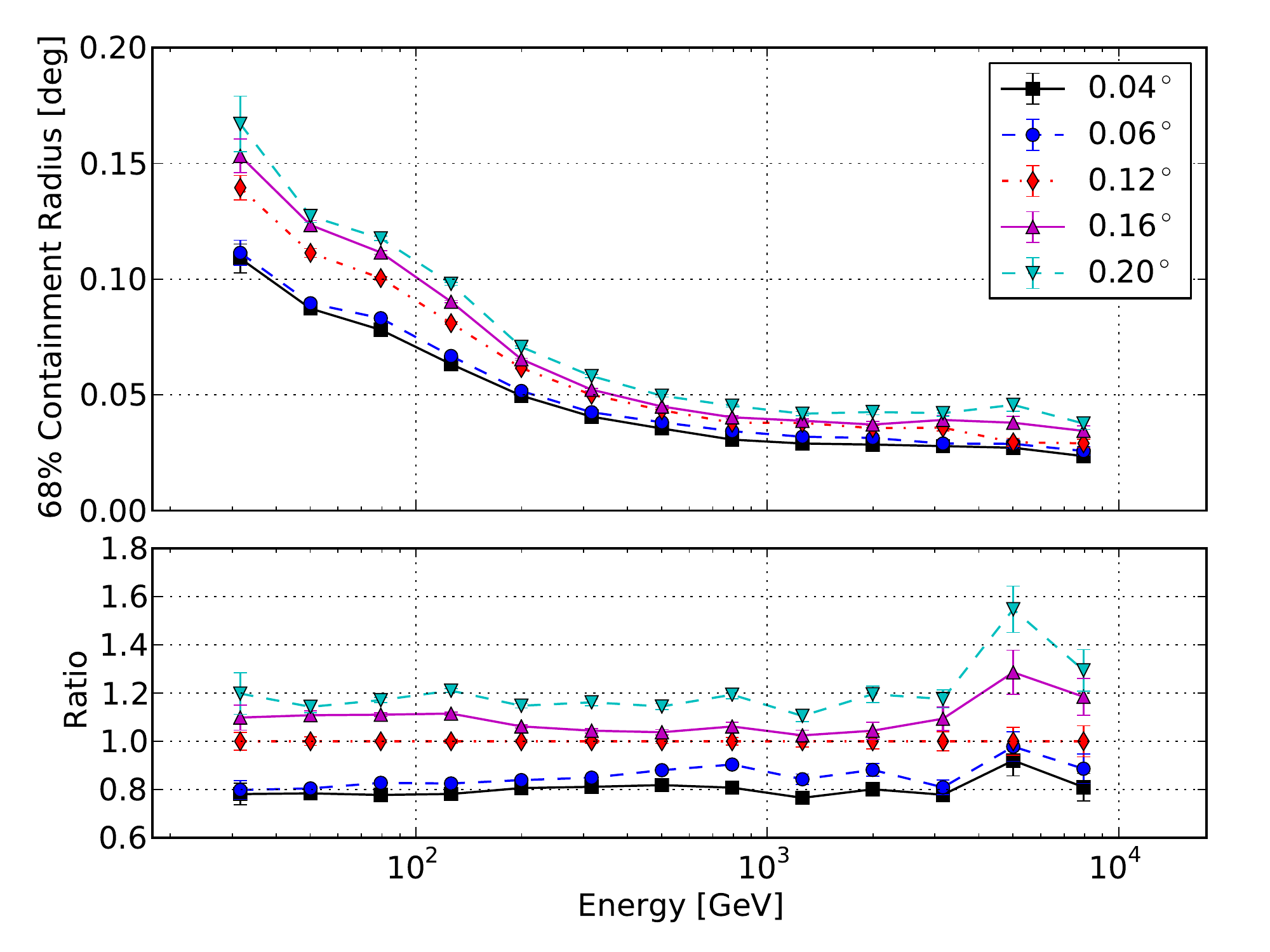}
\caption{\label{FIG:PIXEL_SIZE_SENSITIVITY} Performance of array M61
  simulated with pixel sizes from $0.04^\circ$ to $0.20^\circ$.
  \textbf{Left:} Differential point-source sensitivity for a 50~h
  observation time.  \textbf{Right:} 68\% containment angle of the
  gamma-ray PSF evaluated after \textit{point-source} cuts.}
\end{figure*}

The effect of the pixel size on the differential point-source
sensitivity is shown in Fig.\ref{FIG:PIXEL_SIZE_SENSITIVITY}. The
pixel size has the strongest impact at low energies ($< 100$~GeV)
where a factor of two relative improvement is observed when the pixel
size is reduced from $0.16^{\circ}$ to $0.06^{\circ}$.  At higher
energies the smaller pixel size results in a smaller but still
measurable improvement in point-source sensitivity of 30-40\%.  Above
3~TeV differences between adjacent pixel sizes become
indistinguishable due to the limited background statistics that make
evaluation of small sensitivity differences very difficult.  The
gamma-ray PSF is clearly improved over the complete energy range by
about 50\% as the pixel size is reduced from $0.2^{\circ}$ to
$0.06^{\circ}$. The observed improvement in sensitivity demonstrates
that the intrinsic shower features that can be used for background
suppression and direction reconstruction are still smaller than the
pixel sizes of currently operating Cherenkov telescopes.

\subsection{Light Collection Area} 

\begin{figure*}[tb]
\includegraphics[width=.49\textwidth]{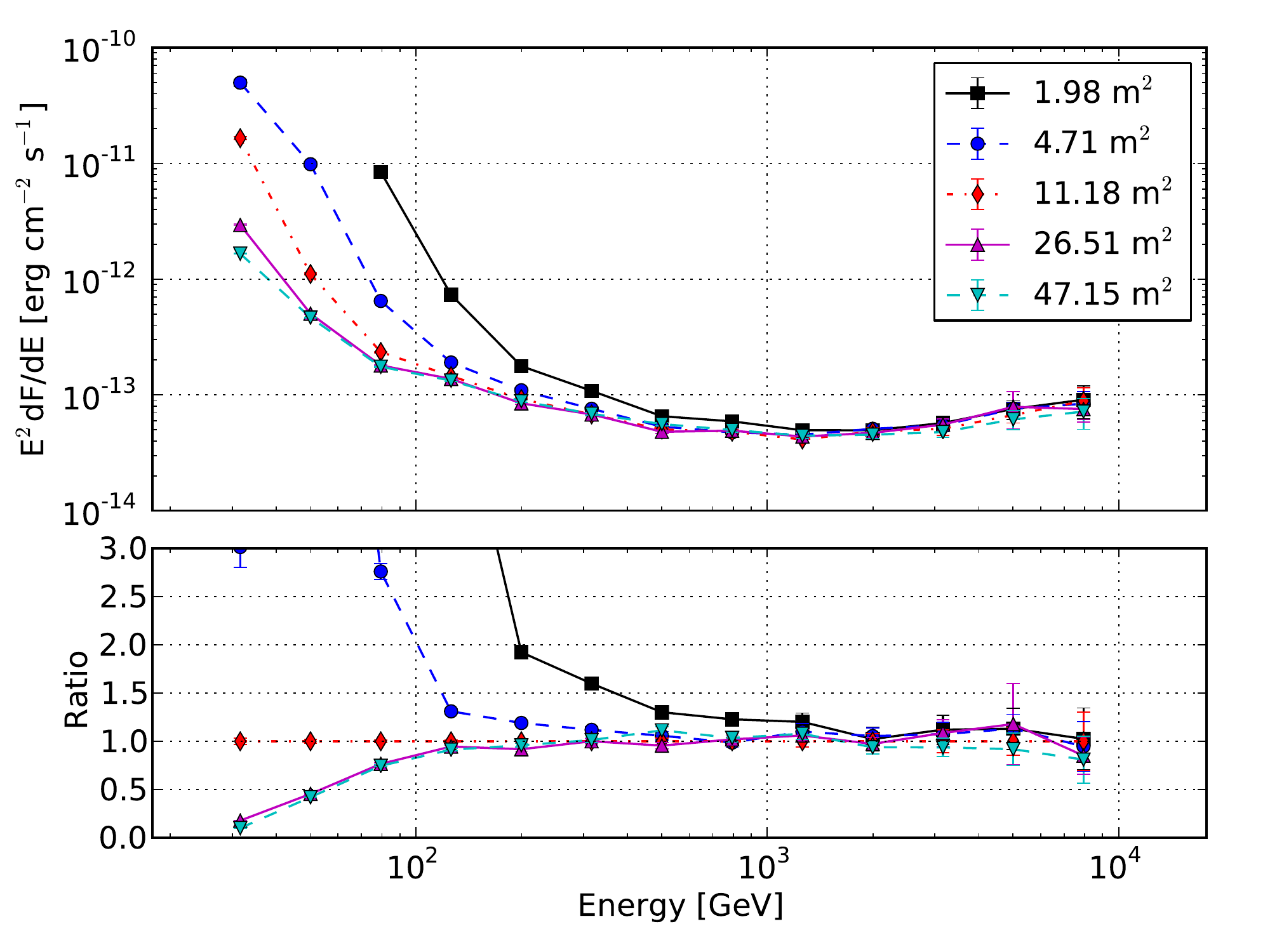}
\includegraphics[width=.49\textwidth]{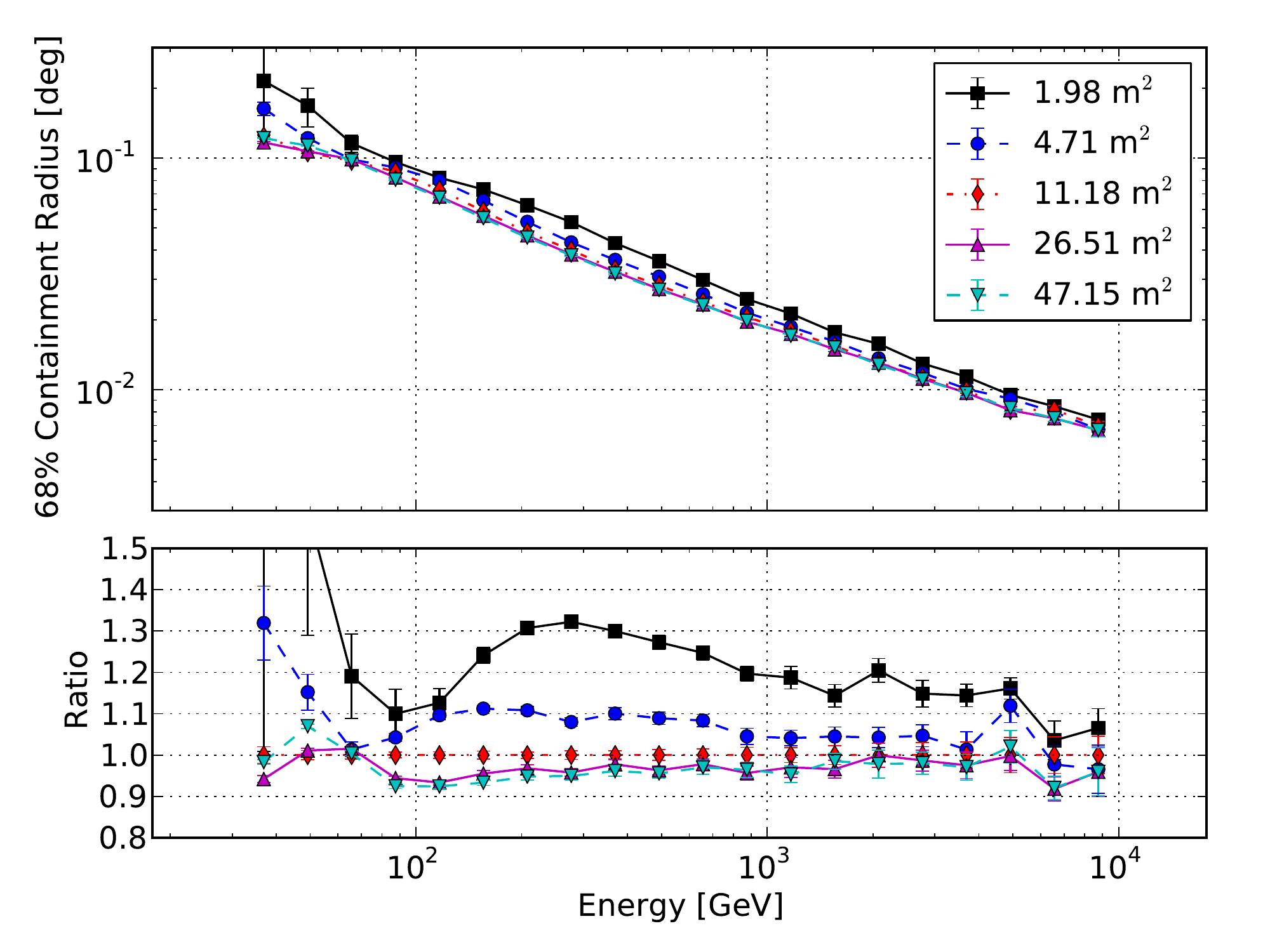}
\caption{\label{FIG:AOPT_SENSITIVITY} Performance of array M61
  simulated with different values of $\aopt$: 1.98~m$^{2}$ (black
  squares), 4.71~m$^{2}$ (blue circles), 11.18~m$^{2}$ (red
  diamonds), 26.51~m$^{2}$ (magenta triangles), and 47.15~m$^{2}$
  (cyan triangles).  \textbf{Left:} Differential point-source
  sensitivity for a 50~h observation time.  \textbf{Right:} 68\%
  containment angle of the gamma-ray PSF after \textit{reconstruction}
  cuts.}
\end{figure*}

The telescope light collection area determines the signal-to-noise
ratio (SNR) of the shower images and the efficiency with which these
images can be recorded by the telescope trigger.  Therefore we expect
that a larger $\aopt$ increases the trigger efficiency and provides
better defined images and hence improves performance of the array.
The role of the $\aopt$ parameter is particularly relevant for the
performance of the array at low energies where the smaller light yield
per image makes reconstruction and analysis of the gamma-ray showers
more challenging.

The assumed design, size, and cost of the proposed telescopes yields
distinct $\aopt$ values.  We studied the effect of the $\aopt$ on the
gamma-ray PSF and point-source sensitivity of the array by examining
the performance of telescope models with $\aopt$ between 2~m$^{2}$ and
50~m$^{2}$.  These models span the range of light collection areas between
SST-like and LST-like telescope designs.
The SST, MST, and LST telescope designs have $\aopt$ of approximately
1--2~m$^{2}$, 5--10~m$^{2}$, and $\sim$50~m$^{2}$ respectively
\citep{2013arXiv1307.4962P,Bernlohr:2013bla}.

Figure \ref{FIG:AOPT_SENSITIVITY} shows the comparison of the
gamma-ray PSF and point-source sensitivity for telescopes with $\aopt$
between 1.98~m$^{2}$ (SST-like) and 47.15~m$^{2}$ (LST-like).
$\aopt$ has only a minor effect on the gamma-ray PSF in most of the
energy range investigated here.  In the middle energy range between
100~GeV and 1~TeV we find an improvement of 5--10 \% when increasing
the telescope light collection area from 11.18~m$^{2}$ to
47.15~m$^{2}$.  The almost insignificant improvement around 100~GeV is
caused by a selection effect of the reconstructed gamma-ray events.
At these low energies, telescopes with smaller $\aopt$ can only
trigger on the brightest showers that convert deep in the atmosphere.
As discussed in Section \ref{subsec:geomag} the larger interaction
depth of these showers lessens the impact of the GF and results in a
more accurate reconstruction of the direction.  Larger telescopes can
efficiently trigger on showers with both large and small interaction
depths which results in a larger effective area but a worsening of the
overall gamma-ray PSF.  This effect reverses at the very lowest
energies (30--50~GeV) where the reduced SNR images recorded by
telescopes with small $\aopt$ dominates the reconstruction quality.

The light collection area has a measurable impact on the point-source
sensitivity only at energies below 300~GeV with telescopes with larger
light collection area yielding better sensitivity. The increase in
sensitivity is most significant below 100~GeV and is a result of the
reduction in the telescope trigger threshold and resulting increase in
the gamma-ray effective area.  The larger light collection area also
yields better SNR in the shower images improving the reconstruction of
low energy events.  At higher energies the impact of light collection
area is significantly reduced as the array becomes fully efficient for
triggering and reconstructing events that impact within the array
boundary.  Improving the image SNR provides little improvement at
these energies because the reconstruction is predominantly limited by
intrinsic shower fluctuations.  Remarkably the improvement in
point-source sensitivity is almost negligible between telescopes with
$26.51~\mathrm{m}^2$ and $47.15~\mathrm{m}^2$ over the whole energy
range.

The observed improvements in array performance above the trigger
threshold are small when considering that light collection area is the
dominant parameter influencing the telescope cost.  Given the small
differences in reconstruction performance, the primary motivation for
choosing a telescope design with larger light collection area is to
reduce the array energy threshold.  However for an array of fixed cost
increasing the light collection area also entails a reduction in the
number of telescopes.  
For gamma-ray energies between 100~GeV and 1~TeV, a telescope with
$\aopt$ of 5--10 m$^{2}$ (MST-like) clearly provides the best
performance to cost ratio.  Array designs that include a small number of
telescopes with larger light collection area can lower the energy
threshold while keeping the cost of the total array within reasonable
limits.  Performance of arrays with different numbers of telescopes
are studied further in Section \ref{subsec:numtels}.

\subsection{Inter-telescope Separation}

\begin{figure*}[tb]
\centering
\includegraphics[width=.49\textwidth]{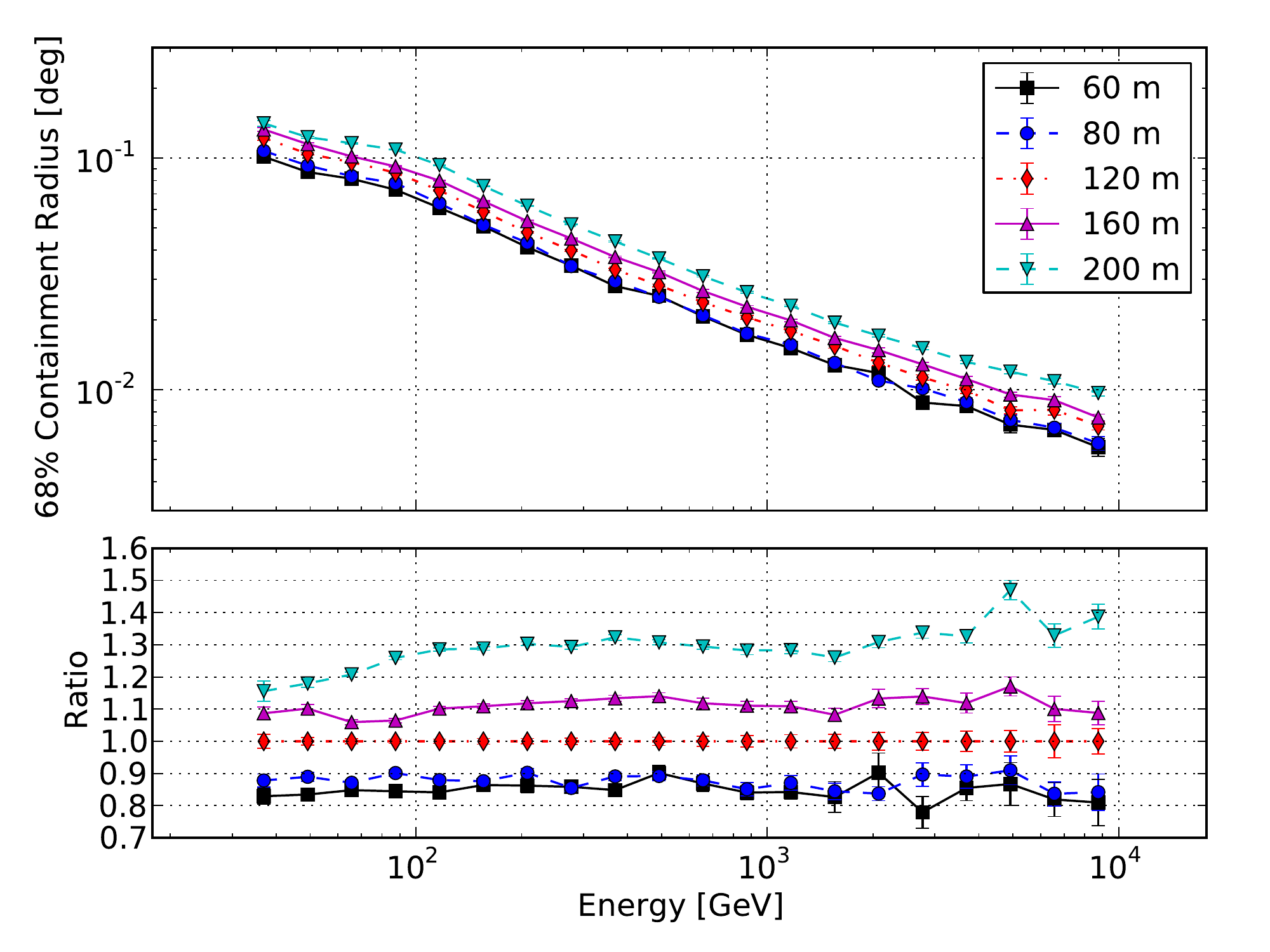}
\includegraphics[width=.49\textwidth]{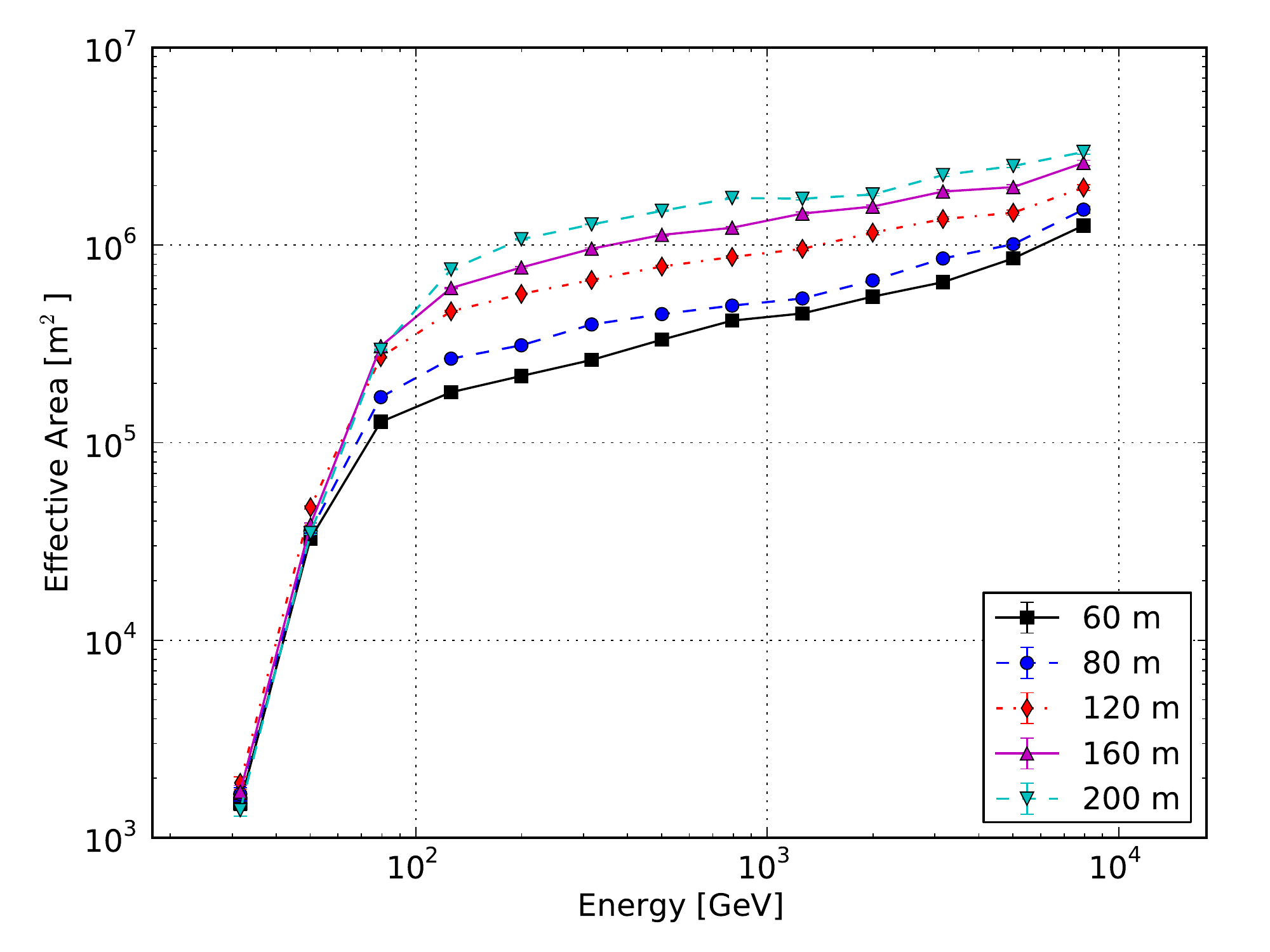}
\includegraphics[width=.49\textwidth]{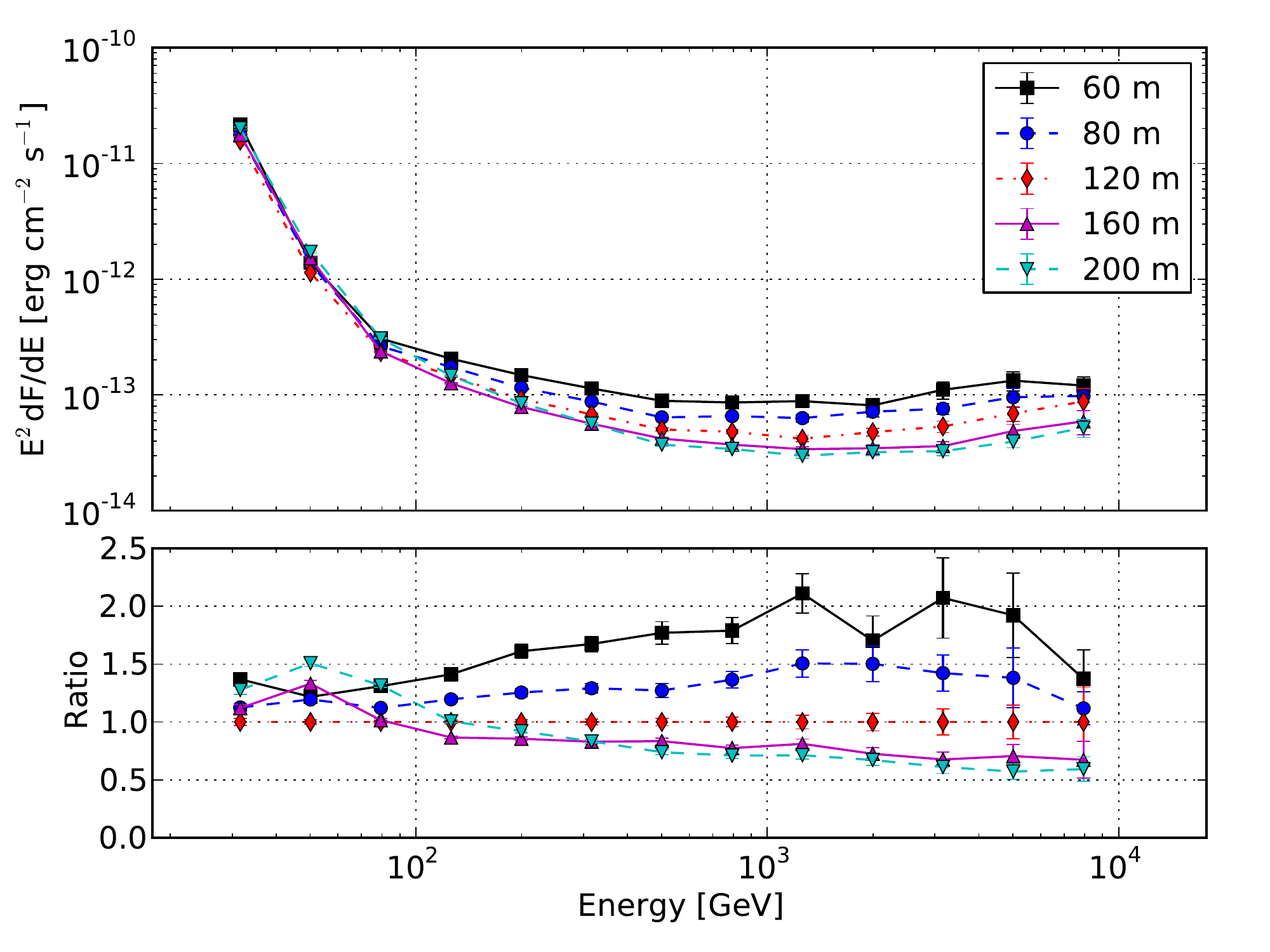}
\includegraphics[width=.49\textwidth]{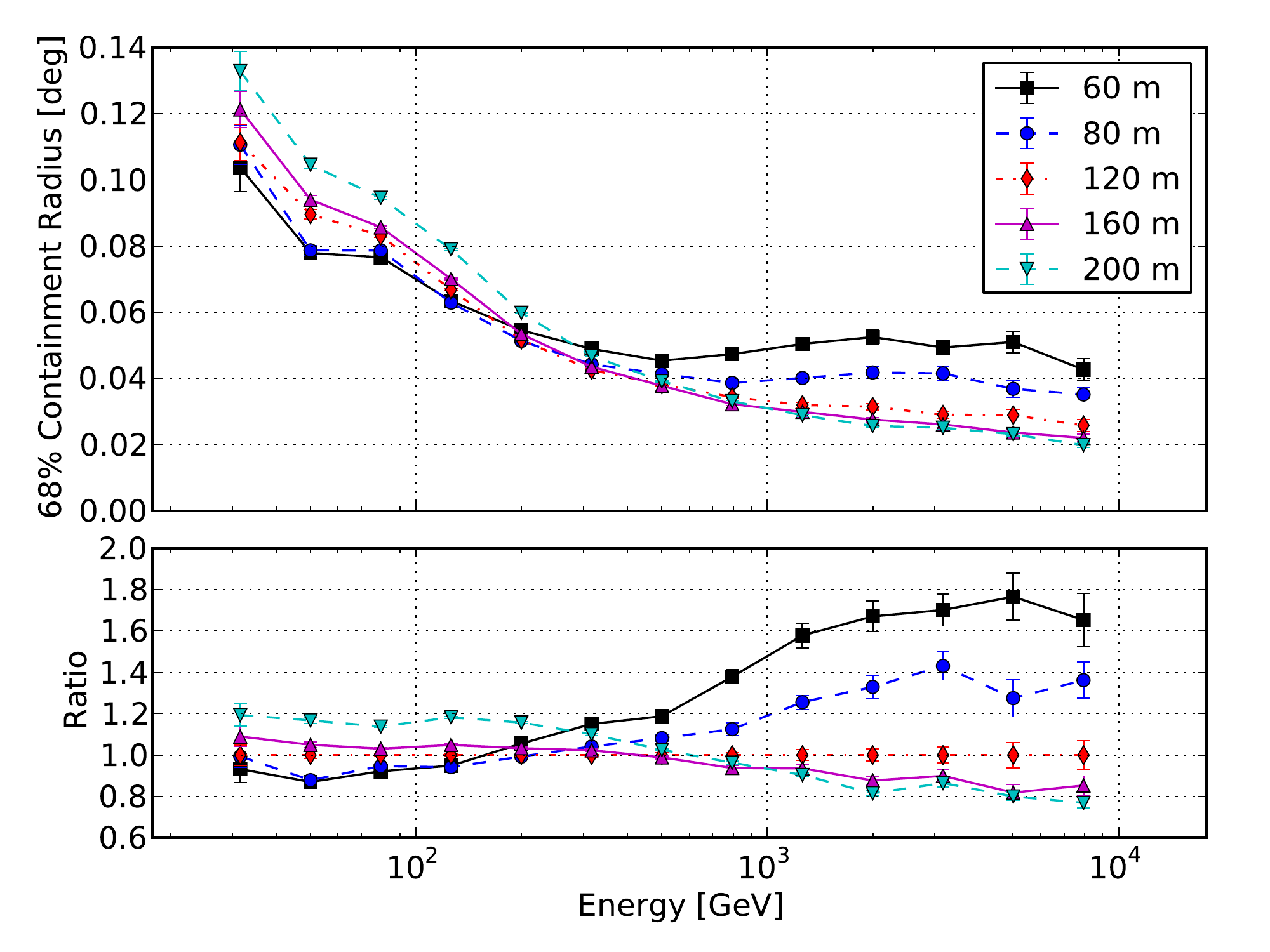}
\caption{\label{FIG:BASELINE_SENSITIVITY} Performance of array M61
  simulated with different inter-telescope separations: 60~m (black
  squares), 80~m (blue circles), 120~m (red diamonds), 160~m (magenta
  triangles) and 200~m (cyan triangles).  \textbf{Top Left:}
  68\% containment angle of the gamma-ray PSF after
  \textit{reconstruction} cuts.  \textbf{Top Right:} Gamma-ray
  effective area after \textit{point-source} cuts.  \textbf{Bottom
    Left:} Differential point-source sensitivity for a 50~h
  observation time.  \textbf{Bottom Right:} 68\% containment angle of
  the gamma-ray PSF after \textit{point-source cuts}.}
\end{figure*}

The inter-telescope separation determines both the physical area of
the array footprint as well as the average number of telescopes that
will participate in the reconstruction of individual showers.  Smaller
telescope separations improve reconstruction quality for contained
showers at the cost of lowering the total effective area of the array.
Larger telescope separations are generally preferred when optimizing
for sensitivity at higher energies since the point-source sensitivity
of IACT arrays at moderate exposures (10--50~hours) is signal limited
above 1--3~TeV.  Another important consideration when optimizing the
telescope separation is the number of telescopes within the Cherenkov
light pool.  Telescopes within the Cherenkov light pool sample light
emitted by higher energy particles in the shower core and provide a
more accurate determination of the shower trajectory.  Telescope
separations that are comparable to the size of the Cherenkov light
pool (100--150~m) ensure that multiple telescopes will sample each
shower within its light pool.  Finally smaller separations may
potentially improve background rejection by increasing the efficiency
for detecting Cherenkov light from hadronic subshowers produced in
cosmic-ray background events.

The impact of the telescope separation on the gamma-ray PSF is
illustrated in the top panel of Fig.~\ref{FIG:BASELINE_SENSITIVITY}
which shows a comparison of arrays with separations between 60~m and
200~m.  In this comparison we consider only showers passing
\textit{reconstruction} cuts with core positions near or within the
array boundary.  These cuts select events with the best PSF and reduce
the differences in performance caused by the finite array size.  The
reduction of the telescope grid spacing from 120~m to 60~m results in
a 20\% improvement of the gamma-ray PSF between 30~GeV and
10~TeV. However this rather small improvement would require a
quadrupling in the number of telescopes to cover a similar area. Thus
the improvement of the gamma-ray PSF from reducing the telescope
spacing has to be compared to the reduction of effective detector area
when fixing the number of available telescopes.

The lower left and right panels of Fig.~\ref{FIG:BASELINE_SENSITIVITY}
show the gamma-ray PSF and point-source sensitivity for the set of
telescope separations evaluated with a selection optimized for
point-source sensitivity.  The increase of effective area
with larger telescope spacing generally outweighs the reduction of
sensitivity due to a worsening of the gamma-ray PSF.  The point-source
sensitivity improves with increasing telescope spacing at energies
above 100~GeV with the best sensitivity achieved with a telescope
spacing of 160--200~m.  When increasing the telescope spacing to 200~m a
noticeable worsening of the sensitivity below 300~GeV is seen because the
number of individual telescopes triggering on each event is reduced
and hence the information available for direction and particle type
reconstruction.

When evaluated with point-source cuts as shown in the bottom right
panel of Fig.~\ref{FIG:BASELINE_SENSITIVITY}, the gamma-ray PSF above
300~GeV becomes worse as the telescope separation is decreased.
Although a smaller separation gives a better reconstruction for
contained events, the smaller array footprint results in a larger
fraction of uncontained events which tend to dominate the PSF at high
energies.  This emphasizes that for most applications where the
maximum sensitivity of the array is required the PSF has a quite
different behavior compared to the theoretically possible
behavior. 
A wider spacing of the MSTs will provide a much better performance for
most science cases compared to a narrow spacing that would be only
beneficial for the very few cases where the gamma-ray PSF is much more
important than sensitivity. Thus the best spacing for the MSTs for all
purposes is about 160~m.

\subsection{Trigger Threshold}

\begin{figure*}[tbp]
\includegraphics[width=.49\textwidth]{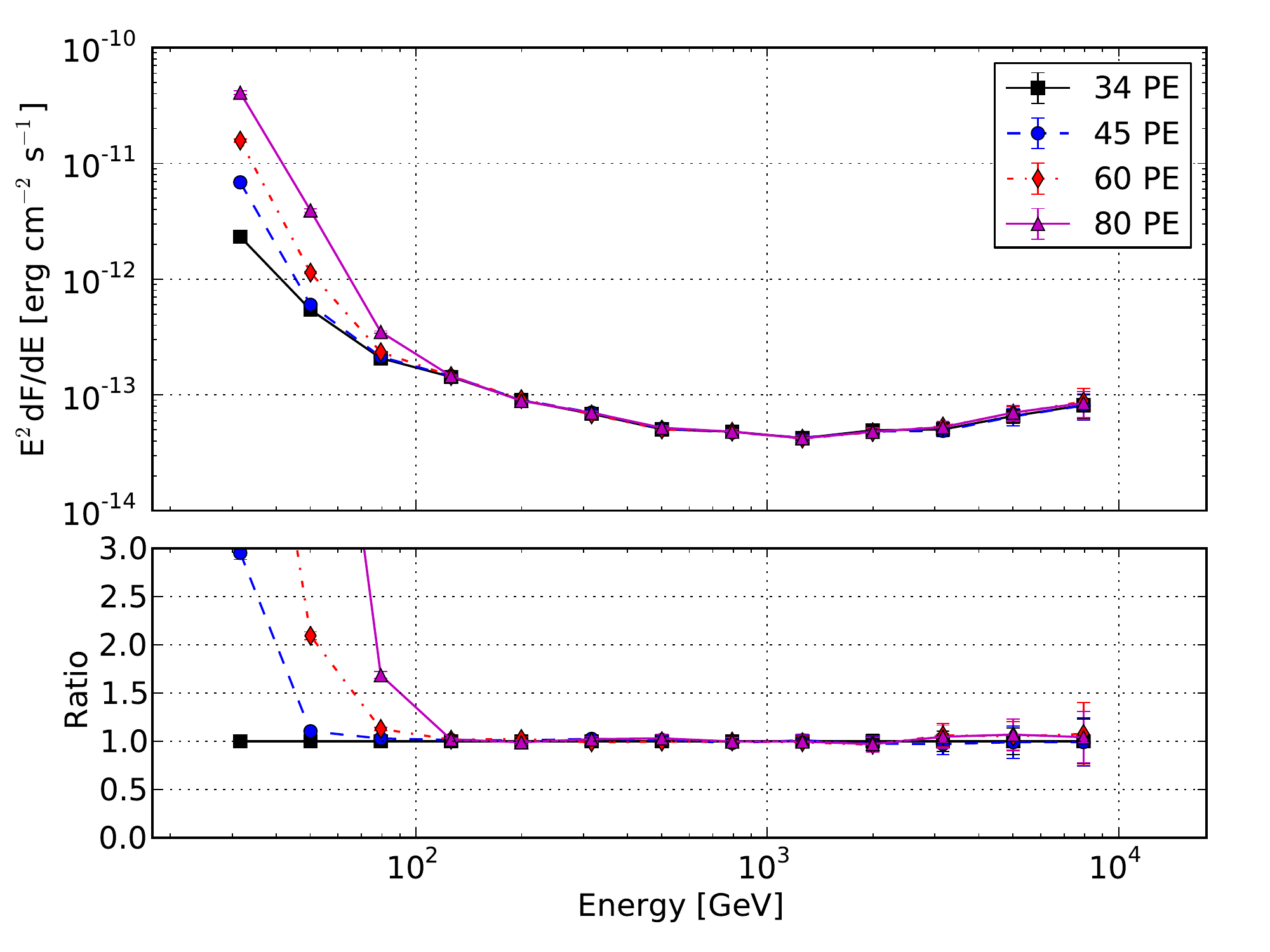}
\includegraphics[width=.49\textwidth]{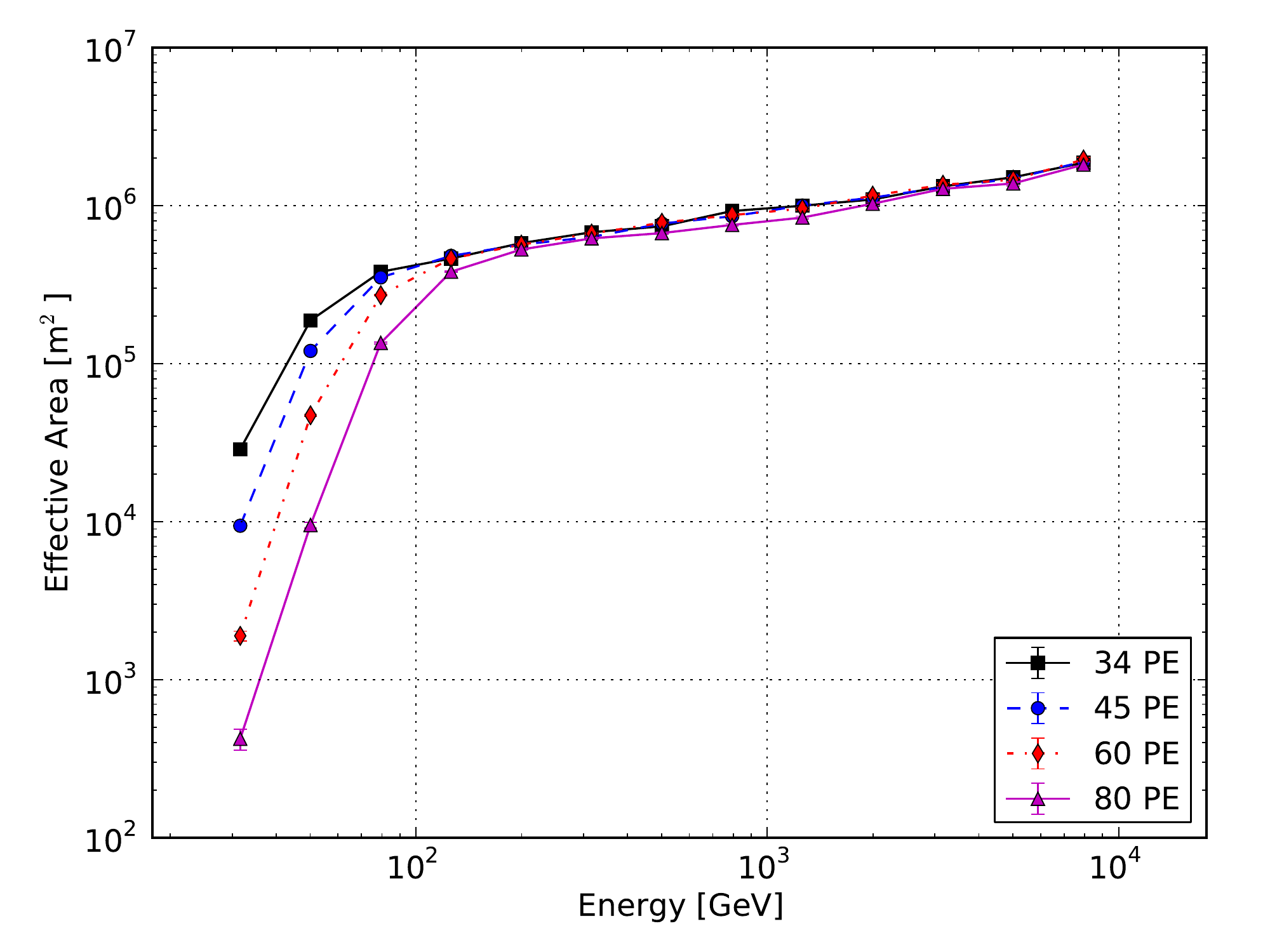}
\caption{\label{FIG:TRIGGER_THRESHOLD}Performance of array M61
  simulated with different camera trigger thresholds: 34~PE (black
  squares), 45~PE (blue circles), 60~PE (red diamonds), 80~PE
  (magenta triangles).  \textbf{Left:} Differential point-source
  sensitivity for a 50~h observation time.  \textbf{Right:} Gamma-ray
  effective area after \textit{point-source} cuts. }
\end{figure*}


The telescope trigger threshold is an important quantity to determine
the accessible energy range by any telescope array.  The impact of the
individual telescope trigger threshold is studied on the differential
point-source sensitivity of the M61 baseline array (see
Fig.~\ref{FIG:TRIGGER_THRESHOLD}).  As expected for an MST-like array
with $\aopt\simeq10$~m$^{2}$ the trigger threshold has little effect on the
sensitivity at energies above 100~GeV.  At higher energies the
telescope trigger becomes fully efficient for showers impacting within
the array and reducing the trigger threshold only increases the
efficiency for showers on the array periphery.  Because these distant
showers are generally not well reconstructed they do not contribute to
the array sensitivity.

Reducing the telescope trigger threshold of Array~M61 is found to
significantly improve the point-source sensitivity below 100~GeV.  A
reduction of the trigger threshold from 80~PE to 34~PE results in a
significant improvement at energies below 100~GeV and reaches up to an
order of magnitude at 30~GeV.  However, in a realistic telescope
design the accidental trigger rate can not be arbitrarily high due to
the limitations on the readout rate that can be sustained by the
telescope data acquisition.  The 60~PE effective trigger threshold
chosen for Array~M61 is a realistic target for a trigger
implementation that follows the same design used by current generation
IACTs.  Lower trigger thresholds may be achievable
by employing more sophisticated designs for the camera- and
array-level triggers such as requiring additional trigger topologies
for individual telescopes or higher multiplicities for the array
trigger.
If further improvements in the performance of the trigger can be
realized then the presented sensitivities at low energies could be
further improved.  Furthermore, it is evident that the likelihood
reconstruction is very efficient at low energies and that any
reduction in trigger threshold is directly translated into an
improvement in sensitivity.  The same statement is not necessarily
true for the moment reconstruction that usually has a higher analysis
threshold compared to the likelihood reconstruction as shown in
Fig.~\ref{FIG:RECON_METHOD}.

\subsection{NSB Rate}

\begin{figure*}[tb]
\includegraphics[width=.49\textwidth, keepaspectratio]{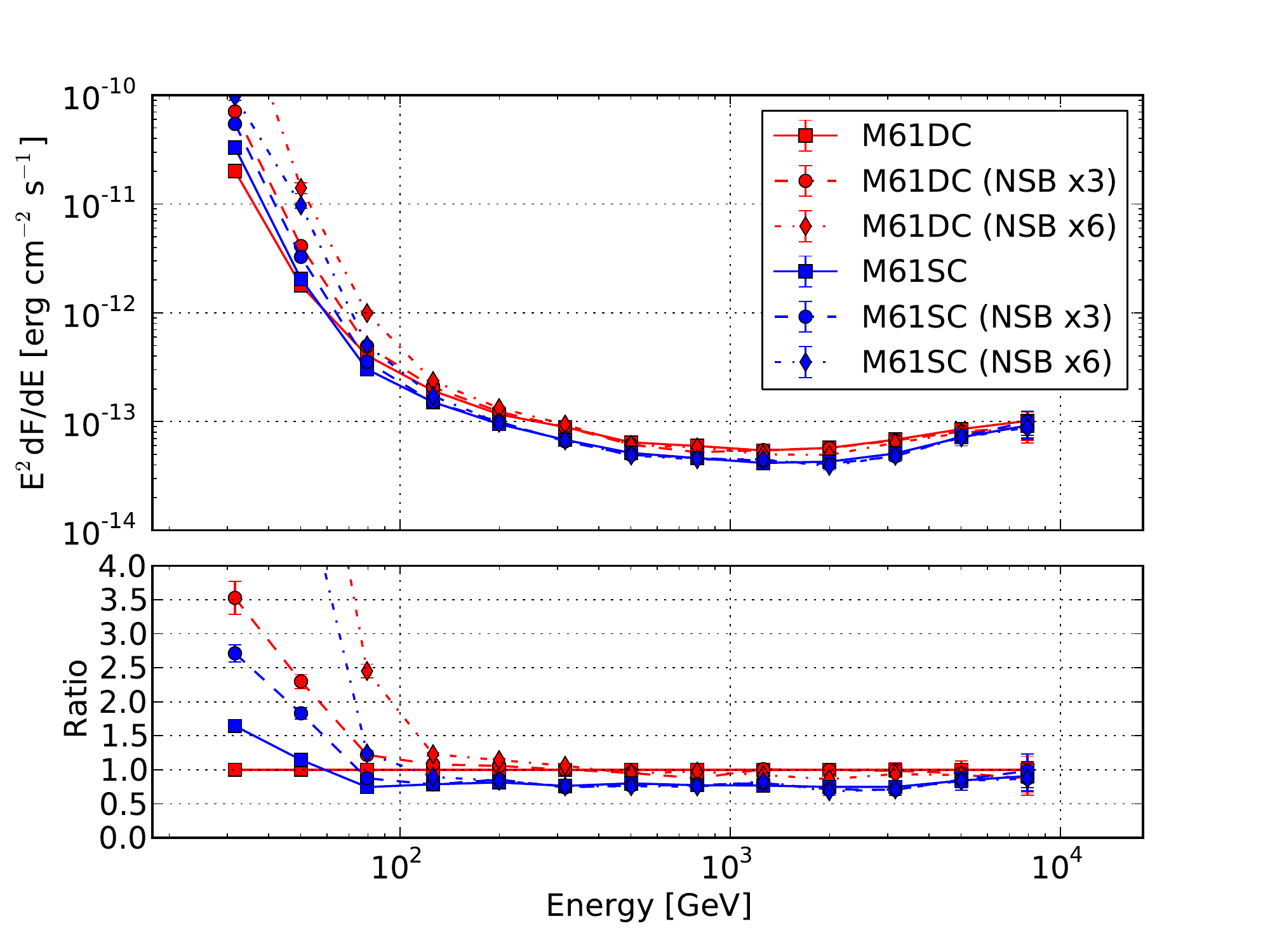}
\includegraphics[width=.49\textwidth, keepaspectratio]{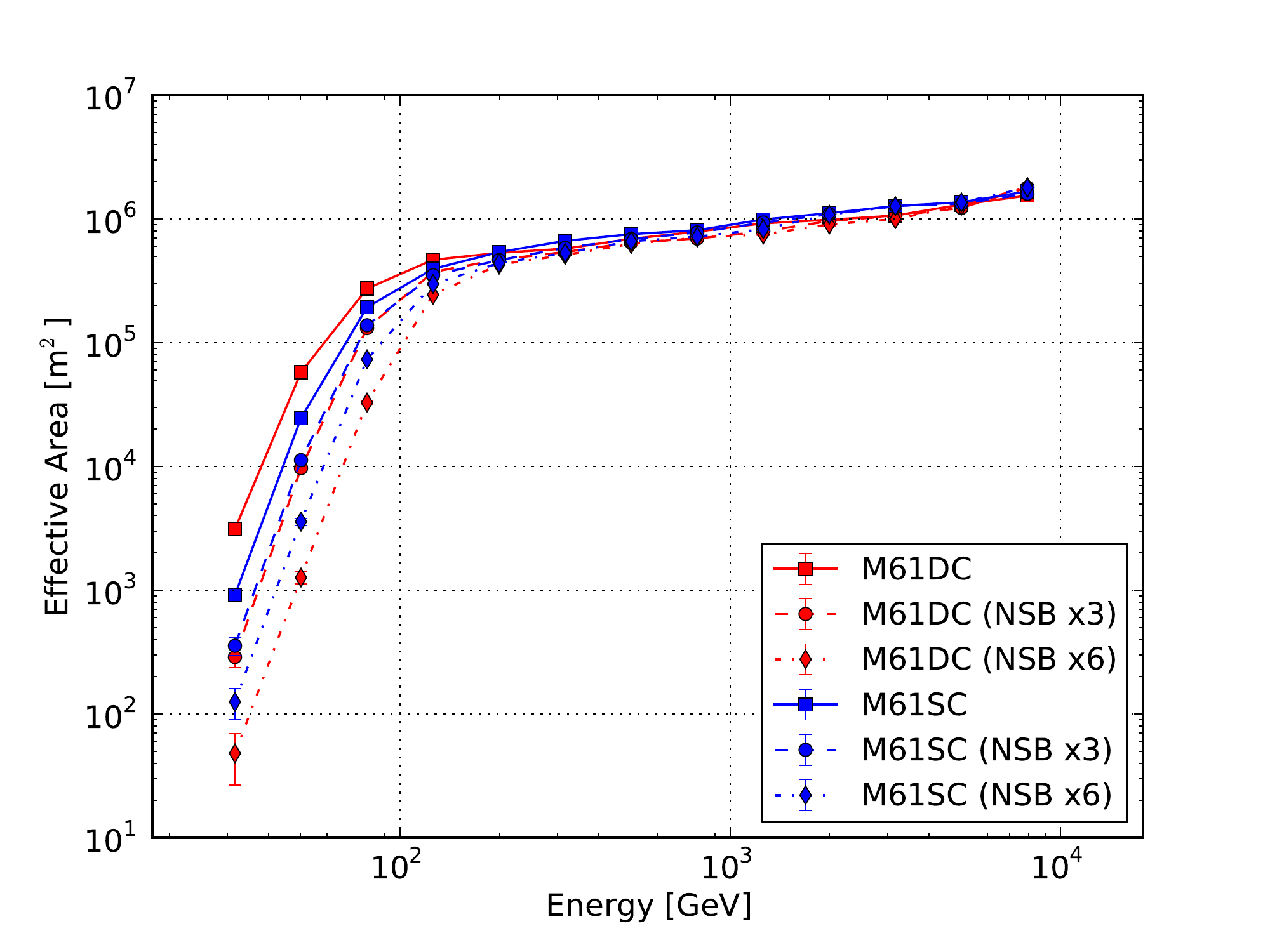}
\caption{\label{FIG:NSB_SENSITIVITY} Performance of arrays M61DC (red)
  and M61SC (blue) simulated with a baseline NSB flux of 365 MHz
  deg$^{-2}$ m$^{-2}$ (circles and solid lines) and an NSB flux that
  is 3 (dashed) and 6 (dash-dotted) times higher than the baseline
  value.  \textbf{Left:} Differential point-source sensitivity for a
  50~h observation time.  \textbf{Right:} Gamma-ray effective area
  after \textit{point-source} cuts.}
\end{figure*}

Night-sky background (NSB) is caused by the presence of light sources
such as stars, the Moon, and artificial light pollution and represents
an irreducible background for the reconstruction and analysis of
gamma-ray air showers.  Because the Cherenkov photons detected in a
single pixel have an intrinsic arrival time dispersion of 3--6~ns,
IACTs can significantly reduce the NSB by integrating the Cherenkov
signal in a narrow time window (typically with $\Delta$T $\sim$
10~ns).  The integrated NSB level thus depends on both the NSB rate as
well as the size of the window used for signal integration.  The need
for a small integration window motivates camera designs with high
bandwidth readout electronics which would allow the integration window
to be made as small as possible.  The impact of the NSB rate on the
sensitivity of the array is also important when considering possible
observatory sites and performing observations during
moonlight. Moonlight observations can considerably increase the duty
cycle of the observatory although the exact amount of observation time
gained depends on the NSB rate that the individual telescope can
handle.

We studied the impact of NSB on the performance of the array by
performing simulations with three NSB flux levels: a baseline flux
level with an integral flux of 365 MHz deg$^{-2}$ m$^{-2}$ and NSB
fluxes that are 3 and 6 times higher than the baseline flux.  As
described in Section~\ref{subsec:det_model}, the baseline flux level
corresponds to the expected night-sky intensity for a dark,
extragalactic field.
The higher NSB fluxes are representative of either a higher NSB rate
due to operation under high night-sky brightness (moonlight) or a
longer effective integration window.  A higher NSB rate also increases
the rate of accidental triggers and would require a higher trigger
threshold in order to maintain the accidental trigger rate at a
constant level.  For this study we kept the trigger threshold fixed at
its nominal value and only examine the impact of the NSB on the pixel
SNR.

Figure~\ref{FIG:NSB_SENSITIVITY} shows the comparison of the
point-source sensitivity and gamma-ray effective area of arrays M61SC
and M61DC simulated at the three NSB levels.  The NSB level only
appreciably affects the sensitivity below 300~GeV where the SNR of the
shower image is lowest.  Most of the reduction in sensitivity is a
result of the lower reconstruction efficiency as low SNR images are
removed at the cleaning stage of the analysis.  Remarkably the
reduction in sensitivity is much more pronounced in the case of larger
pixels (DC-like telescope).  In case of the SC telescope design,
operation at a six times higher NSB rate would only degrade the sensitivity
below about 100 GeV and only up to a factor of two. The DC-like design
would also suffer significant sensitivity loss only below about 100
GeV but to a much greater degree.  Here it should be noted that the
sensitivity advantage of the DC telescopes below 50 GeV under low NSB
is lost in case of three times increased NSB and that the SC design is
better for six times higher NSB at all energies.

\subsection{Number of Telescopes in the Array}\label{subsec:numtels}

\begin{figure*}[tb]
\includegraphics[width=.49\textwidth]{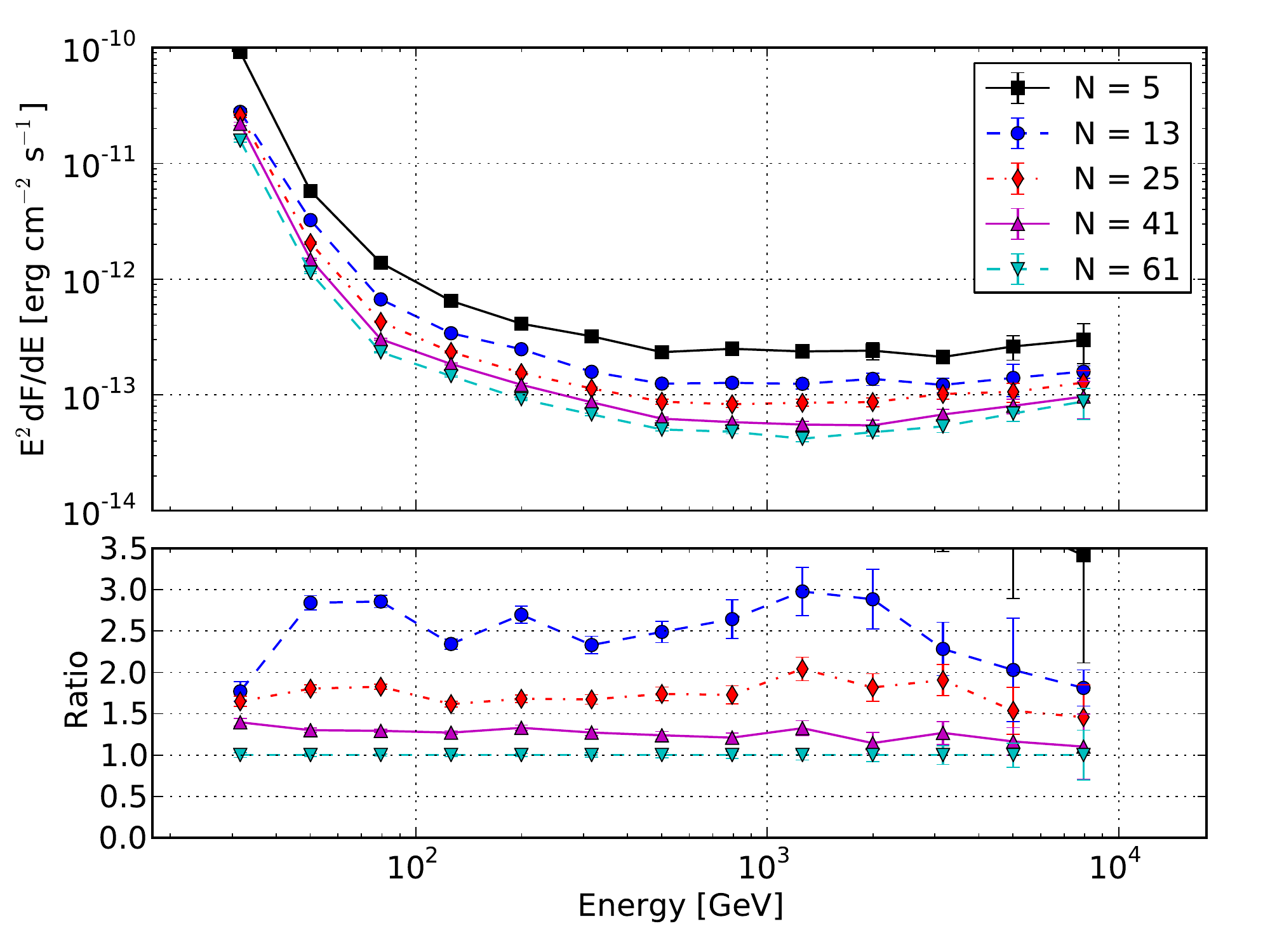}
\includegraphics[width=.49\textwidth]{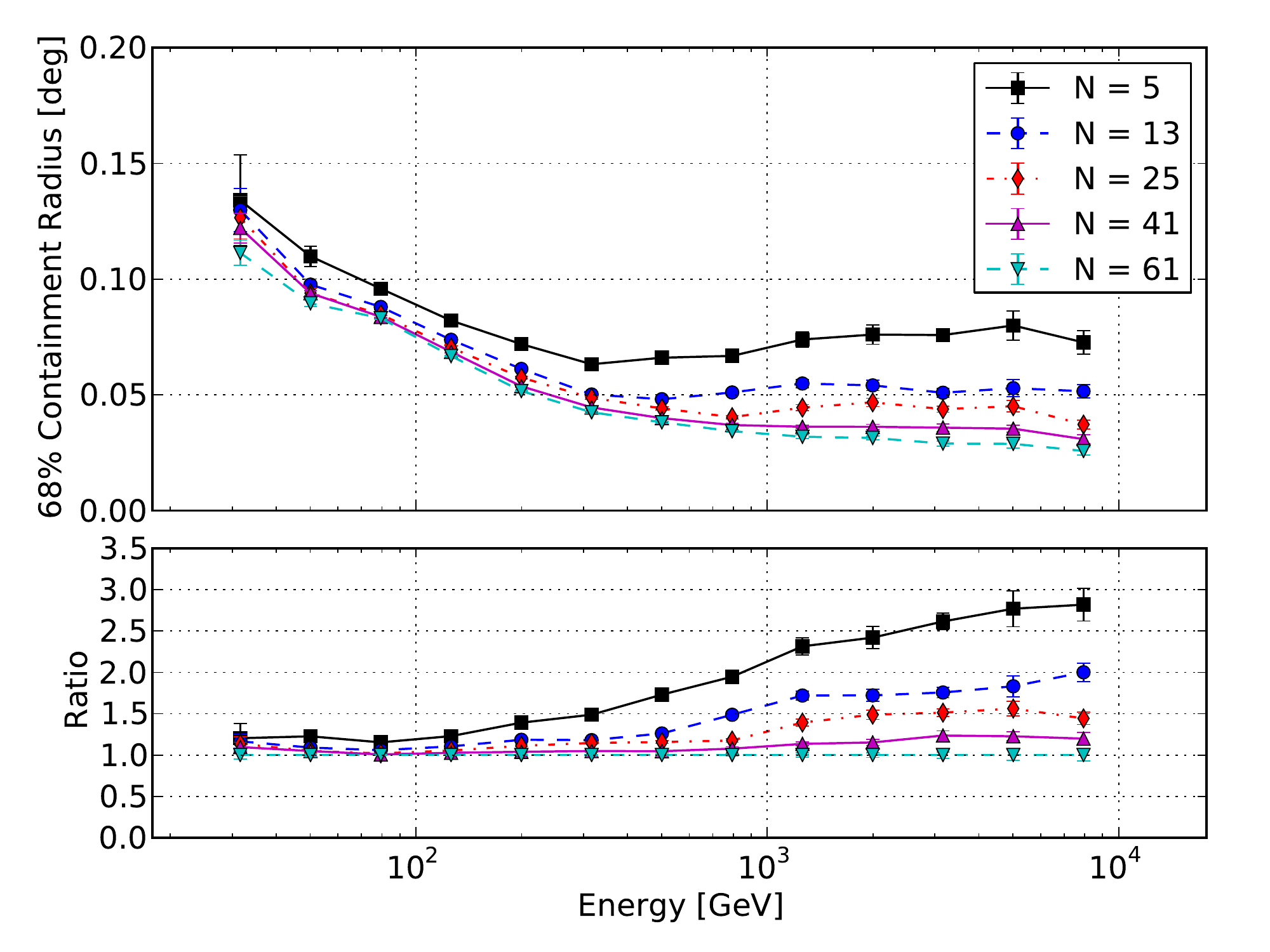}
\caption{\label{FIG:NTELS_SENSITIVITY} \newtext{Performance of array layouts
  with telescope number ($\ntel$) from 5 to 61.  All arrays are
  simulated with a 120~m inter-telescope separation and the same
  telescope model as Array~M61.} \textbf{Left:}
  Differential point-source sensitivity for a 50~h observation time.
  \textbf{Right:} 68\% containment angle of the gamma-ray PSF after
  applying \textit{point-source} cuts.}
\end{figure*}

One of the most important parameters concerning the sensitivity of an
IACT array is the number of telescopes.  A larger number of telescopes
increases both the total effective area for triggering and
reconstructing gamma-ray showers but also increases the average number
of telescopes that participate in the reconstruction of each shower.
Increasing the number of telescopes leads to better point-source
sensitivity and an improved gamma-ray PSF.

Figure~\ref{FIG:NTELS_SENSITIVITY} compares the performance of arrays
with between 5 and 61 telescopes.  We investigate the scaling relation
of the improvement in sensitivity with increasing number of telescope.
In the limit of an infinite array the point-source sensitivity should
scale with the number of telescopes as $\ntel^{1/2}$.  However we
observe an increase of sensitivity that is slightly better than the
$\ntel^{1/2}$ at all energies.  This emphasizes that in the case of
small telescope arrays increasing the number of telescopes yields
larger improvements as compared to the case of extending large arrays.
Adding 36 telescopes to a 25 telescope array improves the sensitivity
by a factor of $\sim$1.7-1.8.

In contrast to the point-source sensitivity, the gamma-ray PSF
improves non-uniformly over energy with increasing telescope
number. The best improvement is seen at larger energies while at $E <
300$~GeV the improvement is only clearly visible between 5 and 13
telescopes. At high energies the curves in
Fig.~\ref{FIG:NTELS_SENSITIVITY} show a clearer separation
demonstrating that more telescopes help to better localize the showers
above 1 TeV. The energy dependency has its origin in the fact that
only high energy showers produce enough light to trigger distant
telescopes. Thus larger arrays with more telescopes benefit at high
energies because the average number of telescopes participating in the
shower reconstruction is increased.  In the case of lower energy
showers, the number of telescopes contributing to the shower analysis
is limited by the telescope spacing and not the absolute number of
telescopes in the array.  Increasing the footprint of the array also
increases the parallax between telescopes observing an uncontained
shower.  The larger parallax yields a better shower direction
reconstruction and further improves the reconstruction performance at
high energies.

\subsection{Comparison of Array Designs for CTA}\label{subsec:benchmark_arrays_compare}

After studying the effect of individual telescope parameters on the
point-source sensitivity and gamma-ray PSF, we now compare realistic
telescope designs against each other to find a suitable array design
for CTA.  To achieve a comprehensive comparison we investigate all the
benchmark arrays defined in Table~\ref{TABLE:BENCHMARK_ARRAYS} and
give a quantitative comparison between the different telescope
layouts. Among the benchmark arrays are also two more theoretically
interesting cases.  Array~L61 is representative of the theoretical
limit for an IACT array if the budget is not limited and only the
number of telescopes is fixed.  In a similar fashion, Array~L5 is
included to study the contribution of an LST subarray with 3--5
telescopes such as currently considered for the baseline configuration
of CTA.

\begin{figure*}[tbp]
\centering
\includegraphics[width=.49\textwidth]{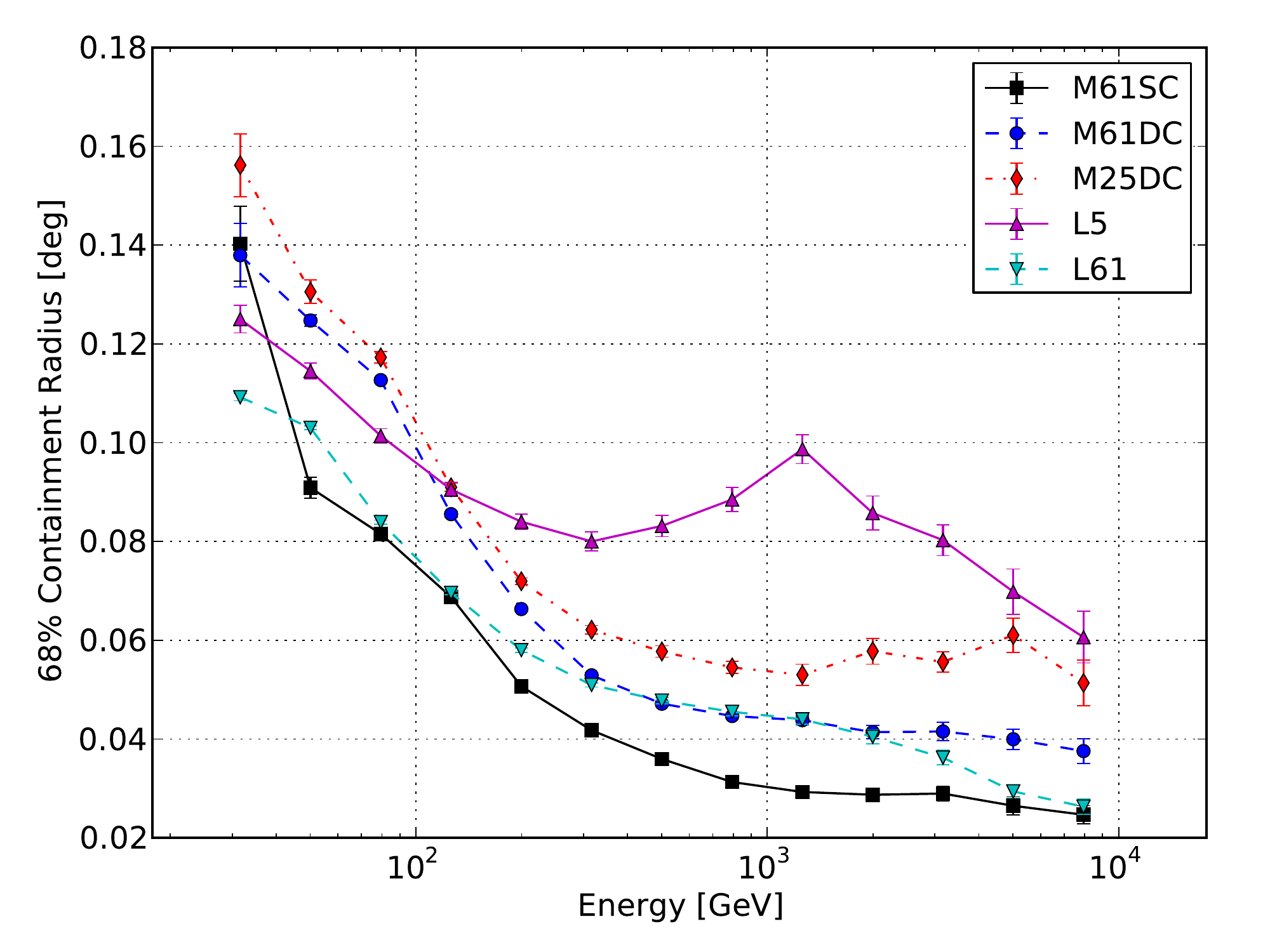}
\includegraphics[width=.49\textwidth]{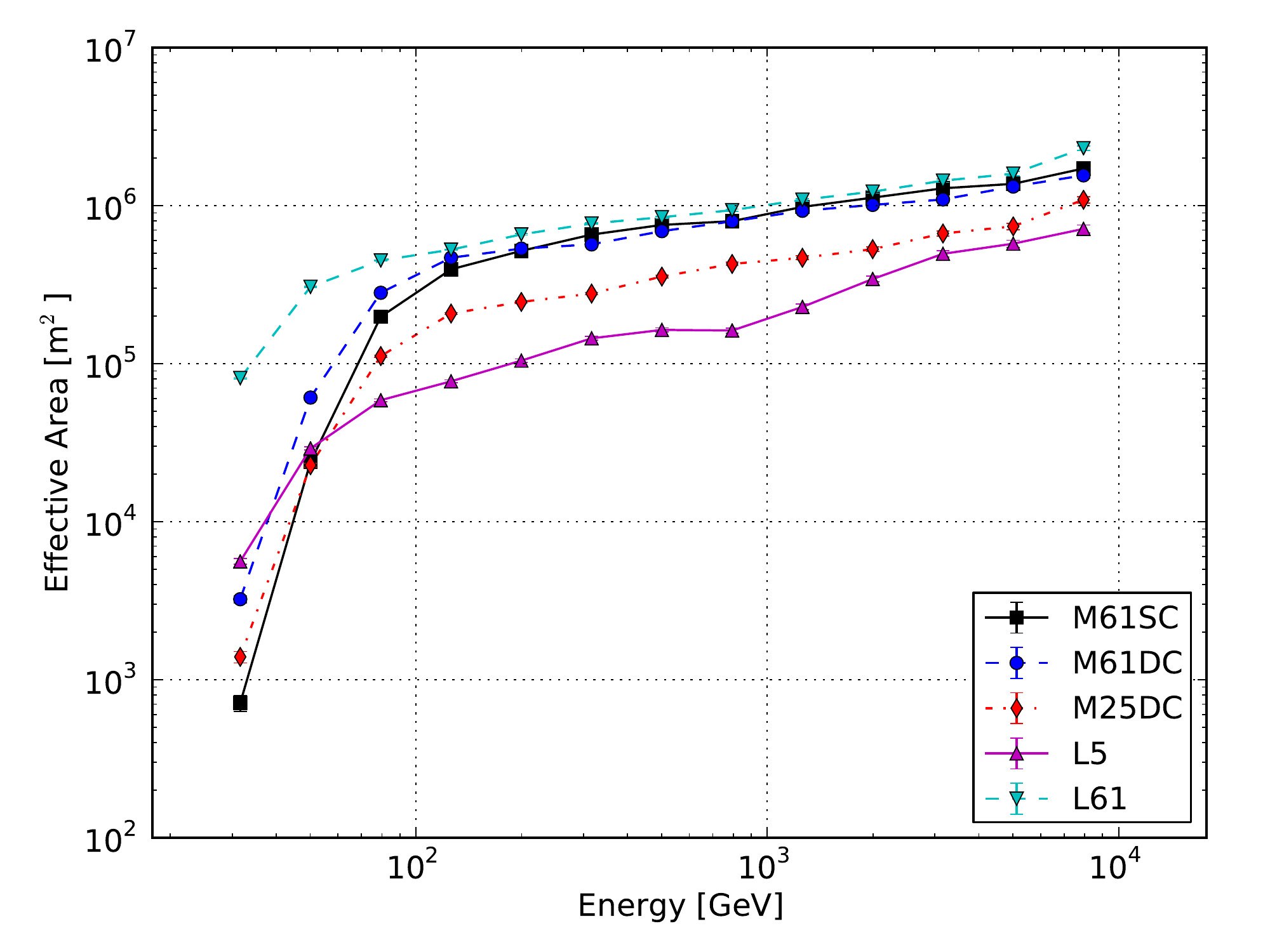}\\
\includegraphics[width=.49\textwidth]{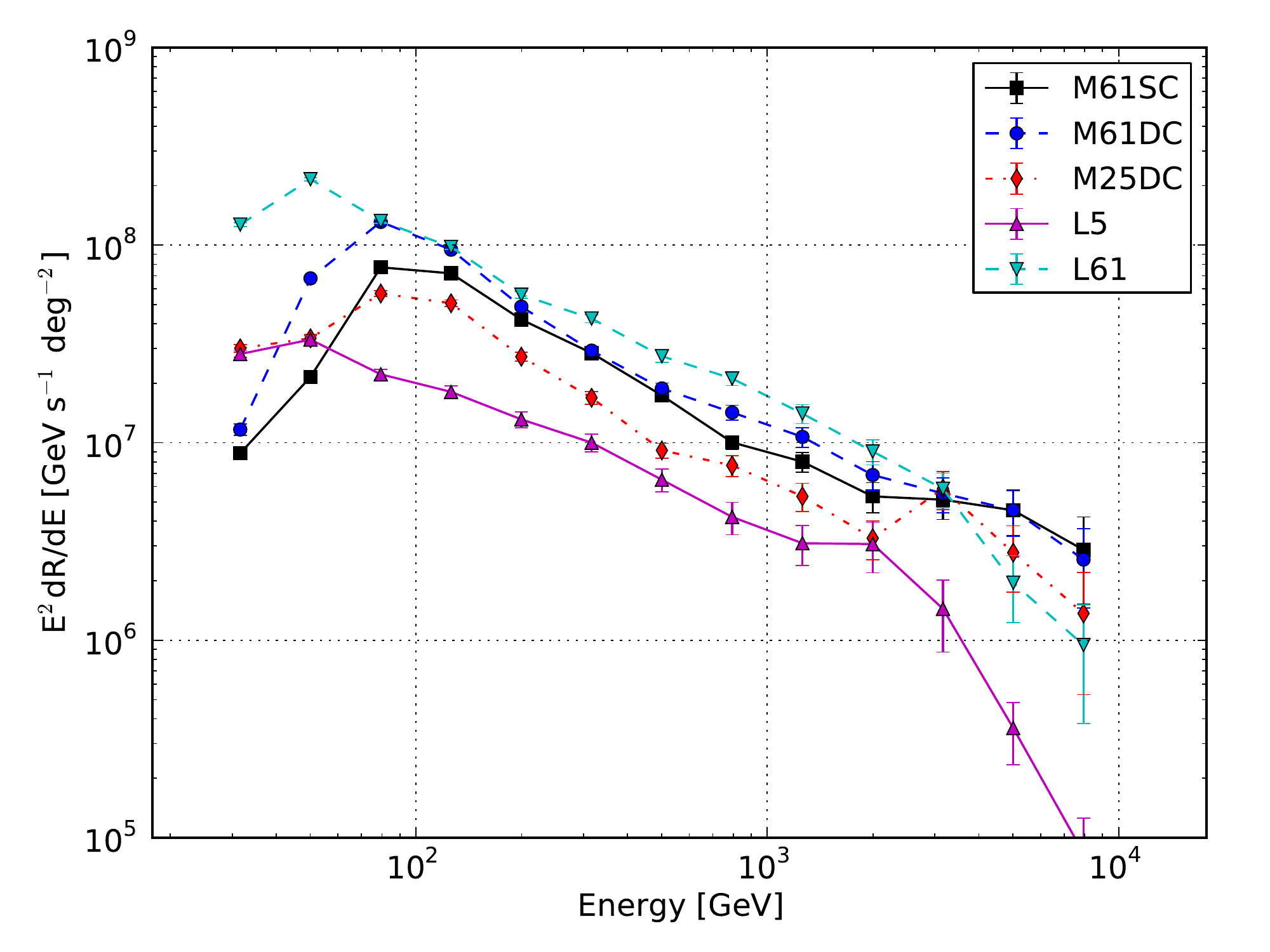}
\includegraphics[width=.49\textwidth]{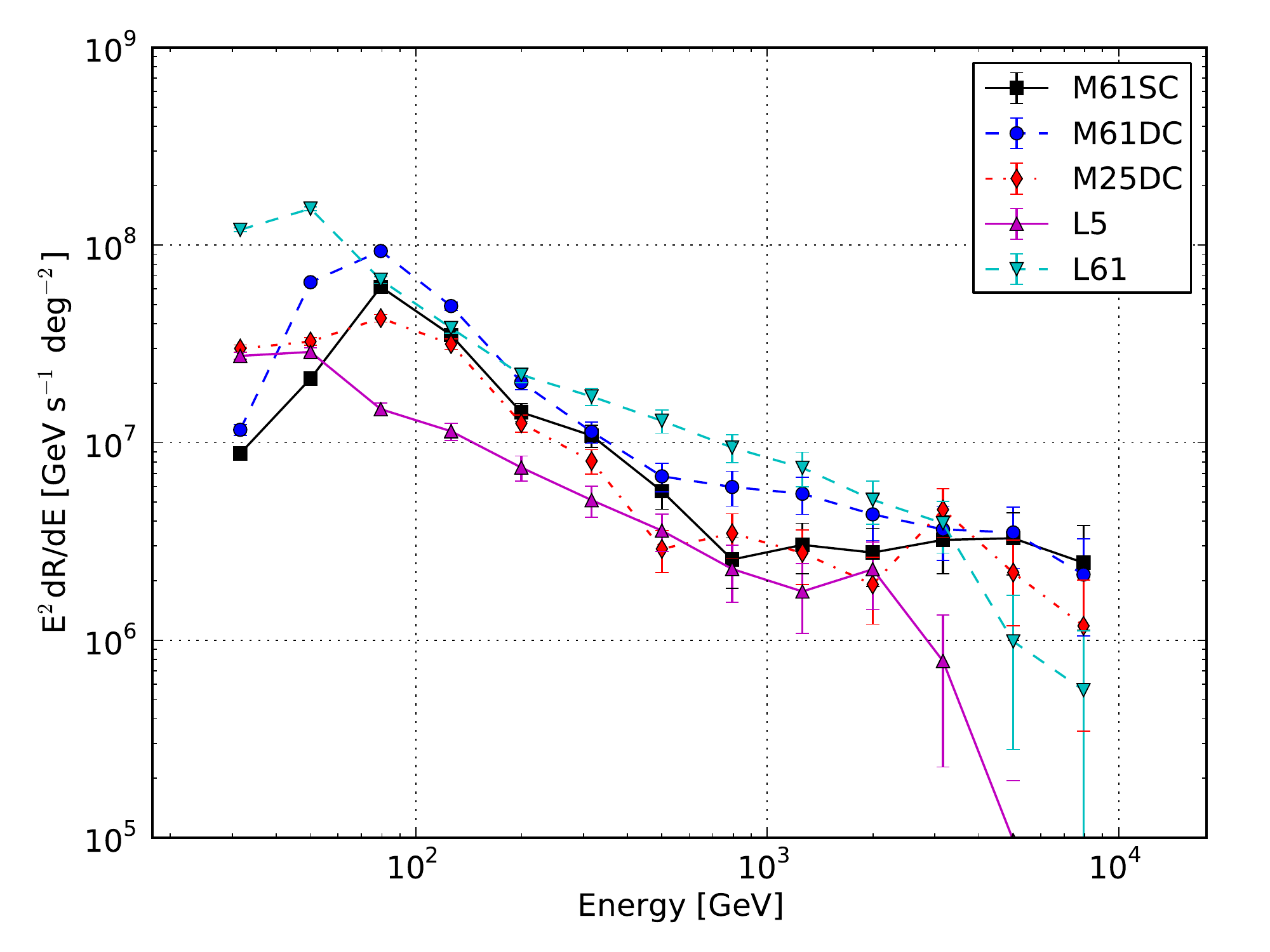}\\
\includegraphics[width=.49\textwidth]{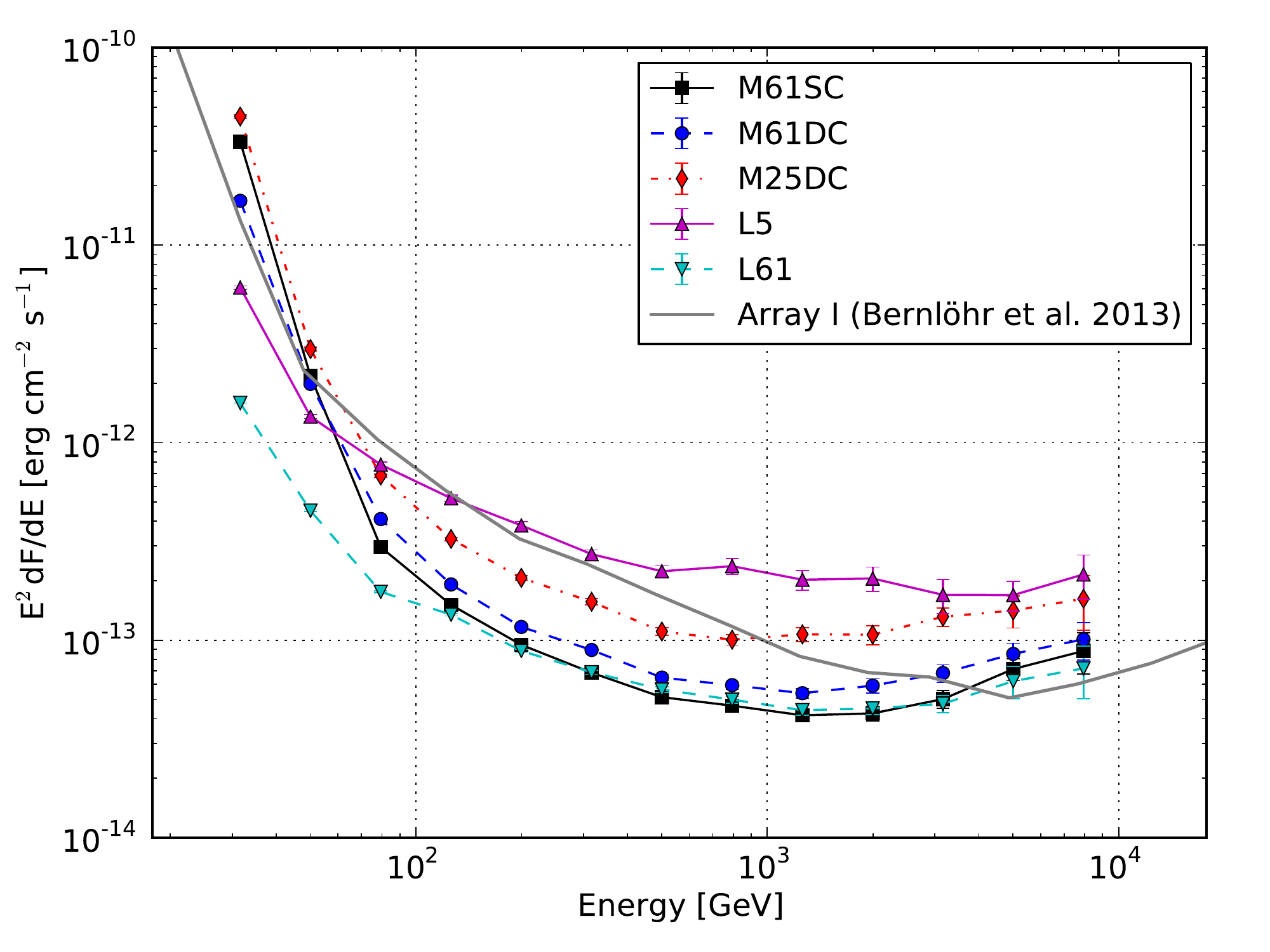}
\includegraphics[width=.49\textwidth]{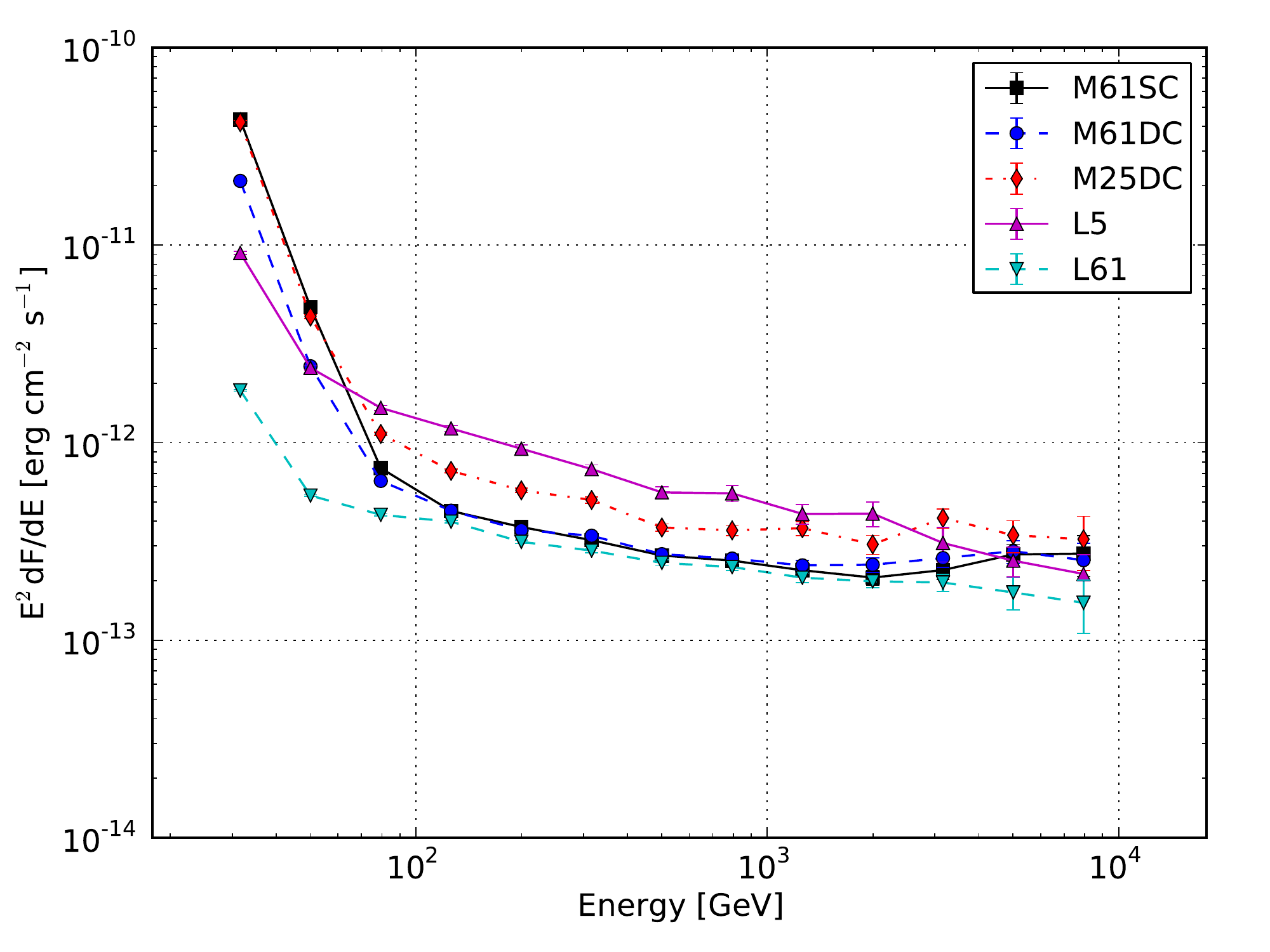}
\caption{\label{FIG:BENCHMARK_SENSITIVITY}Performance of benchmark
  arrays: M61SC, M61DC, M25DC, L5, and L61.  \textbf{Top Left:} 68\%
  containment angle of the gamma-ray PSF after applying
  \textit{point-source} cuts.  \textbf{Top Right:} Gamma-ray effective
  area after \textit{point-source} cuts.  \textbf{Middle Left:}
  Differential rate of the total cosmic-ray background (protons and
  electrons) after \textit{point-source} cuts.  \textbf{Middle Right:}
  Differential rate of protons after \textit{point-source} cuts.
  \textbf{Bottom Left:} Differential point-source sensitivity for a
  50~h observation time.  Shown as the solid gray line is the
  differential sensitivity of Array~I from \cite{2013APh....43..171B}
  evaluated with the most sensitive analysis at each energy from the
  four alternative analyses presented in that work (MPIK, IFAE, SAM,
  and Paris-MVA).  \textbf{Bottom Right:} Differential diffuse-source
  sensitivity ($D = 0.5^\circ$) for a 50~h observation time.}
\end{figure*}

Fig.~\ref{FIG:BENCHMARK_SENSITIVITY} shows that Array~M61SC is more
sensitive than Array~M61DC at all energies above 50 GeV, where the
increase in sensitivity is about 30\%.  In addition to the improvement
in point-source sensitivity, the M61SC array also has a better
gamma-ray PSF at all energies.  The smaller gamma-ray PSF would help
to determine the morphology of extended sources and help to separate
point sources. These additional important effects are difficult to
assess quantitatively because they heavily rely on the source
population and properties in the sky.  The diffuse source is simulated
as an uniformly extended disk with a radius of $0.5^{\circ}$.  The
diffuse-source sensitivity does not show any improvement of the M61SC
array over the M61DC array because the gamma-ray PSF does not help to
reduce the background but still the M61SC would enable for a
non-uniform source to asses the morphology better than
Array~M61DC. The diffuse source sensitivity emphasizes that the
sensitivity gain of the SC array compared to the DC array comes almost
entirely from the PSF improvement while the improvement in the
background rejection power is marginal.

Array~M25DC is representative of \newtext{the MST subset of the CTA
  array design} as it was planned without a US contribution.
Comparing the Array~M61SC and Array~M61DC to the M25DC baseline
configuration, it is obvious that adding MST telescopes will improve
the sensitivity of CTA in the key energy range between 100~GeV and
about 1~TeV by about a factor two regardless of their design. This is
expected from the fact that the sensitivity is improved by the
addition of telescopes, as shown in Fig.~\ref{FIG:NTELS_SENSITIVITY}.

We also compared the point-source sensitivity of our benchmarks arrays
with Array~I from \cite{2013APh....43..171B}.  For Array~I we use a
differential sensitivity curve that is constructed by taking the best
sensitivity in each energy bin from the four alternative analyses
presented in that work (MPIK, IFAE, SAM, and Paris-MVA).  Although the
simulations in this paper were performed with different telescope
models and a different detector simulation package, this array is
representative of the expected performance of the baseline CTA
concept.  In the central energy range from 100~GeV to 3~TeV, Arrays
M61DC and M61SC provide a factor of 3--4 improvement in point-source
sensitivity relative to Array~I.  This improvement can be primarily
attributed to the increase in the number of MSTs from 18 to 61.
Array~I performs better at energies below 50~GeV and above 3 TeV as
compared to Array~M25DC and even Arrays M61DC and M61SC.  This
improvement can be attributed to the inclusion of 56~SSTs and 3~LSTs
in Array~I.  Array~L5 was simulated with five LSTs very similar to the
ones included in Array~I, and the sensitivity curve obtained for L5
matches very well the sensitivity of Array~I at low energies,
demonstrating that the advantage of Array~I at low energies does in
fact come from the LSTs.

Finally Array~L61 yielded only an improvement below 100 GeV, making
such an array impractical based on the large cost differential between
a single MST and LST.  However the performance of this array shows
what is theoretically achievable in the case of no budget constraints.
Array~M61SC provides comparable sensitivity to Array~L61 at all
energies above 100~GeV and thus is very close to the performance of an
ideal array in this energy range.

In case of the diffuse source sensitivity the number of telescopes is
the found to be the most important factor. Again the addition of MSTs
of either type (SC or DC) would result in a considerable improvement
compared to M25DC (similar to Array~I) in the whole energy range.
However the improvement is slightly less significant when
compared to the relative improvement in the point-source sensitivity.

%% file: section5.tex
\section{Conclusions}
\label{sec:conclusions}

This paper describes a new simulation and analysis chain that is used
to study and compare array and telescope design concepts for CTA.  We
specifically focus on the role of MST arrays which are optimized for
performance in the core energy range of CTA between 100~GeV and 1~TeV.   
The simplified detector model used for this study allows for
investigation of a wide range of telescope parameters: effective light
collection area, optical PSF, camera pixel size, effective camera
trigger threshold, and effective integration window in time.  The
simplified telescope description allows us to isolate the most
important telescope design characteristics and fully explore their
influence on the performance of the full array.  Realistic telescope
designs can be mapped to our simplified detector model by choosing
telescope parameters that are matched to the physical characteristics
of each design (mirror area, focal length, photosensor efficiency,
etc.).  This paper also investigates several aspects of the array
geometry optimization including the impact of the number of telescopes
and their separation on array performance.

A benchmark telescope array was used to assess the influence of each
of the telescope and array parameters.  Performance is evaluated for
nominal observing conditions corresponding to a zenith angle of 20$^\circ$
and an NSB rate computed for a dark extragalatic field.  We also
examined the influence of the GF and higher NSB rates.
Under all conditions, an optimized analysis is performed using a
likelihood reconstruction based on simulated image templates and BDTs
for signal extraction.

The likelihood reconstruction based on simulated templates offers a
factor of two improvement in point source sensitivity (30--40\%
improvement in gamma-ray PSF), as well as a reduced energy threshold
relative to image moment-based analysis.  The likelihood
reconstruction takes advantage of the possibility of fully resolving
showers with a finely pixelated camera.  This technique, coupled with
BDTs for event selection, allowed us to compare arrays very close to
their maximum achievable sensitivity.

We find that the substantial improvements in both the gamma-ray
point-source sensitivity and angular resolution of an IACT array can
be realized by telescopes with imaging resolution better than
current-generation IACT designs.  We find a 30--40\% improvement in
the gamma-ray point-source sensitivity between 100~GeV and 3~TeV when
the telescope pixel size is reduced from 0.16$^\circ$ to 0.06$^\circ$.
The gain in point-source sensitivity comes primarily from the
improvement in the gamma-ray angular reconstruction enabled by the
higher resolution imaging of the shower axis.  Over the same energy
range, the performance of an MST array is much less sensitive to the
telescope light collection area and trigger threshold.  We find that
these parameters are important in determining the array energy
threshold but have little influence on the array performance above the
threshold energy.

With higher resolution shower images, the GF becomes more relevant
than ever for the sensitivity of an IACT array.  To determine the
impact of the GF, we compared the same array simulated with values of
$B_\perp$ between $0~\mu\mathrm{T}$ and $20.7~\mu\mathrm{T}$.  For an
MST array, the impact of the GF is largest around 100~GeV where the
point-source sensitivity is reduced by 50\%.  The GF should be an
important factor in selecting a site for future arrays and possibly
for designing an observing strategy.

Increasing the number of telescopes in the array expands the effective
area, improves reconstruction, and increases background rejection
capabilities.  The sensitivity can be improved faster at very low and
very high energies by adding LSTs and SSTs.  However, in the energy
range between a few hundred GeV and tens of TeV, expanding the MST
array efficiently improves the sensitivity, regardless of the
telescope design.  In the limit of a finite array for which
uncontained showers constitute a significant fraction of the total
reconstructed event sample, the improvement in point-source
sensitivity scales faster than the square root of the number of
telescopes between 300~GeV and 3~TeV.  If the baseline CTA design is
expanded to include 36 more MSTs, the point-source sensitivity in the
core energy is improved by a factor of two.

When considering arrays with the same number of telescopes, we find
that the SC telescope design yields a 30-40\% improvement in
point-source sensitivity over the DC telescope design because of its
superior imaging resolution. The DC telescope on the other hand has a
slightly lower energy threshold resulting in better point source
sensitivity below 75 GeV. The improved performance in a wide energy
range from the SC design warrants further investigation.  The improved
sensitivity reduces the total exposure time required for every science
topic, while the smaller gamma-ray PSF additionally helps with source
confusion and morphology studies.  The higher resolution shower images
of the SC-like telescopes are also much less affected by noise from
NSB.  This translates to a much lower energy threshold during brighter
sky conditions, e.g. in the galactic plane.  This may lead to a much
higher effective duty cycle since observations can be continued into
brighter moon phases without sacrificing the low energy regime.

While the SC-like array is more sensitive in comparison to the DC-like
design, no \newtext{SC MST telescope} has yet been built.
Construction of an SC prototype at the site of VERITAS is under way.
This prototype offers a chance to study the performance of the SC
optics in realistic circumstances.  This experience should also
provide a more realistic cost model for the two-mirror systems.

At this point in the design of CTA, it is unlikely that all MST
telescopes would be of the SC design.  If the SC prototype can be
built successfully and cost-efficiently, the baseline CTA array could
be expanded to include an additional number of SC MSTs.  The study of
mixed arrays is ongoing.  No matter which optical design is chosen,
expanding the MST arrays offers significant benefits for the
performance of CTA in the central energy range between 100~GeV and 1~TeV.

%% file: acknowledgements.tex
\section*{Acknowledgements}

We thank Brian Humensky, Emiliano Carmona, and the anonymous referees
for their valuable comments.  This work was supported in part by the
Department of Energy contract DE-AC02-76SF00515.